\documentclass[manuscript,screen,review=false]{acmart}
\AtBeginDocument{%
  \providecommand\BibTeX{{%
    \normalfont B\kern-0.5em{\scshape i\kern-0.25em b}\kern-0.8em\TeX}}}

\setcopyright{acmcopyright}
\copyrightyear{2022}
\acmYear{2022}
\acmDOI{XXXXXXX.XXXXXXX}

\acmJournal{TECS}
\acmPrice{15.00}
\acmISBN{978-1-4503-XXXX-X/18/06}

\usepackage{amsmath}
\usepackage{graphicx}

\usepackage{comment}
\usepackage{enumerate}
\usepackage{amsthm}
\usepackage{amsmath}
\usepackage{xfrac}
\usepackage[misc]{ifsym}
\usepackage{tabularx,booktabs}
\usepackage{times}
\usepackage{soul}
\usepackage{url}
\usepackage[utf8]{inputenc}
\usepackage{caption}
\usepackage{subcaption}
\usepackage{booktabs}
\usepackage{algorithm}
\usepackage{multirow}
\usepackage[absolute,overlay]{textpos}

\newcommand{\inputdomain}{\mathcal{X}}

\newcommand{\network}{{\mathcal M}}

\usepackage{hyperref}

\newtheorem{remark}{Remark}
\newtheorem{assumption}{Assumption}
\newtheorem{definition}{Definition}

\newcommand*\diff{\mathop{}\!\mathrm{d}}
\newcommand*\myExp{{\mathbb E}}
\newcommand*\myVar{{\mathbb V}}

\usepackage[xindy,nonumberlist,nogroupskip]{glossaries}
\makeglossaries
\newacronym{LES}{LES}{Learning-Enabled Systems}
\newacronym{AUV}{AUV}{Autonomous Underwater Vehicles}
\newacronym{RAM}{RAM}{Reliability Assessment Models}
\newacronym{OP}{OP}{Operational Profile}
\newacronym{ACU}{ACU}{Average Cell Unastuteness}
\newacronym{DL}{DL}{Deep Learning}
\newacronym{ML}{ML}{Machine Learning}
\newacronym{VAE}{VAE}{Variational Auto-Encoders}
\newacronym{MLE}{MLE}{Maximum Likelihood Estimation}
\newacronym{AE}{AE}{Adversarial Example}
\newacronym{VnV}{V\&V}{Verification and Validation}
\newacronym{KDE}{KDE}{Kernel Density Estimation}
\newacronym{PDF}{PDF}{Probability Density Function}
\newacronym{RAS}{RAS}{Robotics and Autonomous Systems}
\newacronym{FTA}{FTA}{Fault-Tree Analysis}
\newacronym{HAZOP}{HAZOP}{Hazard and Operability Study}
\newacronym{CLT}{CLT}{Central Limiting Theorem}
\newacronym{GALE}{GALE}{Globally At Least Equivalent}
\newacronym{pmi}{\textit{pmi}}{probability of misclassification per random input}
\newacronym{SMC}{SMC}{Simple Monte Carlo}
\newacronym{ROS}{ROS}{Robot Operating System}
\newacronym{CAE}{CAE}{Claims-Arguments-Evidence}
\newacronym{GSN}{GSN}{Goal Structuring Notation}
\newacronym{ALARP}{ALARP}{As Low As Reasonably Practicable}
\newacronym{SE}{SE}{Substantially Equivalent}
\newacronym{MEM}{MEM}{Minimum Endogenous Mortality}
\newacronym{DVL}{DVL}{Doppler Velocity Log}
\newacronym{IMU}{IMU}{Inertial Measurement Unit}
\newacronym{PID}{PID}{Proportional-Integral-Derivative}
\newacronym{UGV}{UGV}{Unmanned Ground Vehicle}

\begin{document}

\title[Reliability Assessment and Safety Arguments for ML components]{Reliability Assessment and Safety Arguments for Machine Learning Components in System Assurance}

\author{Yi Dong}
\email{yi.dong@liverpool.ac.uk}
\author{Wei Huang}
\authornotemark[0]
\authornote{Both authors contributed equally to this research.}
\email{w.huang23@liverpool.ac.uk}
\affiliation{%
	\institution{Department of Computer Science, University of Liverpool}
	\streetaddress{Ashton Building, Ashton Street}
	\city{Liverpool}
	\country{U.K.}
	\postcode{L69 3BX}
}

\author{Vibhav Bharti}
\email{v.bharti@hw.ac.uk}
\affiliation{%
	\institution{School of Engineering \& Physical Sciences, Heriot-Watt University}
	\city{Edinburgh}
	\country{U.K.}
	\postcode{EH14 4AS}
}

\author{Victoria Cox}
\email{vcox@dstl.gov.uk}
\author{Alec Banks}
\email{abanks@dstl.gov.uk}
\affiliation{%
	\institution{Defence Science and Technology Laboratory}
	\city{Salisbury}
	\country{U.K.}
	\postcode{SP4 0JQX}
}

\author{Sen Wang}
\email{sen.wang@imperial.ac.uk}
\affiliation{%
	\institution{Department of Electrical and
Electronic Engineering, Imperial College London
}
	\city{London}
	\country{U.K.}
	\postcode{SW7 2BX}
}

\author{Xingyu Zhao}
\email{xingyu.zhao@liverpool.ac.uk}
\author{Sven Schewe}
\email{sven.schewe@liverpool.ac.uk}
\author{Xiaowei Huang}
\email{xiaowei.huang@liverpool.ac.uk}
\affiliation{%
	\institution{Department of Computer Science, University of Liverpool}
	\streetaddress{Ashton Building, Ashton Street}
	\city{Liverpool}
	\country{U.K.}
	\postcode{L69 3BX}
}

\renewcommand{\shortauthors}{Dong and Huang, et al.}

\begin{abstract}
 The increasing use of Machine Learning (ML) components embedded in autonomous systems---so-called Learning-Enabled Systems (LESs)---has resulted in the pressing need to assure their functional safety.
 As for traditional functional safety, the emerging consensus within both, industry and academia, is to use assurance cases for this purpose. 
 Typically assurance cases support claims of reliability in support of safety, and can be viewed as a structured way of organising arguments and evidence generated from safety analysis and reliability modelling activities.
 While such assurance activities are traditionally guided by consensus-based standards developed from vast engineering experience, LESs pose new challenges in safety-critical application due to the characteristics and design of ML models.
 In this article, we first present an overall assurance framework for LESs with an emphasis on quantitative aspects, e.g., breaking down system-level safety targets to component-level requirements and supporting claims stated in reliability metrics.
 We then introduce a novel model-agnostic Reliability Assessment Model (RAM) for ML classifiers that utilises the operational profile and robustness verification evidence. 
 We discuss the model assumptions and the inherent challenges of assessing ML reliability uncovered by our RAM and propose solutions to practical use.
 Probabilistic safety argument templates at the lower ML component-level are also developed based on the RAM.
 \textcolor{black}{Finally, to evaluate and demonstrate our methods, we not only conduct experiments on synthetic/benchmark datasets but also scope our methods with case studies on simulated Autonomous Underwater Vehicles and physical Unmanned Ground Vehicles.}
\end{abstract}


\maketitle
\begin{textblock*}{20cm}(1cm,1cm)
	\textcolor{red}{{\large Accepted by ACM Transactions on Embedded Computing Systems 
		\url{https://dl.acm.org/journal/tecs}}}
\end{textblock*}

\begin{table}[H]
\begin{tabular}{lllll}
\hline
\multicolumn{5}{l}{\textbf{\textcolor{black}{LIST OF ACRONYMS}}}                                                            \\ \hline
 & ACU   & Average Cell Unastuteness        & MEM    & Minimum Endogenous Mortality                      \\
 & AE    & Adversarial Example              & ML     & Machine Learning                                  \\
 & ALARP & As Low As Reasonably Practicable & MLE    & Maximum Likelihood Estimation                     \\
 & AUV   & Autonomous Underwater Vehicles   & OP     & Operational Profile                               \\
 & CAE   & Claims Arguments Evidence        & PDF    & Probability Density Function                      \\
 & CLT   & Central Limiting Theorem         & PID    & Proportional-Integral-Derivative                  \\
 & DL    & Deep Learning                    & $pmi$  & probability of misclassification per random input \\
 & DVL   & Doppler Velocity Log             & RAM    & Reliability Assessment Models                     \\
 & FTA   & Fault-Tree Analysis              & RAS    & Robotics and Autonomous Systems                   \\
 & GALE  & Globally At Least Equivalent     & ROS    & Robot Operating System                            \\
 & GSN   & Goal Structuring Notation        & SE     & Substantially Equivalent                          \\
 & HAZOP & Hazard and Operability Study     & SMC    & Simple Monte Carlo                                \\
 & IMU   & Inertial Measurement Unit        & UGV    & Unmanned Ground Vehicle                           \\
 & KDE   & Kernel Density Estimation        & V\&V & Verification and Validation                       \\
 & LES   & Learning-Enabled Systems         & VAE    & Variational Auto-Encoders                         \\ \hline
\end{tabular}%
\end{table}

\section{Introduction}
\label{sec_introduction}

Industry is increasingly adopting AI/\gls{ML}  algorithms to enhance the operational performance, dependability, and lifespan of products and service -- systems with embedded \gls{ML}-based software components. For such \gls{LES}, in safety-related applications high reliability is essential to ensure successful operations and regulatory compliance. For instance, several fatalities were caused by the failures of \gls{LES} built in Uber and Tesla’s cars. IBM’s Watson, the decision-making engine behind the Jeopardy AI success, has been deemed a costly and potentially deadly failure when extended to medical applications like cancer diagnosis. Key industrial foresight reviews have identified that the biggest obstacle to reap the benefits of \gls{ML}-powered \gls{RAS} is the assurance and regulation of their safety and reliability \cite{lane_new_2016}. Thus, there is an urgent need to develop methods that enable the dependable use of AI/\gls{ML} in critical applications and, just as importantly, to \textit{assess} and \textit{demonstrate} the dependability for certification and regulation.

For traditional systems, safety regulation is guided by well-established standards/policies, and supported by mature development processes and \gls{VnV} techniques. The situation is different for \gls{LES}: they are disruptively novel and often treated as a black box with the lack of validated standards/policies \cite{BKCF2019}, while they require new and advanced analysis for the complex requirements in their safe and reliable function.
Such analysis needs to be tailored to fully evaluate the new character of \gls{ML} \cite{alves_considerations_2018,burton_mind_2020,KKB2019}, despite some progress made recently \cite{huang_survey_2020}. This reinforces the need for not only an overall methodology/framework in assuring the whole \gls{LES}, but also innovations in safety analysis and reliability modelling for \gls{ML} components, which motivate our work.

In this article, we first propose an overall assurance framework for \gls{LES}, presented in \gls{CAE} assurance cases \cite{bloomfield_safety_2010}. While inspired by \cite{bloomfield2021safety}, ours is with greater emphasis on arguing for quantitative safety requirements. 
This is because the unique characteristics of \gls{ML} increase apparent non-determinism \cite{johnson_increasing_2018} that explicitly requires \textit{probabilistic claims} to capture the uncertainties in its assurance \cite{zhao_safety_2020,asaadi_quantifying_2020,bloomfield2020assurance}. 
To demonstrate the overall assurance framework as an \textit{end-to-end} methodology, we also consider important questions on how to derive and validate (quantitative) safety requirements and how to break them down to functionalities of ML components for a given \gls{LES}. Indeed, there should not be any generic, definitive, or fixed answers to those hard questions for now, since AI/ML is an emerging technology that is still heavily in flux. That said, we propose a solution that we believe is the most practical for the moment: we exercise the method \cite{hills2022} which extends the \gls{HAZOP} (a systematic hazards identification method) \cite{swann_twenty_five_1995}, quantitative \gls{FTA} (a common probabilistic root-cause analysis) \cite{lee_fault_1985}, and propose to leverage existing regulation principles to validate the acceptable and tolerable safety risk, e.g., \gls{GALE} to non-AI/ML systems or human performance.

Upon establishing safety/reliability requirements on low-level functionalities of ML components, we build dedicated \gls{RAM}. 
In this article, we mainly focus on assessing the reliability of the \textit{classification function} of the ML component, extending our initial \gls{RAM} in \cite{zhao_assessing_2021} with more practical considerations for scalability.
Our \gls{RAM} explicitly takes the \textit{\gls{OP}} information and \textit{robustness evidence} into account, because: (i) software reliability, as a \textit{user-centred} property, depends on the end-users' behaviours \cite{littlewood_software_2000}, and the \gls{OP} (quantifying how the software will be operated \cite{musa_operational_1993}) should therefore be explicitly modelled in the assessment; (ii) a \gls{RAM} without considering robustness evidence is not convincing given \gls{ML} is notorious to be unrobust. To the best of our knowledge, our \gls{RAM} is the first to consider both, the \gls{OP} and robustness evidence.
It is inspired by partition-based testing \cite{hamlet_partition_1990,pietrantuono_reliability_2020}, operational/statistical testing \cite{strigini_guidelines_1997,zhao_assessing_2020} and \gls{ML} robustness evaluation \cite{webb_statistical_2019}.
Our \gls{RAM} is \textit{model-agnostic} and designed for \textit{pre-trained} ML models, yielding estimates of, e.g., expectation or confidence bounds on the metric \textit{\gls{pmi}}. Then, we present a set of safety case templates to support reliability claims\footnote{We deal with probabilistic claims in this part, so ``reliability'' claims are about probabilities of occurrence of failures, and ``safety'' claims are about failures that are safety-relevant. The two kinds do not require different statistical reasoning, thus we may use the two terms safety and reliability interchangeably when referring to the probabilities of safety-relevant failures.} stated in \gls{pmi} based on our new RAM---the ``backbone'' of the probabilistic safety arguments for ML components, where the key argument is over the rigour of the four main steps of the \gls{RAM}.


\textcolor{black}{Finally, we conduct comprehensive case studies based on \gls{AUV} and \gls{UGV} in both simulated and physical environments to demonstrate and validate our methods.
All source code, \gls{ML} models, datasets and experimental results used in this work are publicly available at the our project repository \url{https://github.com/Solitude-SAMR} with video demos at \url{https://youtu.be/akY8f5sSFpY} and \url{https://youtu.be/E95vh5sxs7I}.}

\paragraph{Summary of Contributions} The key contributions of this work include:
\begin{itemize}
	\item  An assurance case framework for \gls{LES} that: (i) emphasises the arguments for quantitative claims on safety and reliability; (ii) with an ``end-to-end'' chain of safety analysis and reliability modelling methods for arguments ranging from the very top safety claim of the whole \gls{LES} to low-level \gls{VnV} evidence of \gls{ML} components.

	\item A first \gls{RAM} evaluating reliability for \gls{ML} software, leveraging \textit{both} the \gls{OP} information and robustness evidence. Moreover, based on the RAM, templates of probabilistic arguments for reliability claims on \gls{ML} software are developed.
	\item Identification of open challenges in building safety arguments for \gls{LES} and highlighting the inherent difficulties of assessing \gls{ML} reliability, uncovered by our overall assurance framework and the proposed \gls{RAM}, respectively. Potential solutions are discussed and mapped onto on-going studies to advance in this research direction.
	\item A prototype tool of our \gls{RAM} and a simulator platform of \gls{AUV} for underwater missions. Besides, real-world case studies based on a physical \gls{UGV} and a healthcare application are conducted to support the proposed approaches.
\end{itemize}

\paragraph{Organisation of this Article} 
\textcolor{black}{The scope and related work are summarised in Section \ref{sec_related_work}.
After presenting preliminaries in Section \ref{sec_preliminary}, we outline our overall \textit{system-level} assurance framework in Section \ref{sec_overall_framework}. 
We focus on the \textit{ML-component level} where the \gls{RAM} is described in detail in Section \ref{sec_ram}. 
We then present case studies in Section \ref{sec_case_study}. Furthermore, in-depth discussions are provided in Section \ref{sec_discussion}. Finally, we conclude and outline plans for future work in Section \ref{sec_conclusion_future_work}.}

\section{Scope and Related Work}\label{sec_related_work}

\begin{figure}[htbp]
    \centering
    \includegraphics[width=\textwidth]{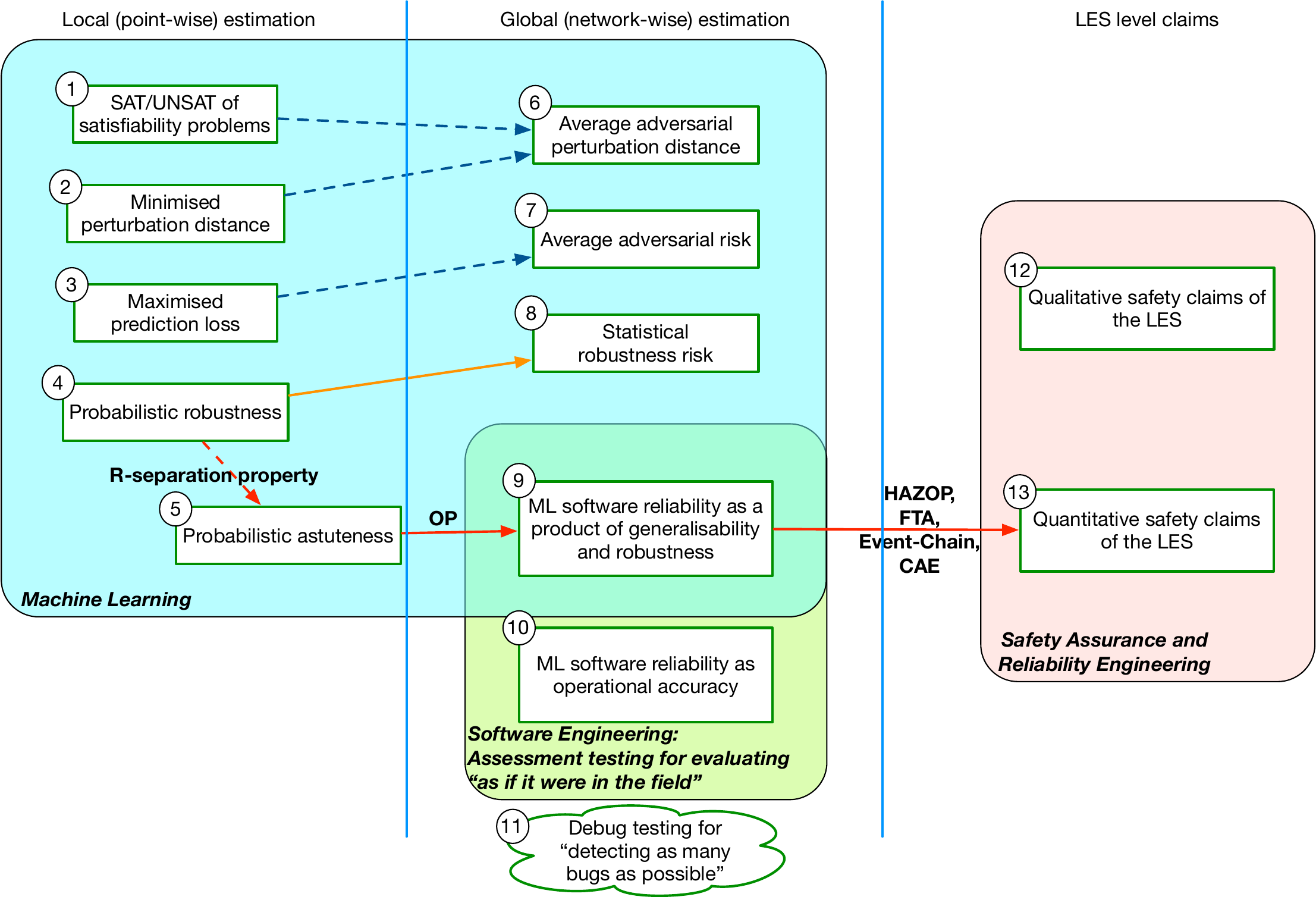}
	\caption{Scope and position: This paper presents a chain of methods in blocks \textcircled{5}, \textcircled{9} and \textcircled{13} across three levels linked by techniques shown on the red edges, which has its roots in the \textit{probabilistic} notion of local robustness (block \textcircled{4}). In stark contrast, other blocks at the local (i.e., point-wise robustness) level concern binary and extreme-cases that are non-probabilistic. Consequently, at the global network-wide level (i.e., the ML software), successor blocks of those non-probabilistic metrics can only consider \textit{averaged worst-case} robustness (blocks \textcircled{6} and \textcircled{7}). At this ML software level, our \gls{RAM} (block \textcircled{9}) agrees with the statistical robustness (block \textcircled{8}) that describes the pervasiveness of AEs in the input space, but with extra considerations on the oracle and \gls{OP} and thus unifies robustness (astuteness, to be exact) and generalisability. Blocks \textcircled{9}+\textcircled{10} and \textcircled{11} are representing two distinct types of testing in software engineering, i.e. \textit{assessment testing} vs \textit{debug testing}, while our RAM falls into the former category making the latter category irrelevant. Unlike ours (Block \textcircled{9}), existing assessment testing methods only consider the OP and ignores the point-wise robustness evidence (Block \textcircled{10}). Finally, at the the whole system level of LES, our work supports \textit{quantitative} safety claims.}
	\label{fig:scope}
\end{figure}

The aim of this work is to bridge the gap between local robustness evaluation evidence to system-level safety claims. By the interdisciplinary nature of the target problem, it spans across different subjects including \gls{ML}, software engineering, safety assurance and reliability engineering.  We refer to Fig. \ref{fig:scope} to highlight the scope and position of this paper (covering blocks \textcircled{4}, \textcircled{5}, \textcircled{9}, \textcircled{13}) as what follows.

\subsection{ML Robustness Evaluation}
Despite the exact definitions of robustness vary in literatures, they all share the intuition that inputs in a small region (usually a norm ball defined in some $L_{p}$-norm distance) have the same prediction label. Inside the local region, if an input is classified differently to the central point by the \gls{ML} model, then it is detected as an \gls{AE}. In recent years, great efforts have been made to study the local robustness of \gls{ML} models, while they formalise the problem in various ways and propose different evaluation metrics accordingly.

As shown in the first column of Fig. \ref{fig:scope}, the most classical type of approaches is by framing the robustness evaluation as a satisfiability problem that usually can be solved by SAT/SMT solvers \cite{huang_safety_2017,katz2017reluplex,gehr2018ai2,singh2019abstract}. Thus, given a local perturbation distance as constraints, a binary metric (i.e., Sat/Unsat) can be evaluated (block \textcircled{1}). The main limitation of such methods is their scalability to the size of \gls{ML} models and input dimensions. A more scalable type of methods is based on adversarial attacks, which normally requires an expensive gradient computation to detect \gls{AE}s and compute a minimised perturbation distance (block \textcircled{2}), e.g., \cite{moosavi2016deepfool,moosavi2017universal,aminifar2020universal,weng2019proven}. Similarly, \cite{WZCYSGHD2018} converts robustness analysis into a local Lipschitz constant estimation problem, deriving a lower bound on the minimal perturbation distance required to craft an \gls{AE}. When considering \textit{adversarial training}, maximised prediction loss (block \textcircled{3}) is often the metric of interest, e.g., \cite{madry2018towards} introduces the adversarial risk which is the maximum prediction loss within norm balls to measure the local robustness. They further apply this metric as the training loss to train \gls{ML} modela that are more resistant to adversarial attacks.

At the local level, all aforementioned methods essentially answer the binary question of whether there exist any \gls{AE}s in some perturbation distance. As argued by \cite{webb_statistical_2019}, they all suffer from two major drawbacks: i) they provide no notion of \textit{how} robust the model is whenever an \gls{AE} can be found; ii) there is a computational problem whenever no or only rare \gls{AE}s exist. To address the shortfalls, \cite{webb_statistical_2019} develops a new measure of \textit{probabilistic} robustness (block \textcircled{4}) under a local distribution over a set of inputs. Arguably, such probabilistic metric is of more practical interest in twofold \cite{webb_statistical_2019}: i) binary verification concerning extreme cases is neither necessary nor realistically achievable, i.e., one actually desires to know the \textit{proportion} of AEs in the input set, not just a binary answer as to whether it is robust or not; ii) all practical applications have acceptable level of risk, so that instead of confirming that this probability is exactly zero, showing the probability of a violation below a required threshold is good enough. We concur with these key insights and extend the probabilistic robustness to probabilistic \textit{astuteness} (block \textcircled{5}) in this work, by casting constraints (i.e., the $r$-separation property in Remark \ref{remark_r_sep}) on the norm ball radius as a principled way of determine the oracle (cf. later Remark \ref{remark_astutenss}). 
\textcolor{black}{More explicitly, in Table \ref{metrics_comp_local} of Appendix \ref{sec_app_ram_comparison}, we highlight the new characteristics of our block \textcircled{5} by comparing with DeepFool \cite{moosavi2016deepfool} and AI2 \cite{gehr2018ai2} locally, showing the differences between their inputs and outputs.}

We usually concern the \textit{overall} robustness of the ML model \cite{wang_statistically_2021}, i.e., its robustness across the range of possible inputs that the ML component will see in the real operation \cite{kurakin2018adversarial,zhao_detecting_2021}. This has motivated ``global'' or ``network-wide'' robustness metrics (second column in Fig. \ref{fig:scope}). Stemming from the corresponding local robustness metrics, the common metrics at this level simply take the average over a set of local robustness evaluation from regions represented by the training dataset, e.g., average minimal adversarial distance \cite{fawzi2018analysis}, average adversarial risk \cite{madry2018towards} (blocks \textcircled{6} and block \textcircled{7} respectively). As discussed by \cite{wang_statistically_2021}, it is often misleading to use those metrics with binary extreme-case meanings at network-wide level---indeed, an ML model can only be perfectly worst-case robust if it is robust to all possible perturbations of all inputs, something which will very rarely be achievable in practice. 
\textcolor{black}{Later in our experiments, we highlight the characteristics of our new block \textcircled{9} compared to block \textcircled{6} (that uses DeepFool \cite{moosavi2016deepfool} locally) in Table~\ref{metrics_comp_global} of Appendix \ref{sec_app_ram_comparison}. We may observe they are different metrics by nature, while arguably ours is of more practical interest.}
To overcome the problem, \cite{wang_statistically_2021} proposes a set of statistical robustness risks (block \textcircled{8}) which assess the overall probabilistic robustness by averaging the point-wise statistical robustness of \cite{webb_statistical_2019}. Aligning with the same idea, we build upon our local probabilistic astuteness metric to get the reliability of the ML component (block \textcircled{9}). Unlike \cite{wang_statistically_2021}, we explicitly incorporate the \gls{OP} and oracle information.

\subsection{OP-based Software Reliability Assessment}
\textcolor{black}{OP-based software testing, also known as statistical/operational testing \cite{strigini_guidelines_1997}, is an established practice and supported by industry standards for conventional systems.
There is a huge body of literature in the traditional software reliability community on OP-based testing and reliability modelling techniques, e.g.,  \cite{bertolino_adaptive_2021,bishop_deriving_2017,pietrantuono_reliability_2020,zhao_assessing_2020}.
In contrast to this, OP-based software testing for ML components is still in its infancy: to the best of our knowledge, there are only two recent works that explicitly consider the OP in testing. Li \textit{et al.} \cite{li_boosting_2019} propose novel stratified sampling based on ML specific information to improve the testing efficiency.
Similarly, Guerriero \textit{et al.} \cite{guerriero_operation_2021} develop a test case sampling method that leverages ``auxiliary information for misclassification'' and provides unbiased testing accuracy estimators. The motivation behind these two works is to leverage advanced sampling techniques to reduce the cost of testing in the operational dataset to yield the operational accuracy (block \textcircled{10}), 
However, neither of them considers robustness evidence in their assessment like our RAM does. }

At the whole LES level, there are reliability studies based on operational and statistical data, e.g., \cite{kalra_driving_2016,zhao_assessing_2019} for self-driving cars, \cite{hereau_testing_2020,zhao_probabilistic_2019} for AUV, and \cite{robert_virtual_2020} for agriculture robots doing autonomous weeding. 
However, knowledge from low-level ML components is usually ignored.
In \cite{zhao_assessing_2020}, we improved \cite{kalra_driving_2016} by providing a Bayesian mechanism to combine such knowledge, but did not discuss \textit{where} to obtain the knowledge.
In that sense, this article also contains follow-up work of \cite{zhao_assessing_2020}, providing the prior knowledge required based on the OP and robustness evidence.

Given that the OP is essentially a distribution defined over the whole input space, a related topic is the \textit{distribution-aware testing for \gls{DL}} (block \textcircled{11}) developed recently. 
For instance, in \cite{DBLP:conf/icse/Berend21}, distribution-guided coverage criteria are developed to guide the generation of new unseen test cases while identifying the validity of errors in DL system tasks.
In \cite{DBLP:conf/icse/DolaDS21}, a generative model is utilised to guide the generation of valid test cases. 
However, all existing distribution-aware testing methods are designed for \textit{detecting as many AEs as possible}, instead of trying to do \textit{reliability assessment} like ours, which are two distinct types of testing \cite{frankl_evaluating_1998,bertolino_adaptive_2017,cotroneo_relai_2016}.

\subsection{Assurance Cases for AI/ML-powered Autonomous Systems} 
\textcolor{black}{Work on safety arguments and assurance cases for AI/ML models and autonomous systems has emerged in recent years.
Burton \textit{et al.} \cite{burton_mind_2020} draw a broad picture of assuring AI and identify/categorise the ``gap'' that arises across the development process.
Alves \textit{et al.} \cite{alves_considerations_2018} present a comprehensive discussion on the aspects that need to be considered when developing a safety case for increasingly autonomous systems that contain ML components. 
Similarly, in \cite{BKCF2019}, an initial safety case framework is proposed with discussions on specific challenges for ML, which is later implemented with more details in \cite{bloomfield2021safety}.
A recent work \cite{javed_towards_2021} also explicitly suggests the combination of HAZOP and FTA in safety cases for identifying/mitigating hazards and deriving safety requirements (and safety contracts) when studying Industry 4.0 systems. 
In \cite{KKB2019}, safety arguments that are being widely used for conventional systems---including conformance to standards, proven in use, field testing, simulation, and formal proofs---are recapped for autonomous systems with discussions on the potential pitfalls. 
Both, \cite{matsuno_tackling_2019} and \cite{ishikawa_continuous_2018}, propose utilising continuously updated arguments to monitor the weak points and the effectiveness of their countermeasures, while a similar mechanism is also suggested in our assurance case, e.g., continuously monitor/estimate key parameters of our RAM---all essentially aligns with the idea of dynamic assurance cases \cite{calinescu_engineering_2018,asaadi_dynamic_2020}.}

\textcolor{black}{Although the aforementioned works have inspired this article, our assurance framework is with greater emphasis on, and thus complements them from, the quantitative aspects (block \textcircled{13}), e.g., reasoning for reliability claims stated in bespoke measures and breaking down system-level safety targets to component-level quantitative requirements.
Also exploring quantitative assurance, Asaadi \textit{et al.} \cite{asaadi_quantifying_2020} identifies dedicated assurance measures that are tailored for properties of aviation systems. }

\section{Preliminaries}
\label{sec_preliminary}

\subsection{Assurance Cases, CAE Notations and CAE Blocks}
\label{sec_preliminary_cae}

Assurance cases are developed to support claims in areas such as safety, reliability and security. They are often called by more specific names like security cases \cite{knight_importance_2015} and safety cases \cite{bishop_methodology_2000}. 
A safety case is a compelling, comprehensive, defensible, and valid justification of the system safety for a given application in a defined operating environment; it is therefore a means to provide the grounds for confidence and to assist decision making in certification \cite{bloomfield_safety_2010}.
For decades, safety cases have been widely used in the European safety community to assure system safety. Moreover, they are mandatory in the regulation for systems used in safety-critical industries in some countries, e.g., in the UK for nuclear energy \cite{uk_office_for_nuclear_regulation_purpose_2019}.
Early research in safety cases has mainly focused on their formulation in terms of claims, arguments, and evidence elements based on fundamental argumentation theories like the Toulmin model \cite{s_toulmin_uses_1958}.
The two most popular notations are CAE \cite{bloomfield_safety_2010} and \gls{GSN} \cite{kelly_arguing_1999}. In this article, we choose the former to present our assurance case templates.

\begin{figure*}[!h]
	\centering
	\includegraphics[width=\textwidth]{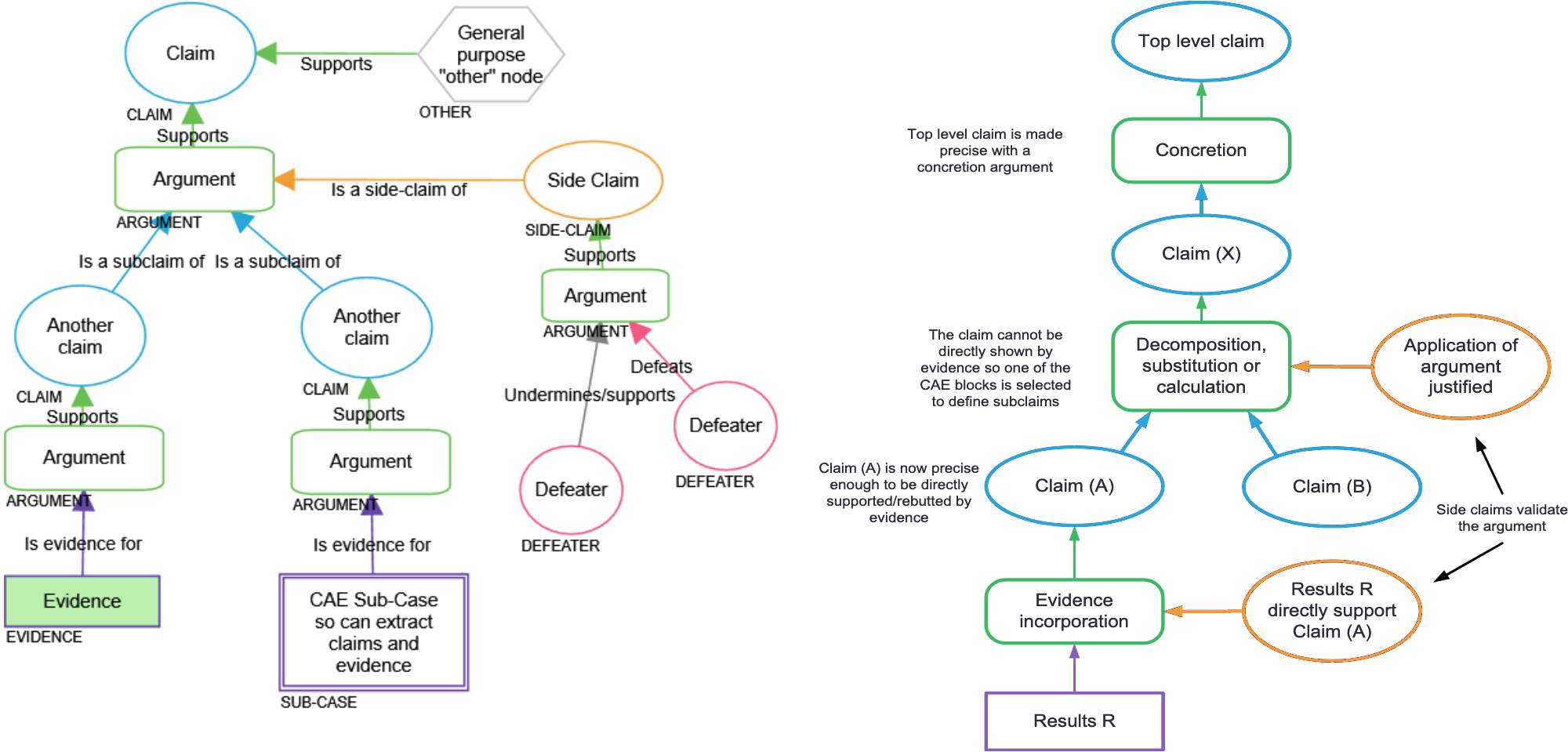}
	\caption{Summary of the CAE notations (lhs) and an example of CAE block use (rhs), cited from \cite{bloomfield2021safety}.}
	\label{fig_cae_notations_blocks}
\end{figure*}

A summary of the CAE notations is provided in Fig. \ref{fig_cae_notations_blocks}. The CAE safety case starts with a top \textit{claim}, which is then supported through an \textit{argument}
by sub-claims.
Sub-claims can be further decomposed until being supported by \textit{evidence}.
A claim may be subject to some context, represented by general purpose \textit{other} nodes, while assumptions (or warranties) of arguments that need to be explicitly justified form new \textit{side-claims}.
A \emph{sub-case} repeats a claim presented in another argument module. Notably,  the basic concepts of CAE are supported by safety standards like ISO/IEC15026-2. Readers are referred to \cite{bloomfield2020assurance,bloomfield2021safety} for more details on all CAE elements.

The CAE framework additionally consists of CAE blocks that provide five common argument fragments and a mechanism for separating inductive and deductive aspects of the argumentation\footnote{The argument strategy can be either inductive or deductive \cite{alves_considerations_2018}. For an inductive strategy, additional analysis is required to ensure that residual risks are mitigated.}. These were identified by empirical analysis of real-world safety cases \cite{bloomfield_building_2014}. The five CAE blocks representing the restrictive set of arguments are:
\begin{itemize}
	\item Decomposition: partition some aspect of the claim---``divide and conquer''.
	\item Substitution: transform a claim about an object into a claim about an equivalent object.
	\item Evidence Incorporation: evidence supports the claim, with emphasis on direct support.
	\item Concretion: some aspect of the claim is given a more precise definition.
	\item Calculation (or Proof): some value of the claim can be computed or proven.
\end{itemize}
An illustrative use of CAE blocks is shown in Fig. \ref{fig_cae_notations_blocks}, while more detailed descriptions can be found in \cite{bloomfield_building_2014,bloomfield2021safety}.

\subsection{\gls{HAZOP} and \gls{FTA}}
\label{sec_preliminary_fta_hazop}

\gls{HAZOP} is a structured and systematic safety analysis technique for risk management, which is used to identify potential hazards for the system in the given operating environment. \gls{HAZOP} is based on a theory that assumes risk events are caused by deviations from design or operating intentions. Identification of such deviations is facilitated by using sets of ``guide words'' (e.g., too much, too little and no) as a systematic list of deviation perspectives. It is commonly performed by a multidisciplinary team of experts during brainstorming sessions. HAZOP is a technique originally developed and used in chemical industries. There are studies that successfully apply it to software-based systems \cite{swann_twenty_five_1995}. Readers will see an illustrative example in later sections, while we refer to \cite{crawley2015hazop} for more details.

\gls{FTA} is a quantitative safety analysis technique on how failures propagate through the system, i.e., how component failures lead to system failures.  The fundamental concept in \gls{FTA} is the distillation of system component faults that can lead to a top-level event into a structured diagram (fault tree) using logic gates (e.g., AND, OR, Exclusive-OR and Priority-AND). We show a concrete example of \gls{FTA} in our case study section, while a full tutorial of developing \gls{FTA} is out of the scope of this article, and readers are referred to \cite{ruijters_fault_2015} for more details.

\subsection{OP Based Software Reliability Assessment}
\label{sec_preliminary_software_ram}
The \textit{delivered reliability}, as an \textit{user-centred} and \textit{probabilistic} property, requires to model the end-users' behaviours (in the operating environments) and to be formally defined by a quantitative metric \cite{littlewood_software_2000}.
Without loss of generality, we focus on \gls{pmi} as a generic metric for ML classifiers, where inputs can, e.g., be images acquired by a robot for object recognition. 
\begin{definition}[\gls{pmi}]
	We denote the unknown \gls{pmi} by a variable $\lambda$, which is formally defined as
	\begin{equation}
		\label{eq_pfd_def}
		\lambda:=\int_{x\in \inputdomain} I_{\{x \text{ causes a misclassification}\}}(x)\mathsf{Op}(x)\diff{x} \ ,
	\end{equation}
	where $x$ is an input in the input domain\footnote{We assume continuous $\inputdomain$ in this article. 
		For discrete $\inputdomain$, the integral in Eqn.~\eqref{eq_pfd_def} reduces to sum and $\mathsf{Op}(\cdot)$ becomes a probability mass function.} $\inputdomain$, and ${I}_{\tt S}(x)$ is an indicator function---it is equal to $1$ when {\tt S} is true and equal to $0$ otherwise.
	The function $\mathsf{Op}(x)$ returns the probability that $x$ is the next random input.
\end{definition}
\textcolor{black}{Intuitively, if we randomly selected an input from the input domain according to the distribution of \gls{OP}, the probability of the event that this input being misclassified is measured by \gls{pmi}. In other words, a ``frequentist'' interpretation of \gls{pmi} is that it is the \textit{limiting relative frequency} of inputs for which the classifier fails in an infinite sequence of independently selected inputs. In this regard, \gls{pmi} is a natural extension of the conventional reliability metric \textit{probability of failure on demand} (\textit{pfd}) \cite{zhao_modeling_2017}, but retrofitted for ML classifiers.}

\begin{remark}[\gls{OP}]
	\label{remark_op}
	The \gls{OP} \cite{musa_operational_1993} is a notion used in software engineering to quantify how the software will be operated. Mathematically, the OP is a \textbf{\gls{PDF}} defined over the whole input domain $\inputdomain$.
\end{remark}
\noindent We highlight this Remark \ref{remark_op}, because we use probability density estimators to approximate the \gls{OP} from the collected operational dataset in our \gls{RAM} developed in Section \ref{sec_ram}.

By the definition of \gls{pmi}, successive inputs are assumed to be independent. It is therefore common to use a Bernoulli process
as the mathematical abstraction of the stochastic failure process, which implies a Binomial likelihood.
For traditional software, upon establishing the likelihood, RAMs on estimating $\lambda$ vary case by case---from the basic \gls{MLE} to Bayesian estimators tailored for certain scenarios when, e.g., seeing no failures \cite{miller_estimating_1992,bishop_toward_2011}, inferring ultra-high reliability \cite{zhao_assessing_2020}, with certain forms of prior knowledge like perfectioness \cite{strigini_software_2013}, with vague prior knowledge expressed in imprecise probabilities \cite{walter_imprecision_2009,zhao_probabilistic_2019}, with uncertain \gls{OP}s \cite{bishop_deriving_2017,pietrantuono_reliability_2020}, etc.

OP based RAMs designed for traditional software fail to consider new characteristics of \gls{ML}, e.g., the lack of robustness and a high dimensional input space. Specifically, it is quite hard to gather the required prior knowledge when taking into account the new ML characteristics in the aforementioned Bayesian RAMs.
At the same time, frequentist RAMs would require a large sample size to gain enough confidence in the estimates due to the extremely large population size (e.g., the high dimensional pixel space for images).


\subsection{\gls{ML} Robustness and the $R$-Separation Property}
\label{sec_preliminary_robustness_r_sep}

\gls{ML} is known not to be robust. Robustness requires that the decision of the \gls{ML} model $\network$ is invariant against small perturbations on inputs. 
That is, all inputs in a region $ \eta \subset \inputdomain$ have the same prediction label, where usually the region $\eta$ is a small norm ball (in an $L_{p}$-norm distance\footnote{Distance mentioned in this article is defined in $L_\infty$ if without further clarification.}) of radius $\epsilon$ around an input $x$.
Inside $\eta$, if an input $x'$ is classified differently to $x$ by $\network$, then $x'$ is an \gls{AE}.
Robustness can be defined either as a binary metric (if there exists any \gls{AE} in $\eta$) or as a probabilistic metric (how likely the event of seeing an \gls{AE} in $\eta$ is).
The former aligns with formal verification, e.g.\ \cite{huang_safety_2017}, while the latter is normally used in statistical approaches, e.g.\ \cite{webb_statistical_2019}. The former ``verification approach'' is the binary version of the latter ``stochastic approach''\footnote{Thus, we use the more general term robustness ``evaluation'' rather than robustness ``verification'' throughout the article.}.

\begin{definition}[robustness]
	Similar to \cite{webb_statistical_2019}, we adopt the more general probabilistic definition on the robustness of the model $\network$ (in a region $\eta$ and to a target label $y$):
	\begin{equation}
		\label{eq_robust_def}
		R_{\network}(\eta, y):=\int_{x \in \eta} I_{\{\network(x) \text{ predicts label } y \}}(x) \mathsf{Op}(x\mid x \in \eta)\diff{x} \ ,
	\end{equation}
	where $\mathsf{Op}(x \mid x\in\eta)$ is the \textit{conditional OP} of region $\eta$ (precisely the ``input model'' used by both \cite{webb_statistical_2019} and \cite{weng2019proven}).
\end{definition}

\begin{figure}[ht]
	\centering
	\includegraphics[width=0.7\linewidth]{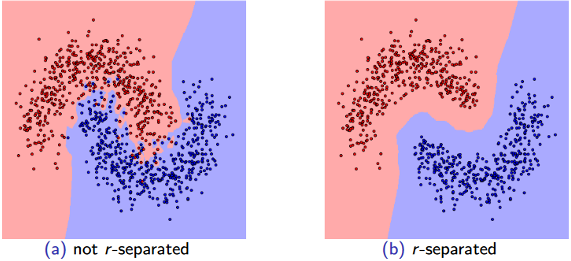}
	\caption{\textcolor{black}{Illustration of the $r$-separation property.}}
	\label{fig_r_sep_demo}
\end{figure}

We highlight the following two remarks regarding robustness:
\begin{remark}[astuteness]
	\label{remark_astutenss}
	Reliability assessment only concerns the robustness to the ground truth label, rather than an arbitrary label $y$ in $R_{\network}(\eta, y)$. When $y$ is such a ground truth, robustness becomes \textbf{astuteness} \cite{yang_closer_2020}, which is also the \textbf{conditional reliability} in the region $\eta$.
\end{remark} 
\noindent Astuteness is a special case of robustness\footnote{Thus, later in this article, we may refer to robustness as astuteness for brevity when it is clear from the context.}. An extreme example showing why we introduce the concept of astuteness is, that a perfectly robust classifier that always outputs ``dog'' for any given input is unreliable.
Thus, robustness evidence cannot directly support reliability claims unless the ground truth label is used in estimating $R_{\network}(\eta, y)$.
\begin{remark}[$r$-separation]
	\label{remark_r_sep}
	For real-world image datasets, any data-points with different ground truth are at least distance $2r$ apart in the input space (pixel space), and $r$ is bigger than a norm ball radius commonly used in robustness studies.
\end{remark}
\noindent The $r$-separation property was first observed by \cite{yang_closer_2020}: real-world image datasets studied by the authors implies that $r$ is normally $3\sim 7$ times bigger than the radius (denoted as $\epsilon$) of norm balls commonly used in robustness studies.
Intuitively it says that, although the classification boundary is highly non-linear, there is a minimal distance between two real-world objects of different classes (cf.\ Fig. \ref{fig_r_sep_demo} for a conceptual illustration). Moreover, such a minimal distance is bigger than the usual norm ball size in robustness studies. 




\section{The Overall Assurance Framework}
\label{sec_overall_framework}

\begin{figure}[htbp!]
    \centering
    \includegraphics[width=0.8\textwidth]{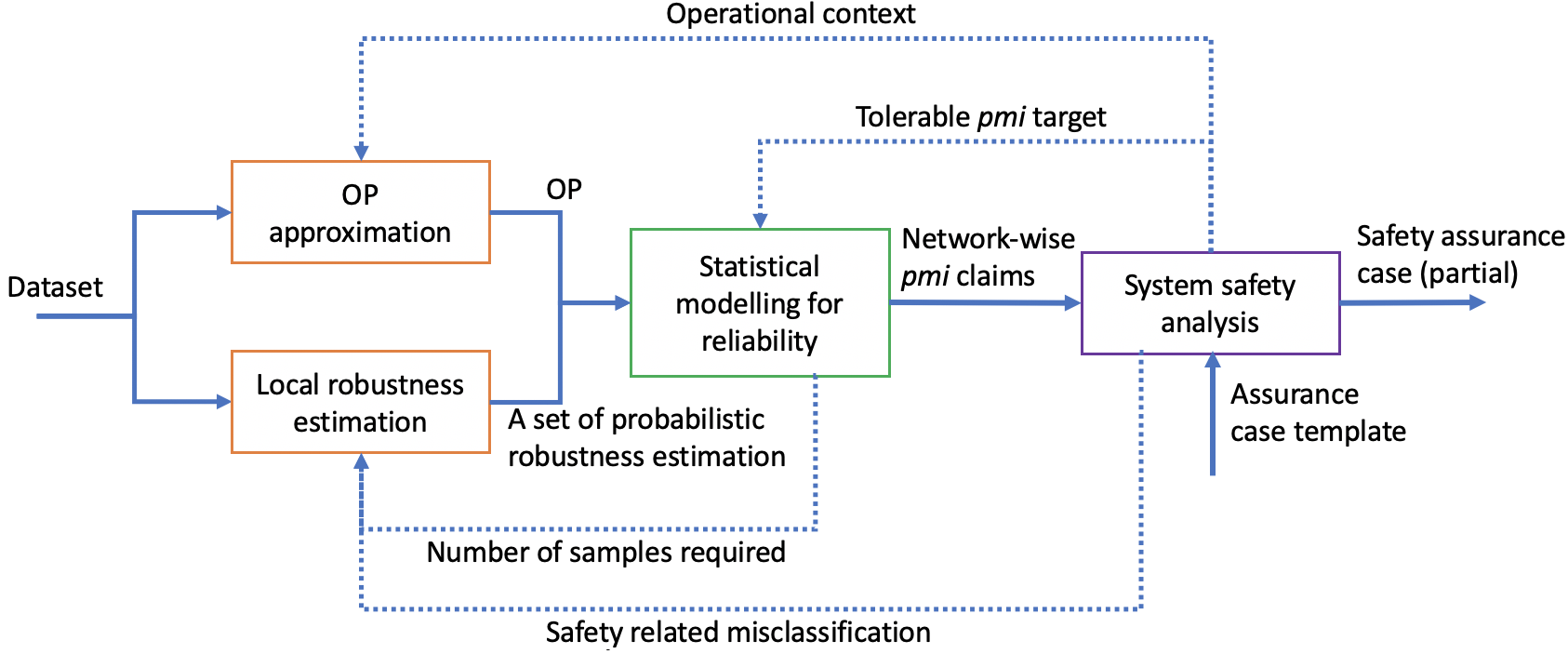}
	\caption{\textcolor{black}{Overview of the proposed assurance framework. Blocks (at different levels are in different colours) represent major assurance activities corresponding to blocks \textcircled{4}, \textcircled{5}, \textcircled{9} and \textcircled{13} in Fig. \ref{fig:scope}. Solid lines show the inputs/output of blocks from lower levels to higher levels, while dotted lines show feedbacks.}}
	\label{fig_overview_of_assurance_framework}
\end{figure}

\textcolor{black}{In this section, we present an overall assurance framework for \gls{LES} (e.g., \gls{AUV}), in which quantitative claims stated in our metric \gls{pmi} act as the bridge between point-wise robustness evidence and the whole system-level safety. As shown in Fig. \ref{fig_overview_of_assurance_framework}, it starts with a dataset, e.g., the dataset used for training the \gls{ML} perception component in \gls{AUV}, to approximate the \gls{OP}. Meanwhile, for each data-point (camera image from the \gls{AUV}) in the dataset, local (point-wise) robustness can be estimated probabilistically. Then, globally at the \gls{ML} component level (network-wise), a statistical model derives the \gls{pmi} claim from the set of local robustness evidence considering the approximated OP. The \gls{pmi} claim further supports the whole system-level safety analysis to form safety arguments in assurance cases (reusing templates for \gls{RAS} extended from \cite{bloomfield2021safety}). Since our work focuses on quantitative safety risks that can be broken down to ML-components doing perception, it \textit{partially} implements the safety case (leaving out claims regarding other components and qualitative safety requirements). Instead of a sequential process, system level safety analysis feedbacks to lower level assurance activities. That is, (i) system operational domain analysis may guide the \gls{OP} approximation by pre-processing the dataset so that the dataset is statistically representing the OP; (ii) not all misclassifications are of the safety concern, thus only safety related ones should be calculated in local robustness estimation; (iii) the statistical inference model for \gls{pmi} informs the required number of point-wise robustness estimations so that the \gls{pmi} estimation uncertainty is sufficiently low; while (iv) the tolerable \gls{pmi} claim is also derived from system-level safety analysis.}


\subsection{Overview of an Assurance Case for \gls{LES}}
\label{sec_overall_overview_ac_les}

\begin{figure*}[!h]
	\centering
	\includegraphics[width=0.8\textwidth]{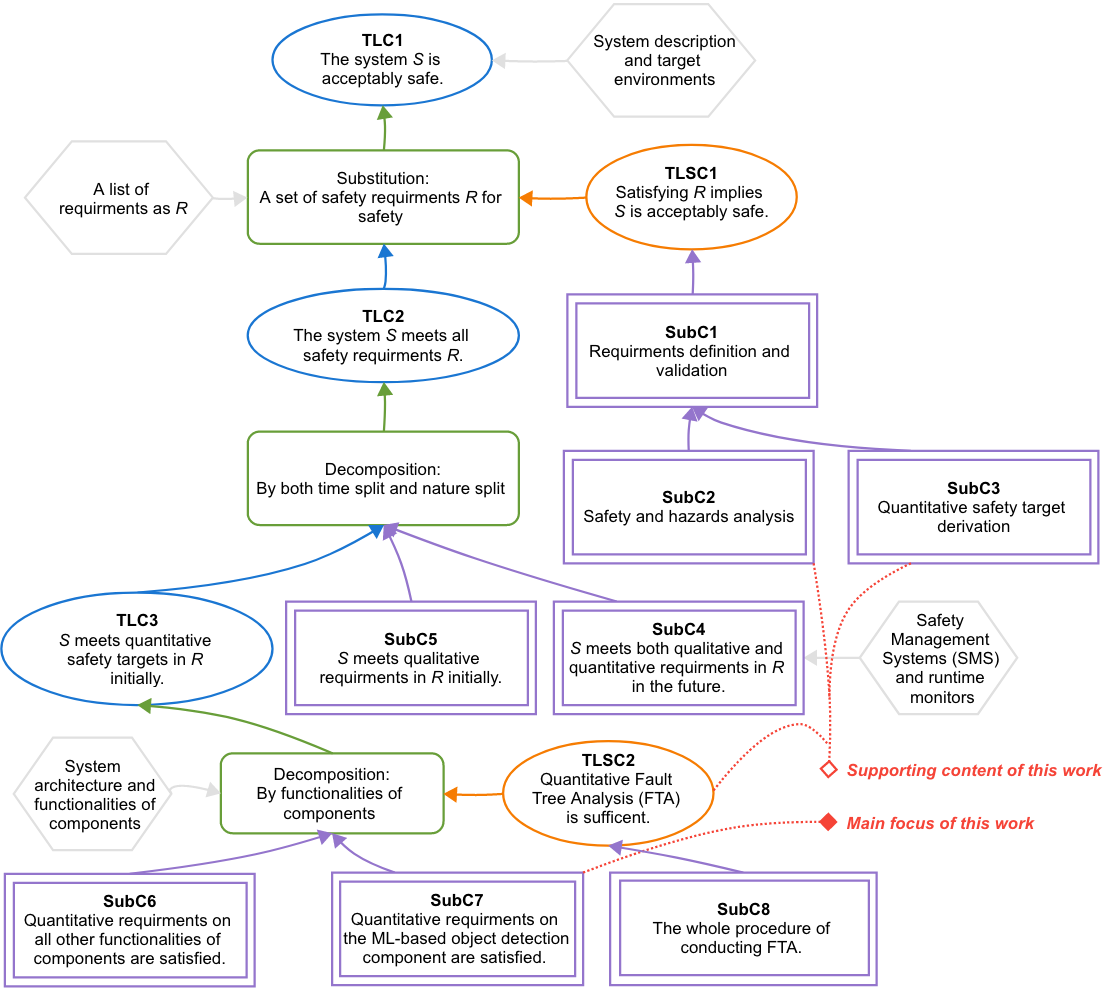}
	\caption{Overview of the proposed safety case template for \gls{LES}, highlighting the main focus and supporting content of this work.}
	\label{fig_overview_ac}
\end{figure*}

\textcolor{black}{The proposed assurance case template is shown in Fig. \ref{fig_overview_ac}, which is an overview of some top-level arguments and eight supporting CAE sub-cases. It also highlights the main focus of this work---a \gls{RAM} for the \gls{ML} component with its probabilistic safety arguments---and all its required supporting analysis to derive the reliability requirements of the low-level \gls{ML} functionalities.} It shows how various kinds of safety and reliability analysis/modelling methods are combined and structured to support our top-level claim \textbf{TLC1}---the \gls{LES} \textit{S} is acceptably safe. Additional information for the top claim should be provided (as for any safety case), describing the system \textit{S} in detail and the target operational environments. The term ``acceptably safe'' is abstract and vague, thus we substitute it with \textbf{TLC2} that all safety requirements \textit{R} are satisfied. This substitution CAE-block/argument needs to be supported by a side-claim \textbf{TLSC1} explaining why satisfying the set of requirements \textit{R} implies \textit{S} is safe enough.

To argue for \textbf{TLSC1}, we refer to the template proposed by \cite[Chap.~5]{bloomfield2021safety} as our sub-case \textbf{SubC1}.
Essentially, in \textbf{SubC1}, we argue \textit{R} is: (i) well-defined (e.g., verifiable, consistent, unambiguous, traceable, etc); (ii) complete that covers all sources (e.g., from hazard analysis and domain-specific safety standards and legislation); and (iii) valid, according to some common risk acceptance criteria/principles in safety regulations of different countries/domains, e.g., ALARP (As Low As Reasonably Practicable). 
Without repeating the content of \cite{bloomfield2021safety}, we only highlight the parts directly supporting the main focus of this work (via the procedure in Fig. \ref{fig_hazop_fta_flow}), which are hazard identification (\textbf{SubC2}) and derivation of quantitative safety target (\textbf{SubC3}).

Similar to \cite{bloomfield2021safety}, we use a decomposition CAE-block/argument to support \textbf{TLC2}. But, in addition to time-split, we also split the claim by the qualitative and quantitative nature, since the main focus of this work, \textbf{SubC7}, concerns the probabilistic reliability modelling of \gls{ML} components.
Further decomposition of the whole system's quantitative requirements into functionalities of individual components (\textbf{TCL3}) is non-trivial, for which we utilise quantitative \gls{FTA}. The decomposition requires a side-claim on the sufficiency of the \gls{FTA} study \textbf{TLSC2}. A comprehensive development \textbf{SubC8} for \textbf{TLSC2} is out of the scope of this work, while we illustrate the gist and an example of the method in later sections. Finally, we reach the main focus of this work \textbf{SubC7} and will develop the full sub-case for it in Appendix \ref{sec_prob_safe_argument}.

\subsection{Deriving Quantitative Requirements for ML Components}
\label{sec_dqrmlc}

\begin{figure*}[!h]
	\centering
	\includegraphics[width=0.6\textwidth]{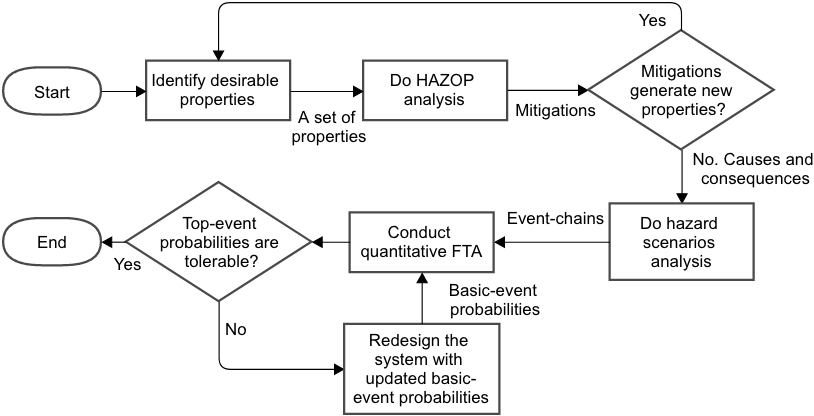}
	\caption{The workflow of combining \gls{HAZOP} and quantitative \gls{FTA} to derive probabilities of basic-events of components.}
	\label{fig_hazop_fta_flow}
\end{figure*}

In this work we are mainly developing low-level probabilistic safety arguments, based on the dedicated \gls{RAM} for ML components developed in Section \ref{sec_ram}. An inevitable question is, \textit{how to quantitatively determine the tolerable and acceptable risk levels of the ML components}? Normally the answer involves a series of well-established safety analysis methods that systematically breaks down the whole-system level risk to low-level components, considering the system architecture \cite{littlewood_reasoning_2012,zhao_safety_2020}. While, the whole-system level risk is determined on a case by case basis through the application of principles and criteria required by the safety regulations extant in the different countries/domains. To align with this best practice, we propose the procedure articulated in Fig. \ref{fig_hazop_fta_flow}, whose major steps correspond to the supporting sub-cases  \textbf{SubC2}, \textbf{SubC3} and \textbf{SubC8}. 

In Fig. \ref{fig_hazop_fta_flow}, for the given \gls{LES}, we first identify a set of safety properties that are desirable to stakeholders. Then, a \gls{HAZOP} analysis is conducted, based on deviations of those properties, to systematically identify hazards (and their causes, consequences, and mitigations). 
New safety properties may be introduced by the mitigations identified by \gls{HAZOP}, thus \gls{HAZOP} is conducted in an iterative manner that forms the first loop in Fig. \ref{fig_hazop_fta_flow}.

Then, we leverage the \gls{HAZOP} results to do \textit{hazard scenario modelling}, inspired by \cite{guo_extended_2015}, so that we may combine \gls{HAZOP} and \gls{FTA} later on. Usually, as noted in \cite{guo_extended_2015}, a property deviation can have several causes and different consequences in \gls{HAZOP} analysis. It is complicated and difficult to directly convert \gls{HAZOP} results into fault trees. Thus, hazard scenario modelling is needed to explicitly link the initial events (causes) to the final events (consequences) with a chain of intermediate events. Such event-chains facilitate the construction of fault trees, specifically in three steps:
\begin{itemize}
	\item The initial events (causes) may or may not be further decomposed at even lower-level sub-functionalities of components to determine the root causes, which are used as basic events (BE) in \gls{FTA}. Thus, BEs are typically failure events of software/hardware components, e.g., different types of misclassifications, failures in different modes of a propeller.
	\item Adding a specific logic gate among all intermediate events (IE) on the same level, which models how failures are propagated, tolerated and/or compounded throughout the system architecture.
	\item Final events (consequences) are used as top events (TE) of the \gls{FTA}. In other words, TEs are violations of system-level safety properties.
\end{itemize}

Upon establishing the fault trees, conventional quantitative \gls{FTA} can be performed to propagate probabilities of BEs to the TE probability, or, reversely, to allocate/break-down TE probability to BEs.  What-if calculations and sensitivity analysis are expected to find the most practical solution of BE probabilities that makes the required TE risk tolerable. Then the practical solution for the BE associated with the ML component of our interest becomes our target reliability claims for which we develop probabilistic safety arguments. Notably, the ML component may need several rounds of retraining/fine-tuning to achieve the required level of reliability. This forms part\footnote{Other non-ML components may be updated as well to jointly make the whole-system risk tolerable.} of the second iterative loop in Fig. \ref{fig_hazop_fta_flow}. We refer readers to \cite{zhao_detecting_2021} for a detailed description on this \textit{debug-retrain-assess} loop for ML software.

Finally, the problem boils down to (i) \textit{how to derive the system-level quantitative safety target}, i.e., assigning probabilities for those TEs of the fault trees; and (ii) \textit{how to demonstrate the component-level reliability is satisfied}, i.e., assessing the BE probabilities for components based on evidence. We address the second question in the next section, while the first question is essentially ``how safe is safe enough?'', for which the general answer depends on the \textit{existing} regulation/certification principles/standards of different countries and industry domains.
Unfortunately, existing safety standards cannot be applied on \gls{LES}, and revisions are still ongoing. Therefore, we currently do not have a commonly acknowledged practice that can be easily applied to certify or regulate LES \cite{BKCF2019,klas2021using}. 
That said, emerging studies on assuring/assessing the safety and reliability of AI and autonomous systems have borrowed ideas from existing regulation principles on risk acceptability and tolerability, to name a few:
\begin{itemize}
	\item \gls{ALARP}: ALARP states that the residual risk after the application of safety measures should be as low as reasonably practicable. The notion of being reasonably practicable relates to the cost and level of effort to reduce risk further. It originally arises from UK legislation and is now applied in many domains like nuclear energy.
	\item \gls{GALE}: is a principle required by French law for railway safety, which indicates the new technical system shall be at least as safe as comparable existing ones.
	\item \gls{SE}: similar to GALE; new medical devices in the US must be demonstrated to be substantially equivalent to a device already on the market. This is required by the U.S. Food \& Drug Administration (FDA).
	\item \gls{MEM}: MEM states that a new system should not lead to a significant increase in the risk exposure for a population with the lowest endogenous mortality. For instance, the rate of natural deaths is a reference point for acceptability.
\end{itemize}
While a complete list of all principles and comparisons between them are beyond the scope of this work, we believe that the common trend is that, for many \gls{LES}, a promising way of determining the system-level quantitative safety target is to argue the acceptable/tolerable risk over the average human-performance. For instance, self-driving cars' targets of being as safe as or two-magnitude safer than human-drivers (in terms of metrics like fatalities per mile) are studied in \cite{kalra_driving_2016,zhao_assessing_2020,liu_how_2019}. In \cite{picardi_pattern_2019}, human-doctors' performance is used as the benchmark in arguing the safety of ML-based medical diagnosis systems.

In summary, we are only presenting the essential steps of combining \gls{HAZOP} and quantitative \gls{FTA} via \textit{hazard scenario modelling} to derive component-level reliability requirements from whole system-level safety targets, while each of those steps with concrete examples can be found in Section \ref{sec_case_study} as part of the \gls{AUV} case study.

\section{Modelling the Reliability of ML Classifiers}
\label{sec_ram}

\subsection{A Running Example of a Synthetic Dataset}
\label{sec_running_exp}
\begin{figure}[ht]
	\centering
	\includegraphics[width=0.6\linewidth]{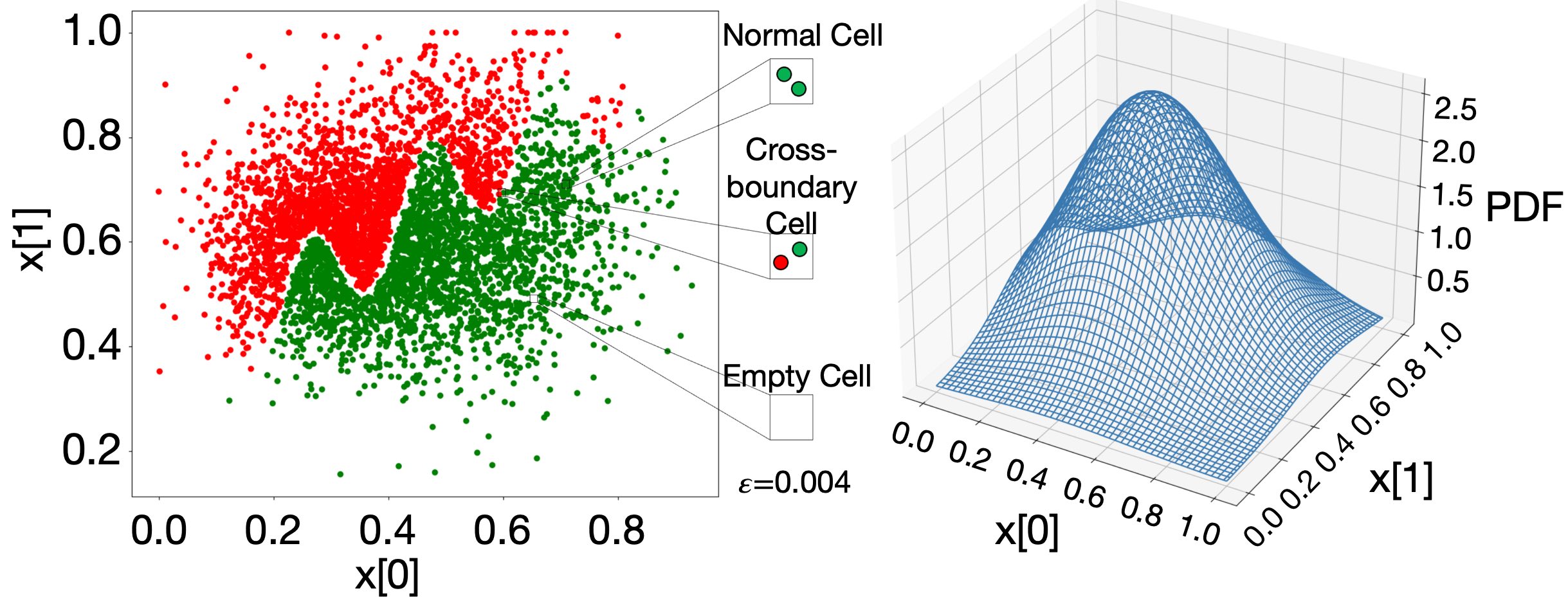}
	\caption{\textcolor{black}{The 2D-point dataset (lhs), and its approximated \gls{OP} (rhs).}}
	\label{fig_running_example}
\end{figure}

To better demonstrate our \gls{RAM}, we take the Challenge of AI Dependability Assessment raised by Siemens Mobility\footnote{
	\url{https://ecosystem.siemens.com/topic/detail/default/33}
} as a running example. The challenge is to firstly train an \gls{ML} model to classify a dataset generated on the unit square $[0,1]^2$ according to some unknown distribution (essentially the unknown \gls{OP}).
The collected data-points (training set) are shown in Fig.~\ref{fig_running_example}-lhs, in which each point is a tuple of two numbers between 0 and 1 (thus called a ``2D-point'').
We then need to build a \gls{RAM} to claim an upper bound on the probability that the next random point is misclassified, i.e., the \textit{pmi}.
If the 2D-points represent traffic lights, then we have 2 types of misclassifications---safety-critical ones, when a red data-point is labelled green, and performance related ones otherwise. For brevity, we consider both types of misclassifications here, while our \gls{RAM} can cope with sub-types of misclassifications.

\subsection{The Proposed RAM}
\label{sec_the_model}

\paragraph{Principles and Main Steps of the \gls{RAM}} 
Inspired by \cite{pietrantuono_reliability_2020}, our \gls{RAM} first partitions the input domain into $m$ small cells\footnote{We use the term ``cell'' to highlight the partition that yields exhaustive and mutually exclusive regions of the input space, which is essentially a norm ball in $L_\infty$. Thus, we use the terms ``cell'' and ``norm ball'' interchangeably in this article when the emphasis is clear from the context.}, subject to the $r$-separation property. Then, for each cell $c_i$ (and its ground truth label $y_i$), we estimate:
\begin{align}
	\label{eq_cell_rel_op}
	\lambda_i&:=1-R_{\network}(c_i, y_i) \quad \mbox{and} \quad \mathsf{Op}_i:=\int_{x\in c_i}\mathsf{Op}(x)\diff{x} \ ,
\end{align}
which are the \textit{unastuteness} and \textit{pooled OP} of the cell $c_i$ respectively---we introduce estimators for both later. Eqn.~\eqref{eq_pfd_def} can then be written as the weighted sum of the \textit{cell-wise} unastuteness (i.e., the conditional \textit{pmi} of each cell%
\footnote{We use ``cell unastuteness'' and ``cell \textit{pmi}'' interchangeably later.}),
where the weights are the pooled OP of the cells: 
\begin{equation}
	\label{eq_pfd_cell}
	\lambda=\sum_{i = 1}^{m} \mathsf{Op}_i \lambda_i  
\end{equation}

Eqn.~\eqref{eq_pfd_cell} captures the essence of our RAM---it shows clearly how we incorporate the OP information and the robustness evidence to claim reliability.
This reduces the problem to: 
(i) \textit{how to obtain the estimates on those $\lambda_i$s and $\mathsf{Op}_i$s} and (ii) \textit{how to measure and propagate the uncertainties in the estimates}. 
These two questions are challenging.
To name a few, for the first question: estimating $\lambda_i$ requires to determine the ground truth label of cell $i$; and estimating $\mathsf{Op}_i$s may require a large amount of operational data.
For the second question, the fact that all estimators are imperfect entails that they need a measure of trust (e.g., the variance of a point estimate), which may not be easy to derive.

In what follows, by referring to the running example, we proceed in four main steps: (i) partition the input space into cells; (ii) approximate the OP of cells (the $\mathsf{Op}_i$s); (iii) evaluate the unastuteness of these cells (the $\lambda_i$s); and (iv) ``assemble'' all cell-wise estimates for $\lambda$ in a way that estimation uncertainties are propagated and compounded.


\paragraph{Step 1: Partition of the Input Domain $\inputdomain$} As per Remark \ref{remark_astutenss}, the astuteness evaluation of a cell requires its ground truth label. To leverage the $r$-separation property and the later Assumption \ref{assumption_single_gt_cell}, we partition the input space by choosing a cell radius $\epsilon$ so that $\epsilon < r$. Although we concur with Remark \ref{remark_r_sep} (first observed by \cite{yang_closer_2020}) and believe that there should exist an \textit{$r$-stable ground truth} (which means that the ground truth is stable in such a cell)
for any real-world \gls{ML} classification applications, it is hard to estimate such an $r$ (denoted by $\hat{r}$) and the best we can do is to assume:
\begin{assumption}
	\label{assumption_r_estiamtes_from_data}
	There is a $r$-stable ground truth
	(as a corollary of Remark \ref{remark_r_sep}) for any real-world classification problems, and the $r$ parameter can be sufficiently estimated from the existing dataset.
\end{assumption}

That said, in the running example, we get $\hat{r}=0.004013$ by iteratively calculating the minimum distance of different labels. Then we choose a cell radius\footnote{We use the term ``radius'' for cell size defined in $L_{\infty}$, which happens to be the side length of the square cell of the 2D running example.} $\epsilon$, which is smaller than $\hat{r}$---we choose $\epsilon=0.004$. With this value, we partition the unit square $\inputdomain$ into $250 \times 250$ cells.

\paragraph{Step 2: Cell OP Approximation}
Given a dataset $(X,Y)$, we estimate the pooled OP of cell $c_i$ to get $\myExp[\mathsf{Op}_i]$ and $\myVar[\mathsf{Op}_i]$. We use the well-established \gls{KDE} to fit a $\widehat{\mathsf{Op}}(x)$ to approximate the OP. 
\begin{assumption}
	\label{assumption_dataset_represents_OP}
	\textcolor{black}{The given dataset $(X,Y)$ is collected and sampled based on the OP, and thus statistically represents the OP.}
\end{assumption}
\noindent This assumption may not hold in practice: training data is normally collected in a \textit{balanced} way, since the \gls{ML} model is expected to perform well in all categories of inputs, especially when the OP is unknown at the time of training and/or expected to change in future. Although our model can relax this assumption in various ways (discussed in Section~\ref{sec_discussion}), we adopt it for brevity in demonstrating the running example. 

Given a set of 
(unlabelled) data-points $(X_1,\dots, X_n)$ from the dataset $(X,Y)$, \gls{KDE} then yields
\begin{equation}
	\label{eq_kde}
	\widehat{\mathsf{Op}}(x) = \frac{1}{n h} \sum_{j = 1}^{n} K(\frac{x-X_j}{h}) \ ,
\end{equation}
where $K$ is the kernel function (e.g. Gaussian or exponential kernels), and $h > 0$ is a smoothing parameter, called the bandwidth, cf. \cite{silverman1986density} for guidelines on tuning $h$. 
The approximated OP\footnote{In this case, the KDE uses a Gaussian kernel and $h=0.2$ that optimised by cross-validated grid-search \cite{bergstra_random_2012}.} is shown in Fig. \ref{fig_running_example}-rhs. 

Since our cells are small and all equal size, instead of calculating $\int_{x\in c_i}\widehat{\mathsf{Op}}(x)dx$, we may approximate $\mathsf{Op}_i$ as
\begin{equation}
	\label{eq_cell_op}
	\widehat{\mathsf{Op}}_i = \widehat{\mathsf{Op}}\left(x_{c_i}\right) v_c
\end{equation}
where $\widehat{\mathsf{Op}}(x_{c_i})$ is the probability density at the cell's central point $x_{c_i}$, and $v_c$ is the constant cell volume ($0.000016$ in the running example).


Now if we introduce new variables $W_j = \frac{1}{h} K(\frac{x-X_j}{h})$, the \gls{KDE} evaluated at $x$ is actually the sample mean of $W_1,\dots,W_n$.
Then by invoking the \gls{CLT}, we have $\widehat{\mathsf{Op}}(x) \sim \mathcal{N}(\mu_W,\frac{\sigma_W^2}{n})$,
where the mean is exactly the value from Eqn.~\eqref{eq_kde}, while the variance of $\widehat{\mathsf{Op}}(x)$ is a known result of:
\begin{align}
	\myVar[\widehat{\mathsf{Op}}(x)] &= \frac{f(x) \int K^2(u) du}{nh} + O(\frac{1}{nh}) \approx \hat{\sigma}^2_B(x) \ ,
	\label{eq_clt_KDE_var}
\end{align}
where the last step of Eqn.~\eqref{eq_clt_KDE_var} says that $\myVar[\widehat{\mathsf{Op}}(x)]$ can be approximated using a bootstrap variance $\hat{\sigma}^2_B(x)$ \cite{chen2017tutorial} (cf.\ Appendix A for details).

Upon establishing Eqn.s~\eqref{eq_kde} and \eqref{eq_clt_KDE_var}, together with Eqn.~\eqref{eq_cell_op}, we know for a given cell $c_i$ (and its central point $x_{c_i}$):
\begin{align}
	\myExp[\mathsf{Op}_i]=v_{c}\myExp[\widehat{\mathsf{Op}}(x_{c_i})],\quad
	\myVar[\mathsf{Op}_i]=v_{c}^2\myVar[\widehat{\mathsf{Op}}(x_{c_i})] \ ,
\end{align}
which are the OP estimates of this cell.

\paragraph{Step 3: Cell Astuteness Evaluation}
As a corollary of Remark \ref{remark_r_sep} and Assumption \ref{assumption_r_estiamtes_from_data}, we may confidently assume:
\begin{assumption}
	\label{assumption_single_gt_cell}
	If the radius of $c_i$ is smaller than $r$, all data-points in the cell $c_i$ share a single ground truth label.
\end{assumption}

Now, to determine such ground truth label of a cell $c_i$, we can classify our cells into three types:
\begin{itemize}
	\item Normal cells: a normal cell contains data-points from the existing dataset. These data-points from a single cell are sharing a same ground truth label, which is then determined as the ground truth label of the cell.
	\item Empty cells: a cell is ``empty'' in the sense that it contains no data-points from the dataset of already collected data.
	Some of the empty cells will eventually become non-empty as more future operational data being collected, while most of them will remain empty forever---once cells are sufficiently small, only a small share of cells will refer to physically plausible images, and even fewer are possible in a given application.
	For simplicity, we do not further distinguish these two types of empty cells in this paper. 
 
 Due to the lack of data, it is hard to determine an empty cell's ground truth. For now, we do voting based on labels predicted (by the \gls{ML} model) for random samples from the cell, making the following assumption.
	\begin{assumption}
		\label{assumption_empty_cell_label}
		The accuracy of the \gls{ML} model is better than a classifier doing random classifications in any given cell.
	\end{assumption}
	This assumption essentially relates to the oracle problem of \gls{ML} testing, for which we believe that recent efforts (e.g. \cite{guerriero_reliability_2020}) and future research may relax it.
	
	\item Cross-boundary cells: our estimate of $r$ based on the existing dataset is normally \textit{imperfect}, e.g., due to noise in the dataset and the dataset size is not large enough.
	Thus, we may still observe data-points with different labels in a single cell (especially when new operational data with labels is collected). Such cells are crossing the classification boundary.
	If our estimate on $r$ is sufficiently accurate, they will be very rare. Without the need to determine the ground truth label of a cross boundary cell, we simply and \textit{conservatively} set the cell unastuteness to 1.
\end{itemize}

So far, the problem is reduced to: given a normal or empty cell $c_i$ with the known ground truth label $y_i$, evaluate the misclassification probability upon a random input $x \in c_i$, $\myExp[{\lambda_i}]$, and its variance $\myVar[{\lambda_i}]$. 
This is essentially a statistical problem that has been studied in \cite{webb_statistical_2019} using Multilevel Splitting Sampling, while we use the \gls{SMC} method for brevity in the running example:
$$
\hat{\lambda}_i = \frac{1}{n} \sum_{j = 1}^n I_{\{M(x_j) \neq y_i\}}
$$
The \gls{CLT} tells us $\hat{\lambda}_i \sim \mathcal{N}(\mu, \frac{\sigma^2}{n})$ when $n$ is large, where $\mu$ and $\sigma^2$ are the population mean and variance of $I_{\{\network(x_j) \neq y_i\}}$.
They can be approximated with sample mean $\hat{\mu}_n$ and sample variance $\hat{\sigma}_{n}^2/n$, respectively.
Finally, we get
\begin{align}
	\myExp[{\lambda_i}]&= \hat{\mu}_n = \frac{1}{n} \sum_{j = 1}^n I_{\{\network(x_j) \neq y_i\}} 
	\\ 
	\myVar[{\lambda_i}] &= \frac{\hat{\sigma}_{n}^2}{n} = \frac{1}{(n-1)n} \sum_{j = 1}^n (I_{\{\network(x_j) \neq y_i\}} - \hat{\mu}_{n})^2
\end{align}

Notably, to solve the above statistical problem with sampling methods, we need to assume how the inputs in the cell are distributed, i.e., a distribution for the conditional OP $\mathsf{Op}(x \mid x\in c_i)$. Without loss of generality, we assume:
\begin{assumption}
	\label{assumption_conditonal_OP_uniform}
	The inputs in a small region like a cell are uniformly distributed.
\end{assumption}
\noindent
This assumption is not uncommon (e.g., it is made in \cite{webb_statistical_2019,weng2019proven}) and can be replaced by other distributions, provided there is supporting evidence for such a change.

\paragraph{Step 4: Assembling of the Cell-Wise Estimates}
Eqn.~\eqref{eq_pfd_cell} represents an ideal case in which we know those $\lambda_i$s and $\mathsf{Op}_i$s with certainty. In practice, we can only estimate them with imperfect estimators yielding, e.g., a point estimate with variance capturing the measure of trust\footnote{This aligns with the traditional idea of using \gls{FTA} (and hence the assurance arguments around it) for future reliability assessment.}. To assemble the estimates of $\lambda_i$s and $\mathsf{Op}_i$s to get the estimates on $\lambda$, and also to propagate the confidence in those estimates, we assume:
\begin{assumption}
	\label{assumption_op_lambda_indep}
	All $\lambda_i$s and $\mathsf{Op}_i$s are independent unknown variables under estimations.
\end{assumption}
\noindent Then, the estimate of $\lambda$ and its variance are:

\begin{align}
	\label{eq_expected_pfd}
	\myExp[\lambda] &= \sum_{i = 1}^m \myExp[\lambda_i \mathsf{Op}_i]= \sum_{i = 1}^m \myExp[\lambda_i] \myExp[\mathsf{Op}_i]
	\\
	\myVar[\lambda] &= \sum_{i=1}^m \myVar[\lambda_i \mathsf{Op}_i] = \sum_{i=1}^m \myExp[\lambda_i]^2 \myVar{[\mathsf{Op}_i]} + \myExp[\mathsf{Op}_i]^2 \myVar[\lambda_i]
	+ \myVar[\lambda_i] \myVar[\mathsf{Op}_i] 
	\label{eq_varaince_pfd}
\end{align}
Note, for the variance, the covariance terms are dropped due to the independence assumption.

Depending on the specific estimators adopted, certain parametric families of the distribution of $\lambda$ can be assumed, from which any quantile of interest (e.g., 95\%) can be derived as our confidence bound in reliability. For the running example, we might assume $\lambda \sim \mathcal{N}(\myExp[\lambda], \myVar[\lambda])$ as an approximation by invoking the (generalised) \gls{CLT}\footnote{Assuming $\lambda_i$s and $\mathsf{Op}_i$s are all normally and independently but not identically distributed, the product of two normal variables is approximately normal while the sum of normal variables is exactly normal, thus the variable $\lambda$ is also approximated as being normally distributed (especially when the number of sum terms is large).}. Then, an upper bound with $1-\alpha$ confidence is
\begin{equation}
	\label{eq_pfd_ci}
	\mathit{Ub}_{1-\alpha} =  \myExp[\lambda] + z_{1-\alpha} \sqrt{\myVar[\lambda]} \ ,
\end{equation}
where $Pr(Z \leq z_{1-\alpha}) = 1-\alpha$, and $Z \sim \mathcal{N}(0,1)$ is a standard normal distribution.

\textcolor{black}{
\paragraph{Complexity Analysis on RAM}
The computation complexity of RAM mainly comes from the estimation of $\lambda_i$ and $\mathsf{Op}_i$. For each cell $i$, the SMC requires simulation of $n_1$ samples, while $\mathsf{Op}_i$ is estimated by KDE, trained with $n_2$ collected operational data. The complexity of cell-wise estimation is $\mathcal{O}(n_1 + n_2)$. To get the final estimation of DL model's $\lambda$, we assemble $m$ cells' estimates. This results in the complexity of RAM being $\mathcal{O}(m*(n_1 + n_2))$.
}

\subsection{Extension to High-Dimensional Dataset}
\label{sec_ram_gen_to_hd_ds}

In order to better convey the principles and main steps of our proposed \gls{RAM}, we have demonstrated a ``low-dimensional'' version of our \gls{RAM} which is tailored for the running example (a synthetic 2D-dataset). However, real-world applications normally involve high-dimensional data like images, exposing the presented ``low-dimensional'' \gls{RAM} to scalability challenges. 
In this section, we investigate how to extend our \gls{RAM} for high-dimensional data, and take a few practical solutions to tackle the scalability issues raised by ``the curse of dimensionality''.


\paragraph{Approximating the \gls{OP} in the Latent Feature Space Instead of the Input Pixel Space} The number of cells yielded by the previously discussed way of partitioning the input domain (pixel space) is exponential in the dimensionality of data. Thus, it is hard to accurately approximate the \gls{OP} due to the relatively sparse data collected: the number of cells is usually significantly larger than the number of observations made.
However, for real-world data (say an image), what really determines the label is its \textit{features} rather than the pixels.
Thus, we envisage some latent space, e.g.\ compressed by \gls{VAE}, that captures only the \textit{feature-wise} information; this latent space can be explored for high-dimensional data. 
That is, instead of approximating the \gls{OP} in the input pixel space, we (i) first
encode/project each data-point into the compressed latent space, reducing its dimensionality, (ii) then fit a ``latent space \gls{OP}'' with \gls{KDE} based on the compressed dataset, and (iii) finally ``map'' data-points (paired with the learnt OP) in
the latent space back to the input space.
\begin{remark}[mapping between feature and pixel spaces]
	\label{remark_map}
	Depending on which data compression technique we use and how the ``decoder'' works, the ``map'' action may vary case by case.
	For the \gls{VAE} adopted in our work, we decode one point from the latent space as a ``clean'' image (with only feature-wise information), and then add perturbations to generate a norm ball (with a size determined by the $r$-separation distance, cf.\ Remark \ref{remark_r_sep}) in the input pixel space.
\end{remark}

\paragraph{Applying Efficient Multivariate \gls{KDE} for Cell OP Approximation} We may encounter technical challenges when fitting the \gls{PDF} from high-dimensional datasets. There are two known major challenges when applying \textit{multivariate} \gls{KDE} to high-dimensional data: 
i) the choice of bandwidth $H$ represents the covariance matrix that mostly impacts the estimation accuracy; and
ii) scalability issues in terms of storing intermediate data structure (e.g., data-points in hash-tables) and querying times made when estimating the density at a given input.
For the first challenge, the optimal calculation of the bandwidth matrix can refer to some rule of thumb \cite{silverman1986density,scott2015multivariate} and the cross-validation \cite{bergstra_random_2012}. 
There is also dedicated research on improving the efficiency of multivariate \gls{KDE}, e.g., \cite{backurs2019space} presents a framework for multivariate \gls{KDE} in provably sub-linear query time with linear space and linear pre-processing time to the dimensions.

\paragraph{Applying Efficient Estimators for Cell Robustness}
We have demonstrated the use of \gls{SMC} to evaluate cell robustness in our running example.
It is known that \gls{SMC} is not computationally efficient to estimate rare-events, such as \gls{AE}s in the high-dimensional space of a robust ML model. 
We therefore need more advanced and efficient sampling approaches that are designed for rare-events to satisfy our need.
We notice that the Adaptive Multi-level Splitting method has been retrofitted in \cite{webb_statistical_2019} to statistically estimate the model's local robustness, which can be (and indeed has been) applied in our later experiments for image datasets.
In addition to statistical approaches, formal method based verification techniques might also be applied to assess a cell's \textit{pmi}, e.g., \cite{huang_safety_2017}.
They provide formal guarantees on whether or not the \gls{ML} model will misclassify any input inside a small region. Such ``robust region'' proved by formal methods is normally smaller than our cells, in which case the $\hat{\lambda}_i$ can be conservatively set as the proportion of the robust region covered in cell $c_i$ (under Assumption \ref{assumption_conditonal_OP_uniform}).

\paragraph{Assembling a Limited Number of Cell-Wise Estimates with Informed Uncertainty}
The number of cells yielded by current way of partitioning the input domain is exponential to the dimensionality of data, thus it is impossible to explore all cells for high-dimensional data as we did for the running example. We may have to limit the number of cells under robustness evaluation due to the limited budget in practice. Consequently, in the final ``assembling'' step of our \gls{RAM}, we can only assemble a limited number of cells, say $k$, instead of all $m$ cells. 
In this case, we refer to the estimator designed for weighted average based on samples \cite{bevington_data_1993}.
Specifically, we proceed as what follows:
\begin{itemize}
	\item Based on the collected dataset with $n$ data-points, the \gls{OP} is approximated in a latent space, which is compressed by VAE. 
	Then we may obtain a set of $n$ norm balls (paired with their OP) after mapping the compressed dataset to the input space (cf. Remark \ref{remark_map}) as the sample frame\footnote{While the population is the set of (non-overlapping) norm balls covering the whole input space, i.e. the $m$ cells mentioned in the ``lower-dimensional'' version of the RAM.}.
	\item We define weight $w_i$ for each of the $n$ norm balls according to their approximated \gls{OP}, $w_i:=\myExp[\mathsf{Op}_i]$.
	\item Given a budget that we can only evaluate the robustness of $k$ norm balls,
	$k$ samples are randomly selected (with replacement) and fed into the robustness estimator to get $\myExp[\lambda_i]$. 
	\item We may invoke the unbiased estimator for weighted average \cite[Chapter 4]{bevington_data_1993} as
	\begin{align}
		\label{eq_weighted_avg_est_mean}
		\myExp[\lambda]&=\frac{\sum_{i=1}^{k} w_i \myExp[\lambda_i] }{\sum_{i=1}^{k} w_i} \ \mbox{and}
		\\
		\myVar[\lambda]&=\frac{1}{k-1}\left( \frac{\sum_{i=1}^{k} w_i \left(\myExp[\lambda_i]\right)^2 }{\sum_{i=1}^{k} w_i} - (\myExp[\lambda])^2\right) \ .
		\label{eq_weighted_avg_est_var}
	\end{align}
	Moreover, a confidence upper bound of interest can be derived from Eqn.~\eqref{eq_pfd_ci}.
\end{itemize}
Note that there is no variance terms of $\lambda_i$ and $\mathsf{Op}_i$ in Eqn.s~\eqref{eq_weighted_avg_est_mean} and \eqref{eq_weighted_avg_est_var}, implying the following assumption:
\begin{assumption}
	\label{assumption_k_is_the_major_source_uncertainty}
	The uncertainty informed by Eqn.~\eqref{eq_weighted_avg_est_var} is sourced from the sampling of $k$ norm balls, which is assumed to be the major source of uncertainty.
	This makes the uncertainties contributed by the robustness and \gls{OP} estimators (i.e. the variance terms of $\lambda_i$ and $\mathsf{Op}_i$) negligible.
\end{assumption}

\textcolor{black}{\paragraph{Complexity Analysis on RAM Extension to High-Dimensional Dataset} For high dimensional data, RAM still adopts the KDE, fitted with $n_2$ operational data projected into low dimensional latent feature space. The main difference is the use of more efficient estimators for cell robustness. We refer to the Adaptive Multi-level Splitting method, an advanced Monte Carlo Simulation, that has been used for local robustness estimation in \cite{webb_statistical_2019} and our experiments. If SMC requires the number of simulations at the \textit{order} of $n_1$ for accurate estimation of rare events with probability $1/n_1$ (while omitting the coefficient, cf.~\cite{littlewood_validation_1993} for detailed analytical results), the Adaptive Multi-level Splitting method (cf. \cite{webb_statistical_2019} for more algorithm details) utilises the product of conditional probability with $\log{}n_1$ levels (the quantile $\rho$ is normally set to 0.1). If $n_3$ samples are simulated by SMC for calculating each conditional probability, the computation cost of Adaptive Multi-level Splitting method is $n_3\log{}n_1 \ll n_1$. Finally, we invoke the weighted sampling of $k$ cells for cell-wise estimates assembling, the complexity of which is $\mathcal{O}(k*(n_3\log{}n_1 + n_2))$. 
}

\subsection{Evaluation on the Proposed \gls{RAM}}
\label{sec_ram_evaluation}

\begin{figure}[ht]
	\centering
	\includegraphics[width=0.7\linewidth]{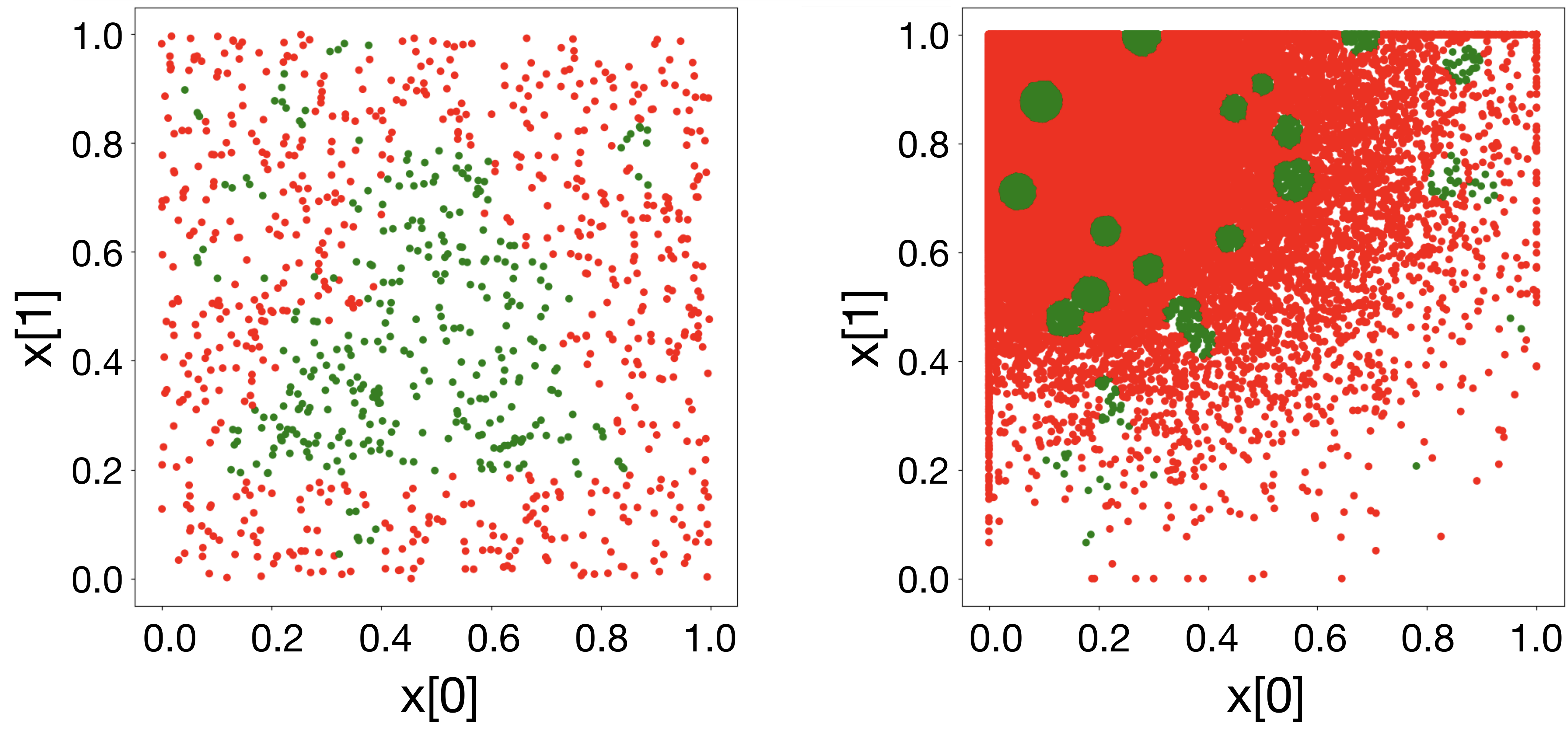}
	\caption{\textcolor{black}{Synthetic datasets DS-1 (lhs) and DS-2 (rhs) representing relatively sparse and dense training data respectively.}}
	\label{fig_two_extra_datasets}
\end{figure}

In addition to the running example, we conduct experiments on two more synthetic 2D-datasets, as shown in Fig. \ref{fig_two_extra_datasets}.
They represent scenarios with relatively sparse and dense training data, respectively.
Moreover, to gain insights on how to extend our RAM for high-dimensional datasets, we also conduct experiments on the popular MNIST and CIFAR10 datasets,
as articulated in Section \ref{sec_ram_gen_to_hd_ds}. 
All modelling details and results after applying our \gls{RAM} on those datasets are summarised in Table \ref{table_model_details}, where we compare the testing error, \gls{ACU} defined by Definition \ref{def_acu}, and our \gls{RAM} results (of the mean $\myExp[\lambda]$, variance $\myVar[\lambda]$ and a 97.5\% confidence upper bound $Ub_{97.5\%}$).

\begin{definition}[\gls{ACU}]
	\label{def_acu}
	Stemmed from the Definition \ref{eq_robust_def} and Remark \ref{remark_astutenss}, the unastuteness $\lambda_i$ of a region $c_i$ is consequently $1-R_{\network}(c_i, y_i)$ where $y_i$ is the ground truth label of $c_i$ (cf. Eqn.~\ref{eq_cell_rel_op}). Then we define the \gls{ACU} of the ML model as:
	\begin{equation}
		\label{eq_acu_def}
		ACU:=\frac{1}{m}\sum_{i=1}^{m} \lambda_i
	\end{equation}
	where $m$ is the total number of regions.
\end{definition}

\label{sec_evaluation}
\begin{table*}[h]
	\centering
	\caption{Modelling details \& results of applying the \gls{RAM} on five datasets. Time is in seconds per cell.}
	\resizebox{\textwidth}{!}{
		\begin{tabular}{c|c|c|c|c|c|c|c|c|c}
			\hline
			&   train/test error  & $r$-separation & radius $\epsilon$ & \# of cells & \gls{ACU}& $\myExp[\lambda]$ & $\myVar[\lambda]$ & $Ub_{97.5\%}$ & time \\ \hline\hline
			The run. exp. &  0.0005/0.0180  &   0.004013          &      0.004       &    $250\times250$    & 0.002982 &   0.004891       &          0.000004      & 0.004899  & 0.04       \\
			Synth. DS-1 & 0.0037/0.0800  &  0.004392 &      0.004      &     $250\times250$   &   0.008025   &  0.008290    &      0.000014       &     0.008319 &  0.03 \\
			Synth. DS-2 &  0.0004/0.0079&     0.002001       &         0.002               &        $500\times500$       &   0.004739  &   0.005249          &           0.000002        &       0.005252   &  0.04  \\ \hline
			Norm. MNIST    &  0.0051/0.0235   & 0.369          &      0.300          &        $k$       &  Fig.~\ref{fig_pmi_acu_mnist}(b)    &     Fig.~\ref{fig_pmi_acu_mnist}(a)      &           Fig.~\ref{fig_pmi_acu_mnist}(a)   &     Fig.~\ref{fig_pmi_acu_mnist}(a)   &   0.43 \\
			Adv. MNIST    &  0.0173/0.0212   & 0.369          &      0.300          &        $k$       &  Fig.~\ref{fig_pmi_acu_mnist}(d)    &     Fig.~\ref{fig_pmi_acu_mnist}(c)      &           Fig.~\ref{fig_pmi_acu_mnist}(c)   &     Fig.~\ref{fig_pmi_acu_mnist}(c)   &   0.43 \\ \hline
			Norm. CIFAR10       &     0.0190/0.0854    &    0.106   &          0.100              &       $k$        &       Fig.~\ref{fig_pmi_acu_cifar10}(b)     &   Fig.~\ref{fig_pmi_acu_cifar10}(a)  &      Fig.~\ref{fig_pmi_acu_cifar10}(a)         &       Fig.~\ref{fig_pmi_acu_cifar10}(a) & 6.74      \\
			Adv. CIFAR10       &     0.0013/0.1628    &    0.106   &          0.100              &       $k$        &       Fig.~\ref{fig_pmi_acu_cifar10}(d)     &   Fig.~\ref{fig_pmi_acu_cifar10}(c)  &      Fig.~\ref{fig_pmi_acu_cifar10}(c)         &       Fig.~\ref{fig_pmi_acu_cifar10}(c) & 6.74      \\
			\hline
	\end{tabular}}
	\label{table_model_details}
\end{table*}

In the running example, we first observe that the \gls{ACU} is much lower than the testing error, which means that the underlying \gls{ML} model is a robust one. 
Since our \gls{RAM} is largely based on the robustness evidence, its results are close to \gls{ACU}, but not exactly the same because of the nonuniform OP, cf.\ Fig. \ref{fig_running_example}-rhs.
\begin{remark}[ACU is a special case of \textit{pmi}]
	\label{remark_acu_equals_ram_when_op_flat}
	When the \gls{OP} is ``flat'' (uniformly distributed), \gls{ACU} and our \gls{RAM} result regarding \textit{pmi} are equal, which can be seen from Eqn.~\ref{eq_pfd_cell} by setting all $\mathsf{Op}_i$s equally to $\frac{1}{m}$.
\end{remark} 
\noindent Moreover, from Fig. \ref{fig_running_example}-lhs, we know that the classification boundary is near the middle of the unit square input space where misclassifications tend to happen (say, a ``buggy area''), which is also the high density area on the OP. 
Thus, the contribution to unreliability from the ``buggy area'' is weighted higher by the OP, explaining why our \gls{RAM} results are worse than the \gls{ACU}.
In contrast, because of the relatively ``flat'' OP for the DS-1 (cf.\ Fig. \ref{fig_two_extra_datasets}-lhs), our \gls{RAM} result is very close to the \gls{ACU} (cf.\ Remark \ref{remark_acu_equals_ram_when_op_flat}). 
With more dense data in DS-2, the $r$-distance is much smaller and leads to smaller cell radius and more cells. 
Thanks to the rich data in this case, all three results (testing error, ACU, and the RAM) are 
more consistent than in the other two cases.
We note that, given the nature of the three 2D-point datasets, \gls{ML} models trained on them are much more robust than image datasets. 
This is why all \gls{ACU}s are better than test errors, and our \gls{RAM} finds a middle point representing reliability according to the OP.
Later we apply the \gls{RAM} on unrobust (by normal training) and robust (by adversarial training) \gls{ML} models trained on image datasets, where the \gls{ACU}s are worse and better than the test error, respectively; it confirms our aforementioned observations.



\begin{figure*}[ht]
	\centering
	\includegraphics[width=\linewidth]{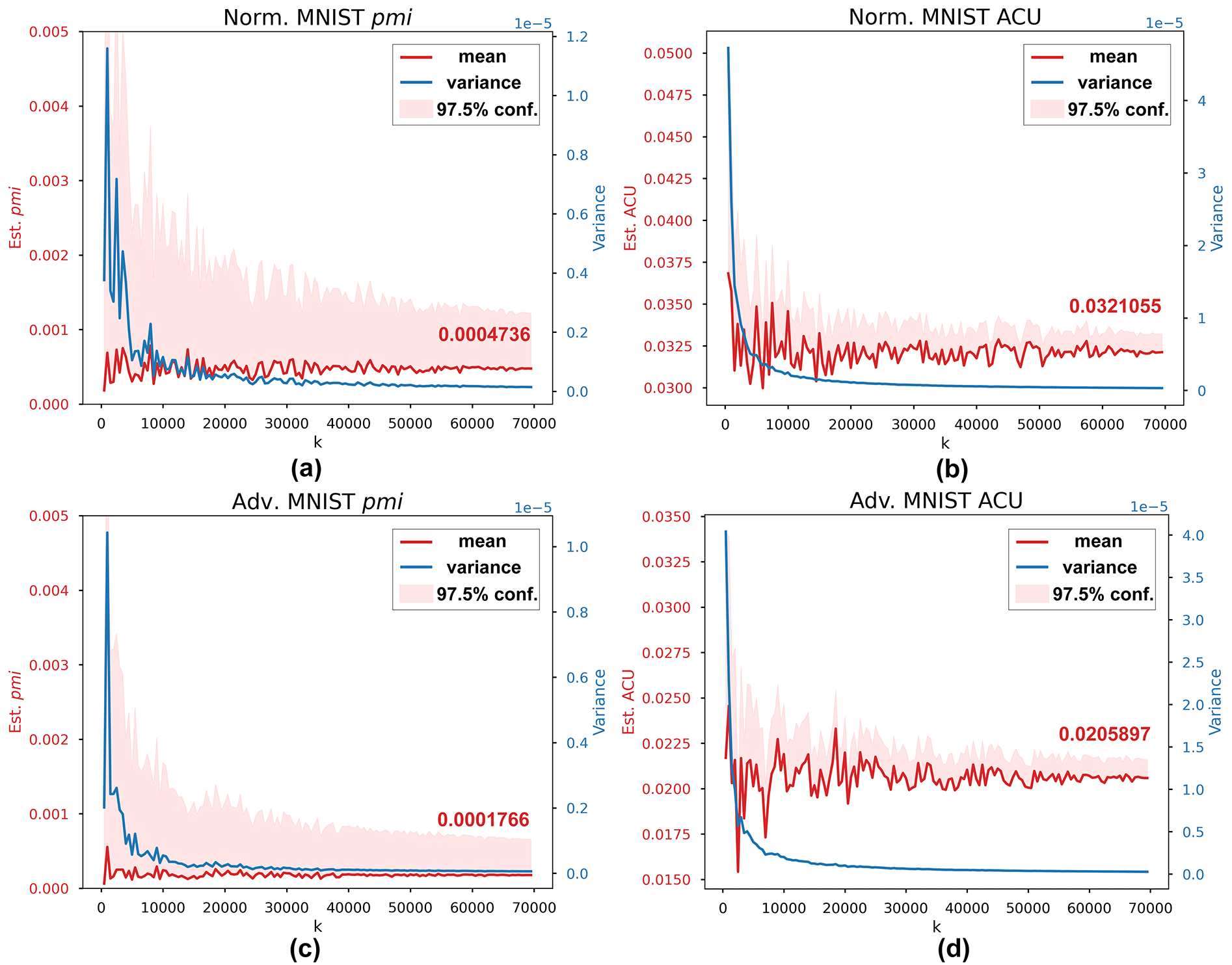}
	\caption{\textcolor{black}{The mean, variance and 97.5\% confidence upper bound of \textit{pmi} and ACU as functions of $k$ sampled norm ball, estimated on MNIST dataset with normally and adversarially trained models.}}
	\label{fig_pmi_acu_mnist}
\end{figure*}

\begin{figure*}[ht]
	\centering
	\includegraphics[width=\linewidth]{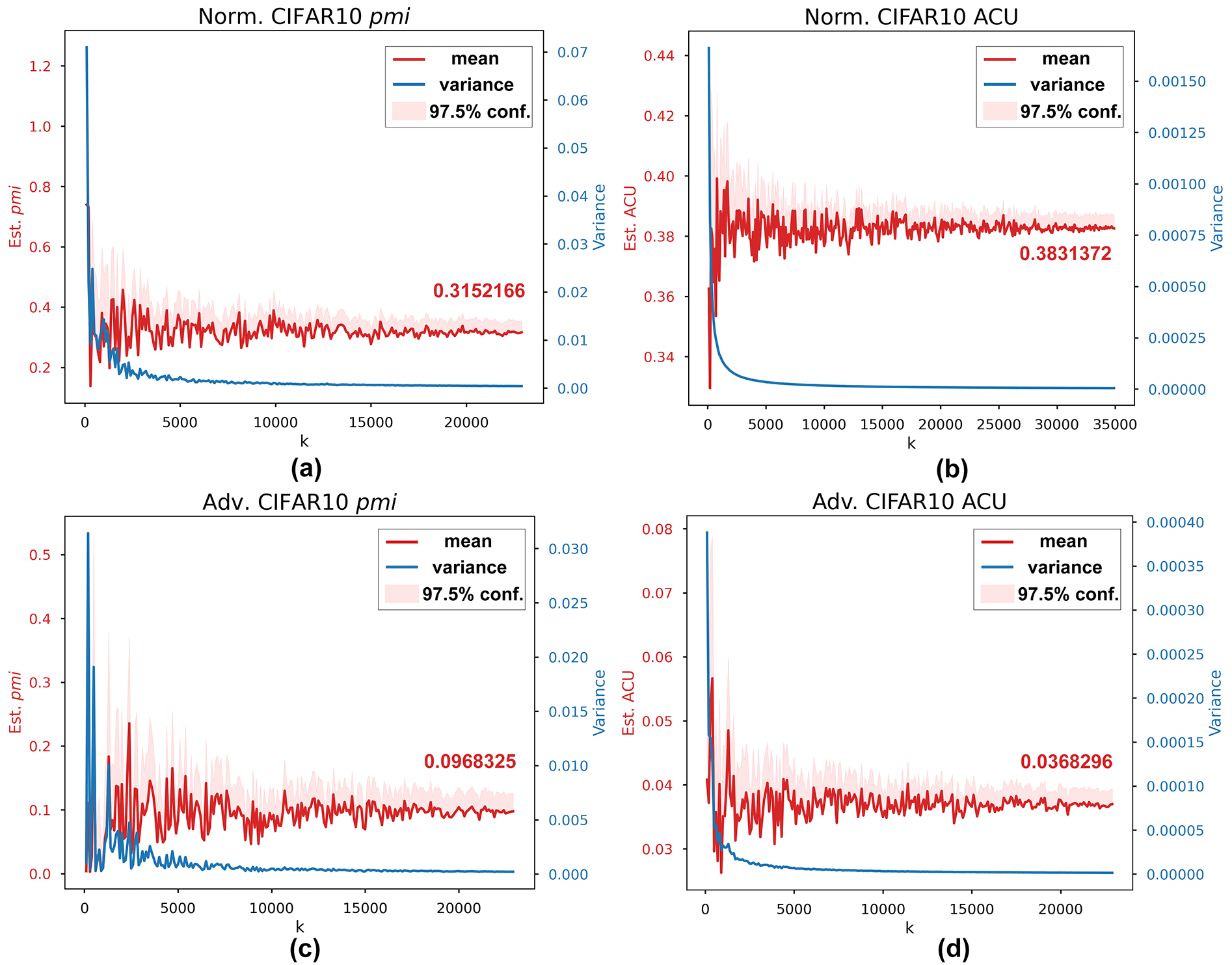}
	\caption{\textcolor{black}{The mean, variance and 97.5\% confidence upper bound of \textit{pmi} and ACU as functions of $k$ sampled norm ball, estimated on CIFAR10 dataset with normally and adversarially trained models.}}
	\label{fig_pmi_acu_cifar10}
\end{figure*}

\textcolor{black}{Regarding the MNIST and CIFAR10 datasets, all the experiment codes for this running example are publicly available at \url{https://github.com/havelhuang/ReAsDL}, and the model details are presented in Appendix \ref{sec:mnist_cifar_model_details}.} In this section, we first train \gls{VAE}s on them and compress the datasets into the low dimensional latent spaces of \gls{VAE}s with 8 and 16 dimensions, respectively.
We then fit the compressed dataset with KDE to approximate the OP. 
Each compressed data-point is now associated with a weight representing its OP. 
Consequently, each norm ball in the pixel space that corresponds to the compressed data-point in the latent space (after the mapping, cf.\ Remark \ref{remark_map}) is also weighted by the OP.
Taking the computational cost into account---say only the astuteness evaluation on a limited number of $k$ norm balls is affordable---we do random sampling, invoke the estimator for \textit{weighted average} Eqn.s~\eqref{eq_weighted_avg_est_mean} and \eqref{eq_weighted_avg_est_var}. \textcolor{black}{We training two DL models with normal training strategy and PGD-based adversarial training strategy \cite{madry2018towards}, respectively,} and plot our RAM results for both models as functions of $k$ in (a) and (c) of Figures \ref{fig_pmi_acu_mnist} and \ref{fig_pmi_acu_cifar10}.
For comparison, we also plot the ACU results\footnote{As per Remark \ref{remark_acu_equals_ram_when_op_flat}, ACU is a special case of \textit{pmi} with equal weights. 
	Thus, ACU results in Fig. \ref{fig_pmi_acu_mnist}, \ref{fig_pmi_acu_cifar10} are also obtained by Eqn.s \eqref{eq_weighted_avg_est_mean} and \eqref{eq_weighted_avg_est_var}.} in (b) and (d) of Figures \ref{fig_pmi_acu_mnist} and \ref{fig_pmi_acu_cifar10}.

In Figures \ref{fig_pmi_acu_mnist} and \ref{fig_pmi_acu_cifar10}, we first observe that both the ACU results (after converging) of normally trained MNIST and CIFAR10 models are worse than their test errors (in Table \ref{table_model_details}), unveiling again the robustness issues of ML models when dealing with image datasets (while the ACU of CIFAR10 is even worse, given that CIFAR10 is indeed a generally harder dataset than MNIST).
For MNIST, the mean \textit{pmi} estimates are much lower than ACU, implying a very ``unbalanced'' distribution of weights (i.e.\ OP).
Such unevenly distributed weights are also reflected in both, the oscillation of the variance and the relatively loose 97.5\% confidence upper bound.
On the other hand, the OP of CIFAR10 is flatter, resulting in closer estimates of \textit{pmi} and ACU (Remark \ref{remark_acu_equals_ram_when_op_flat}). \textcolor{black}{For adversarially trained models, the robustness of which is improved significantly at the cost of accuracy drop shown in Table~\ref{table_model_details}. It is still effective to reduce the $pmi$ and ACU of DL models.}

\textcolor{black}{In summary, for real-world image datasets, our RAM may effectively assess the robustness of the ML model and its generalisability based on the shape of its approximated OP, which is much more informative than either the test error or ACU alone. Finally, based on the RAM, templates of probabilistic arguments for reliability claims on ML components are developed, cf. Appendix \ref{sec_prob_safe_argument}.}

\section{Case Studies}
\label{sec_case_study}

In this section, a case study based on a simulated \gls{AUV} that performs survey and asset inspection missions is conducted.
We first describe the scenario in which the mission is performed, details of the AUV under test, and how the simulator is implemented. Then, corresponding to Section \ref{sec_overall_framework}, we exercise the proposed assurance activities for this AUV application, i.e., HAZOP, hazards scenarios modelling, FTA, and discussions on deriving the system-level quantitative safety target for this scenario.
Finally, we apply our RAM on the image dataset collected from a large amount of statistical testing. 

\subsection{Scenario Design}
\gls{AUV} are increasingly adopted for marine science, offshore energy, and other industrial applications in order to increase productivity and effectiveness as well as to reduce human risks and offshore operation of crewed surface support vessels \cite{lane_new_2016}.
However, the fact that AUVs frequently operate in close proximity to safety-critical assets (e.g., offshore oil rigs and wind turbines) for inspection, repair and maintenance tasks leads to challenges on the assurance of their reliability and safety, which motivates the choice of AUV as the object of our case study.

\subsubsection{Mission Description and Identification of Mission Properties}

Based on industrial use cases of autonomous underwater inspection, we define a test scenario for AUVs that need to operate autonomously and carry out a survey and asset inspection mission, in which an AUV follows several way-points and terminates with autonomous docking.
During the mission, it needs to detect and recognise a set of underwater objects (such as oil pipelines and wind farm power cables) and inspect assets (i.e., objects) of interest, while avoiding obstacles and keeping the required safe distances to the assets.

Given the safety/business-critical mission, different stakeholders have their own interests on a specific set of hazards and safety elements. 
For instance, asset owners (e.g., wind farm operators) focus more on the safety and health of the assets that are scheduled to be inspected, whereas inspection service providers tend to have additional concerns regarding the safety and reliability of their inspection service and vehicles. 
In contrast, regulators and policy makers may be more interested in environmental and societal impacts that may arise when a failure unfortunately happens.
By keeping these different safety concerns in mind, we identify a set of desirable \textbf{mission properties}, whose violation may lead to unsuccessful inspection missions, compromise the integrity of critical assets, or damage of the vehicle itself.

While numerous high-level mission properties are identified based on our engineering experience, references to publications (e.g., \cite{hereau_testing_2020}) and iterations of hazard analysis, we focus on a few that are instructive for the ML classification function in this article (cf.\ the project website for a complete list):
\begin{itemize}
	\item No miss of key assets: the total number of correctly recognised assets/objects should be equal to the total number of assets that are required to be inspected during the mission.
	\item No collision: during the full mission, the AUV should avoid all obstacles perceived without collision.
	\item Safe distancing: once an asset is detected and recognised, the Euclidean distance between the AUV and the asset must be kept to be at least the defined minimal safe operating distance.
	\item Autonomous docking: safe and reliable docking to the docking cage.
\end{itemize}
Notably, such an initial set of desirable mission properties forms the starting point of our assurance activities, cf.\ Fig. \ref{fig_hazop_fta_flow} and Section \ref{sec_assurance_activity_auv_case_study}.

\subsubsection{The AUV Under Test}

\begin{figure*}[h]
	\centering
	\includegraphics[width=0.8\textwidth]{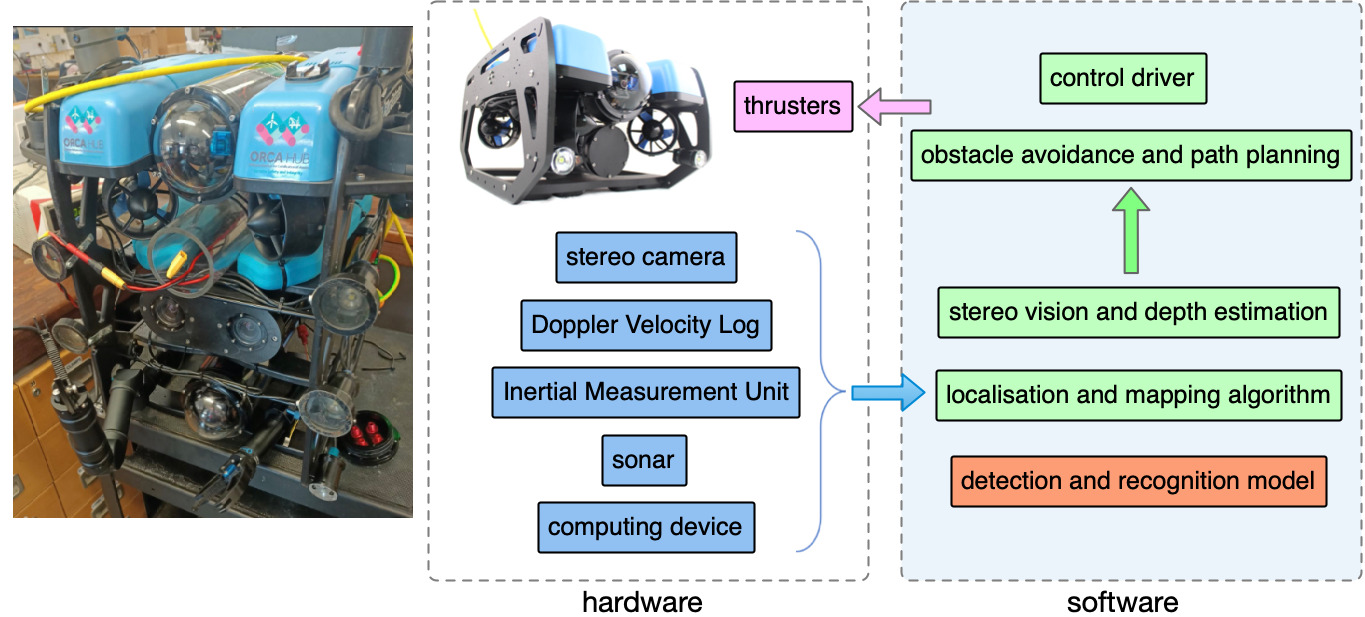}
	\caption{Hardware--software architecture \& key modules for autonomous survey \& inspection missions.}
	\label{fig_software_archi}
\end{figure*}

\paragraph{Hardware} 
Although we are only conducting experiments in simulators at this stage, our trained ML model can be easily deployed to real robots and the experiments are expected to be reproducible in real water tanks. Thus, we simulate the AUV in our laboratory---a customised BlueROV2, which has 4 vertical and 4 horizontal thrusters for 6 degrees of freedom motion. 
As shown in Fig. \ref{fig_software_archi}-lhs, it is equipped with a custom underwater stereo camera designed for underwater inspection. A Water Linked A50 \gls{DVL} is installed for velocity estimation and control. The AUV also carries an \gls{IMU}, a depth sensor and a Tritech Micron sonar. The AUV is extended with an on-board Nvidia Jetson Xavier GPU computer and a Raspberry Pi 4 embedded computer. An external PC can also be used for data communication, remote control, mission monitoring, and data visualisation of the AUV via its tether.

\paragraph{Software Architecture} With the hardware platform, we develop a software stack for underwater autonomy based on the \gls{ROS}. The software modules that are relevant to the aforementioned AUV missions are (cf.\ Fig. \ref{fig_software_archi}):
\begin{itemize}
	\item Sensor drivers. All sensors are connected to on-board computers via cables, and their software drivers are deployed to capture real-time sensing data.
	\item Stereo vision and depth estimation. This is to process stereo images by removing its distortion and enhancing its image quality for inspection. After rectifying stereo images, they are used for estimating depth maps that are used for 3D mapping and obstacle avoidance.
	\item Localisation and mapping algorithm. In order to navigate autonomously and carry out a mission, we need to localise the vehicle and build a map for navigation. We develop a graph optimisation based underwater simultaneous localisation and mapping system by fusing stereo vision, DVL, and IMU. 
	It also builds a dense 3D reconstruction model of structures for geometric inspection.
	\item Detection and recognition model. This is one of the core modules for underwater inspection based on ML models. It is designed to detect and recognise objects of interest in real-time. Based on the properties of detected objects--- in particular the underwater assets to inspect---the AUV makes decisions on visual data collection and inspection.
	\item Obstacle avoidance and path planning. The built 3D map and its depth estimation are used for path planning, considering obstacles perceived by the stereo vision. Specifically, a local trajectory path and its way-points are generated in the 3D operating space based on the 3D map built from the localisation and mapping algorithm. Next the computed way-point is passed to the control driver for trajectory and way-point following.
	\item Control driver. We have a back seat driver for autonomous operations, enabling the robot to operate as an AUV. Once the planned path and/or a way-point is received, a \gls{PID} based controller is used to drive the thrusters following the path and approaching to the way-point. The controller can also be replaced by a learning based adaptive controller. While the robot moves in the environment, it continues perceiving the surrounding scene and processing the data using the previous software modules.
\end{itemize}

\paragraph{ML Model Doing Object Detection}

In this work, the state-of-the-art Yolo-v3 \gls{DL} architecture \cite{redmon2018yolov3} is used for object detection. Its computational efficiency and real-time performance are both critical for its application for underwater robots, as they mostly have limited on-board computing resources and power.
The inference of Yolo can be up to 100 frames per second. 
Yolo models are also open source and built using the C language and the library is officially supported by OpenCV, which makes its integration with other AUV systems not covered in this work straightforward.
Most \gls{DL}-based object detection methods are extensions of a simple classification network. 
The object detection network usually generates a set of proposal bounding boxes; they might contain an object of interest and are then fed to a classification network.
The Yolov3 network is similar in operation to, and is based on, the \textit{darknet53} classification network.

The process of training the Yolo networks using the Darknet framework is similar to the training of most \gls{ML} models, which includes data collection, model architecture implementation, and training. 
The framework consists of configuration files that can be set to match the number of object classes and other network parameters.
Examples of training and testing data are described in Section \ref{sec_simulator} for simulated version of the model.
The model training can be summarised by the following steps:
i) define the number of object categories; 
ii) collect sufficient data samples for each category;
iii) split the data into training and validation sets; and
iv) use the Darknet software framework to train the model.

\subsubsection{The Simulator}
\label{sec_simulator}

\begin{figure*}[h!]
	\centering
	\includegraphics[width=\textwidth]{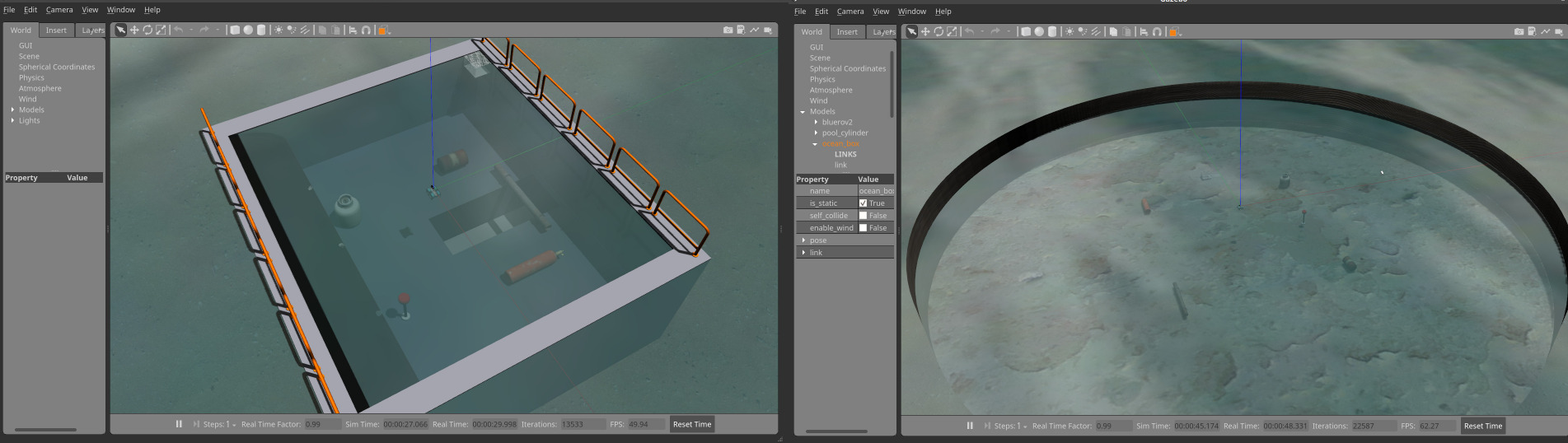}
	\caption {A wave-tank for simulated testing and a simulated pool for collecting the training data.}
	\label{fig_wavetank_gazebo}
\end{figure*}

The simulator uses the popular Gazebo robotics simulator in combination with a simulator for underwater dynamics. 
The scenario models can be created/edited using Blender 3D software. 
We have designed the Ocean Systems Lab's wave tank model (cf.\ Fig. \ref{fig_wavetank_gazebo}-lhs) for the indoor simulated demo, using BlueROV2 within the simulation to test the scenarios. 
The wave tank model has the same dimension as our real tank.
To ensure that the model does not overfit the data, we have designed another scenario with a bigger pool for collecting the training data. The larger size allows for more distance between multiple objects, allowing both to broaden the set training scenarios and to make them more realistic.
The simulated training environment is presented in Fig. \ref{fig_wavetank_gazebo}-rhs.

\begin{figure*}[h!]
	\centering
	\includegraphics[width=\textwidth]{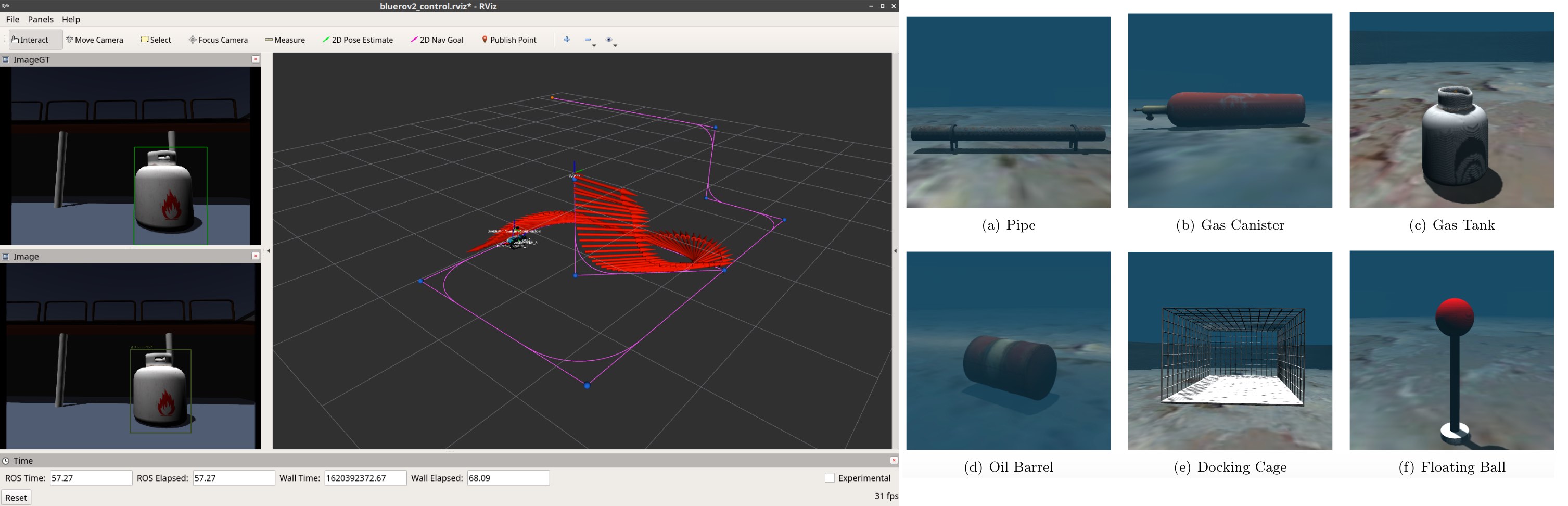}
	\caption {Simulated AUV missions following way-points and the six simulated objects.}
	\label{fig_sim_waypoints_objects}
\end{figure*}

Our simulator creates configuration files to define an automated path using Cartesian way-points for the vehicle to follow autonomously, which can be visualised using Rviz. 
The pink trajectory is the desirable path and the red arrows represent the vehicle poses following the path, cf.\ Fig. \ref{fig_sim_waypoints_objects}-lhs.
There are six simulated objects in the water tank. They are a pipe, a gas tank, a gas canister, an oil barrel, a floating ball, and the docking cage, as shown in Fig. \ref{fig_sim_waypoints_objects}-rhs. 
The underwater vehicle needs to accurately and timely detect them during the mission.
Notably, the mission is also subject to random noise factors, so that repeated missions will generate different data that is processed by the learning-enabled components.

\subsection{Assurance Activities for the AUV}
\label{sec_assurance_activity_auv_case_study}

\paragraph{Hazard Analysis via HAZOP}


\textcolor{black}{Given the AUV system architecture (cf.\ Fig. \ref{fig_software_archi}) and control/data flow among the nodes, we conducted a HAZOP analysis that yields a complete version of Table \ref{tab_hazop_partial_results} in \cite{hills2022}. 
For this work, we only present partial HAZOP results and highlight a few hazards that are due to misclassification.}

\begin{table*}[!h]
	\centering
	\resizebox{0.9\textwidth}{!}{
		\begin{tabular}{@{}|c|c|c|c|c|c|@{}}
			\toprule
			\begin{tabular}[c]{@{}c@{}}HAZOP item:\\ node/flow\end{tabular}                                                                  & \begin{tabular}[c]{@{}c@{}}Process \\ parameter \\ or attribute\end{tabular} & Guide-word  & Cause              & Consequence                                                                                                                           & Mitigation                                                                                                                                                                          \\ \midrule
			\multirow{5}{*}{\begin{tabular}[c]{@{}c@{}}flow from object\\ detection to \\obstacle avoidance\\ and path \\ planning\end{tabular}} & \multirow{2}{*}{data flow}                                              & too late    & ...                & ...                                                                                                                                   & ...                                                                                                                                                                                 \\ \cmidrule(l){3-6} 
			&                                                                         & ...         & ...                & ...                                                                                                                                   & ...                                                                                                                                                                                 \\ \cmidrule(l){2-6} 
			& \multirow{3}{*}{data value}                                             & wrong value & misclassification & \begin{tabular}[c]{@{}c@{}}erratic navigation; \\ unsafe distance to assets; \\collision to assets;\\  failed inspection.\end{tabular} & \begin{tabular}[c]{@{}c@{}}acoustic guidance;\\  minimum DL-classifier \\ reliability for critical\\ objects;  maximum safe\\ distance  maintained \\if uncertain;\\ ...\end{tabular} \\ \cmidrule(l){3-6} 
			&                                                                         & no value    & ...                & ...                                                                                                                                   & ...                                                                                                                                                                                 \\ \cmidrule(l){3-6} 
			&                                                                         & ...         & ...                & ...                                                                                                                                   & ...                                                                                                                                                                                 \\ \midrule
			...                                                                                                                              & ...                                                                     & ...         & ...                & ...                                                                                                                                   & ...                                                                                                                                                                                 \\ \bottomrule
	\end{tabular}}
	\caption{Partial HAZOP results, highlighting the cause of misclassification (NB, entries of ``...'' are intentionally left blank, \textcolor{black}{cf. \cite{hills2022} and our public project repository for a complete version)}.}
	\label{tab_hazop_partial_results}
\end{table*}

\begin{figure*}[!h]
	\centering
	\includegraphics[width=0.8\textwidth]{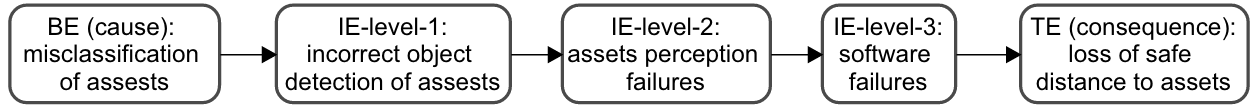}
	\caption{An event-chain based on the hazard scenario modelling, linking causes to consequences.}
	\label{fig_event_chain}
\end{figure*}

\begin{figure*}[!h]
	\centering
	\includegraphics[width=1\textwidth]{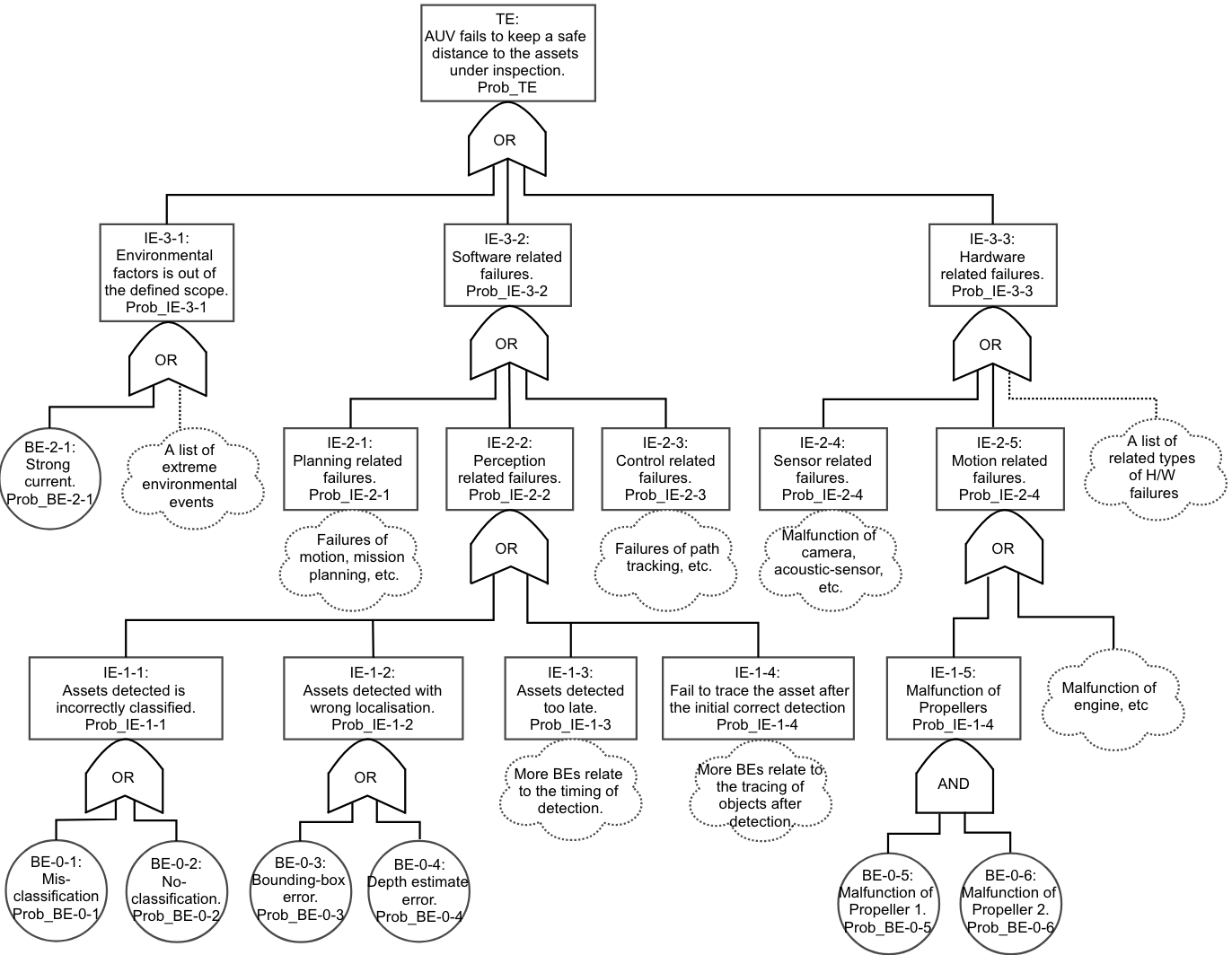}
	\caption{A partial fault tree for the TE of loss of a safe distance to assets. NB, the ``cloud'' notation represents omitted sub-trees.}
	\label{fig_fta}
\end{figure*}

\paragraph{Hazard Scenarios Modelling} Inspired by \cite{guo_extended_2015}, we develop the hazard scenarios as chains of events that link the causes to consequences identified by HAZOP. 
Again, for illustration, a single event-chain is shown in Fig. \ref{fig_event_chain}, which propagates the event of misclassification on assets via the system architecture to the violation of mission property of keeping a safe distance to assets.
Later, readers will see this event-chain forms one path of a fault tree in the FTA in Fig. \ref{fig_fta}.

\paragraph{Quantitative FTA}


We first construct fault trees for each hazard (as TE) identified by HAZOP, by extending and combining (via logic gates) the IEs modelled by hazard scenario analysis.
Each event-chain yielded by the hazard scenario analysis then forms one path in a fault tree.
For instance, the event-chain of Fig. \ref{fig_event_chain} eventually becomes the path of \textbf{BE-0-1 $\rightarrow$ IE-1-1 $\rightarrow$ IE-2-2 $\rightarrow$ IE-3-2 $\rightarrow$ TE} in Fig. \ref{fig_fta}.
Finally, knowing the probabilities of BEs and logic gates allows for the calculation of the TE probability.
As shown by the second iteration loop in Fig. \ref{fig_hazop_fta_flow}, several rounds of what-if calculations, sensitivity analysis and updates of the components are expected to yield the most practical solution of BE probabilities that associates with a given tolerable risk of the TE. 


\paragraph{Deriving Quantitative System Safety Target}


Based on the experience of relatively more developed safety-critical domains of AI, such as self-driving cars and medical devices (cf. Section~\ref{sec_dqrmlc} for some examples), we believe that
referring to the average performance of human divers and/or human remote control operators is a promising way of determining the high-level quantitative safety target for our case of an AUV.
It is presumed that, prior to the use of an AUV for assets inspection, human divers and remotely controlled robots need to conduct the task regularly. This is also similar to how the safety targets were developed in the civil aircraft sector where they refer to acceptable historical accident rates as the benchmark.
In our case, referring to the human-divers/operators' performance as a target for an AUV's safety risk can be potentially impeded by the lack of historical/statistical data on such performance. 
Given the fact that ML for AUV is a relatively novel technique and still developing and transforming to its practical uses, an urgent lesson learnt for all AUV stakeholders (especially manufacturers, operators and end users) from this work is to collect and summarise such data.

\subsection{Reliability Modelling of the AUV's Classification Function}

\begin{figure*}[ht]
	\centering
	\includegraphics[width=\linewidth]{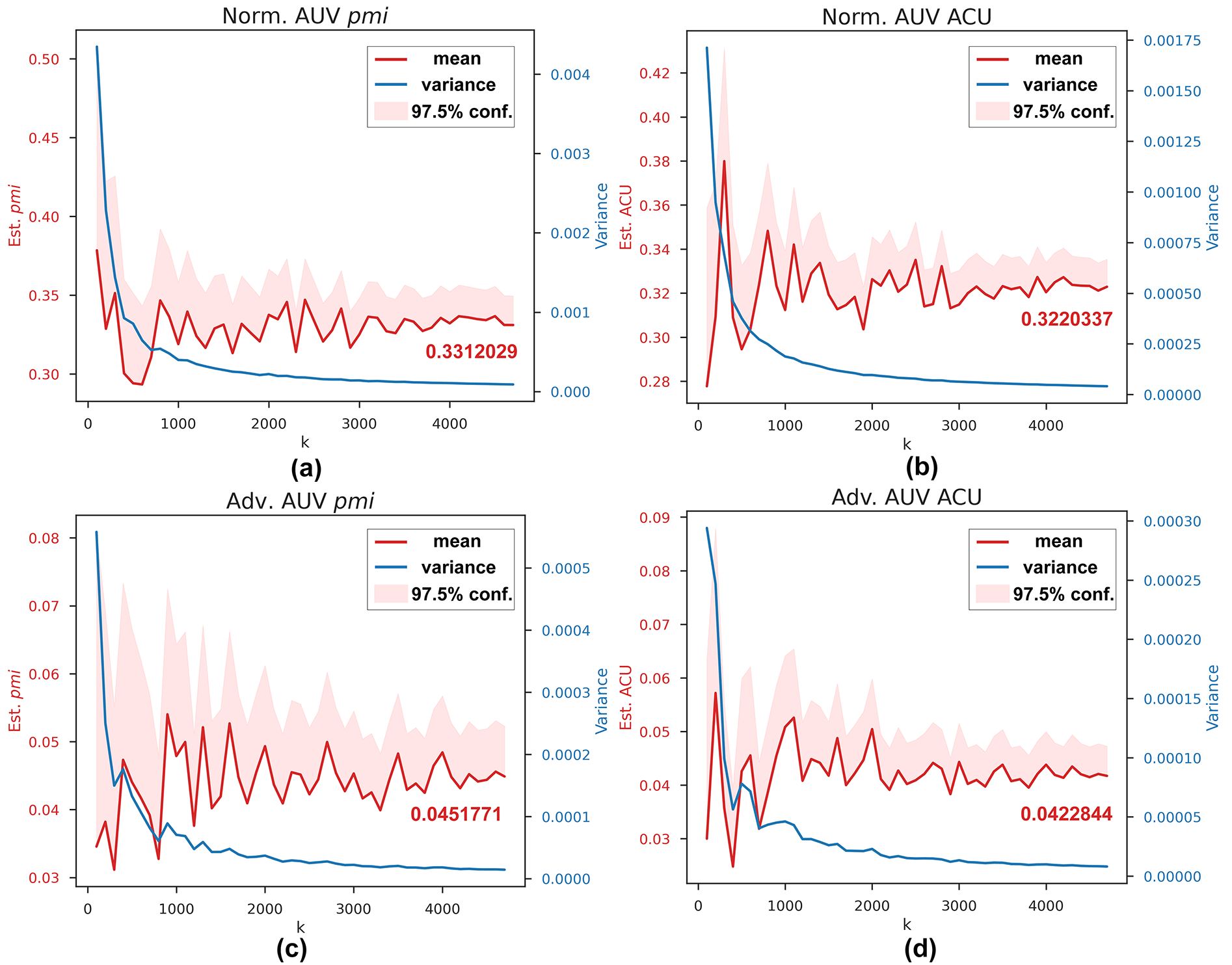}
	\caption{\textcolor{black}{The mean, variance and 97.5\% confidence upper bound of AUV's \textit{pmi} and ACU as functions of $k$ sampled norm balls.}}
	\label{fig_pmi_acu_auv}
\end{figure*}

Details of the Yolo3 model trained in this case study is presented in Table \ref{tab_yolov3_performance}, \ref{sec_app_B}. 
We adopt the practical solutions discussed in Section \ref{sec_ram_gen_to_hd_ds} to deal with the high dimensionality of the collected operational dataset (256*256*3) by first training a VAE model and compressing the dataset into a new space with a much lower dimensionality of 8.
While training details of the VAE model are summarised in Table \ref{tab_VAE_performance}, four sets of examples are shown in Fig. \ref{fig_vae_examples}, from which we can see that the reconstructed images are preserving the essential features of the objects (while blurring the less important background).
We then choose a norm ball radius $\epsilon=0.06$ according to the $r$-separation distance\footnote{Because more than one object may appear in a single image, the label of the ``dominating'' object (e.g., the object with the largest bounding box and/or with higher priority) can be used in the calculation of $r$. For simplicity, we first preprocess the dataset by filtering out images with multiple labels, and then determine the $\epsilon$ based on an estimated $r$.} and invoke the \gls{KDE} and robustness estimator \cite{webb_statistical_2019} for $k$ randomly selected norm balls. 
Individual estimates of the $k$ norm balls are then fed into the estimator for weighted average, Eqn.s~\eqref{eq_weighted_avg_est_mean} and \eqref{eq_weighted_avg_est_var}. For comparison, we also calculate the ACU by assuming equal weights (i.e., a flat OP) in Eqn.s~\eqref{eq_weighted_avg_est_mean} and \eqref{eq_weighted_avg_est_var}. 
Finally, the reliability claims on \textit{pmi} and ACU are plotted as functions of $k$ in Fig. \ref{fig_pmi_acu_auv}. Interpretation of the results is similar as before for CIFAR10, where the OP is also relatively flat. \textcolor{black}{From the comparison results, it can be seen that the adversarial train can effectively improve the robustness of DL models and be captured by our RAM method.}

\subsection{An Extended Case Study on Real World Unmanned Ground Vehicles}
\label{sec_app_D}
\begin{figure}[htbp]
     \centering
     \begin{subfigure}[b]{0.4\textwidth}
         \centering
         \includegraphics[width=\textwidth]{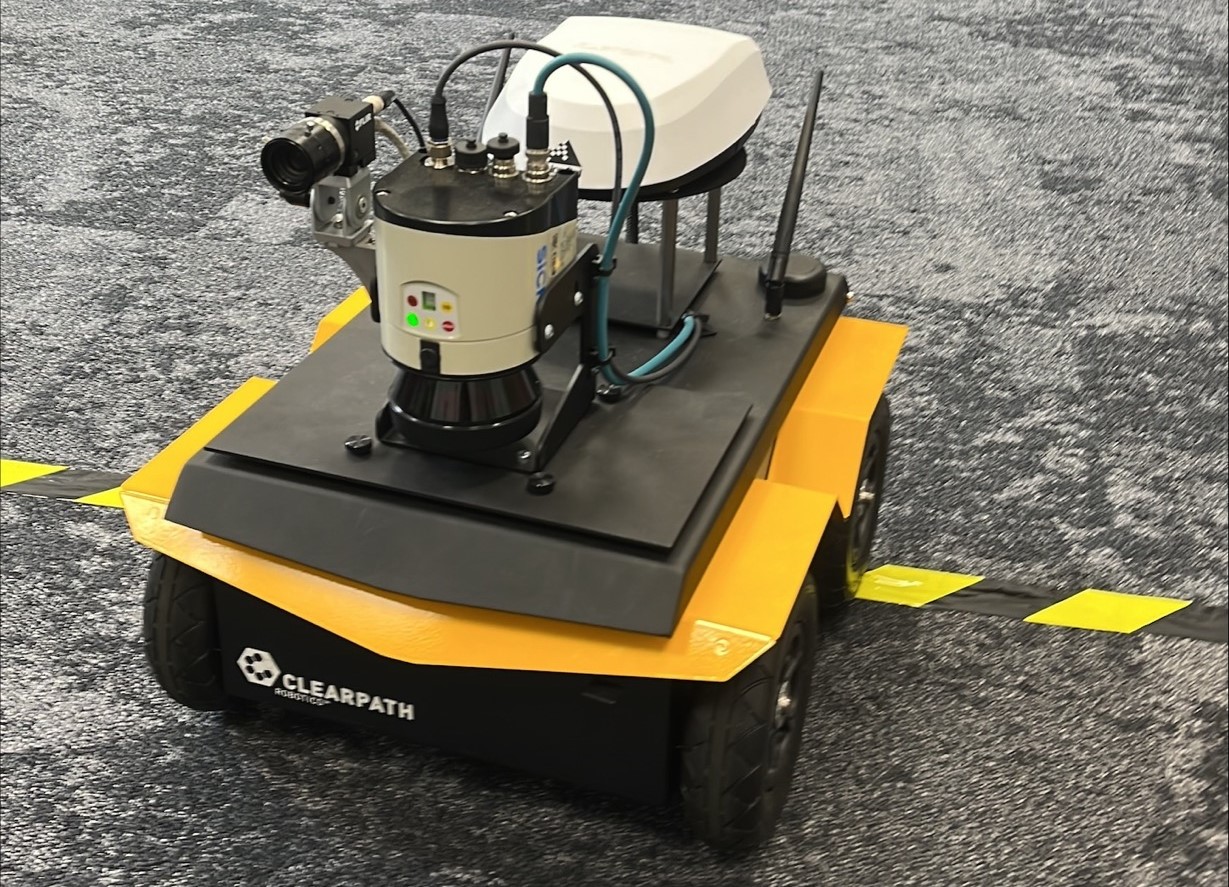}
         \caption{Jackal Robot}
         \label{fig:jackal}
     \end{subfigure}
     \hfill
     \begin{subfigure}[b]{0.52\textwidth}
         \centering
         \includegraphics[width=\textwidth]{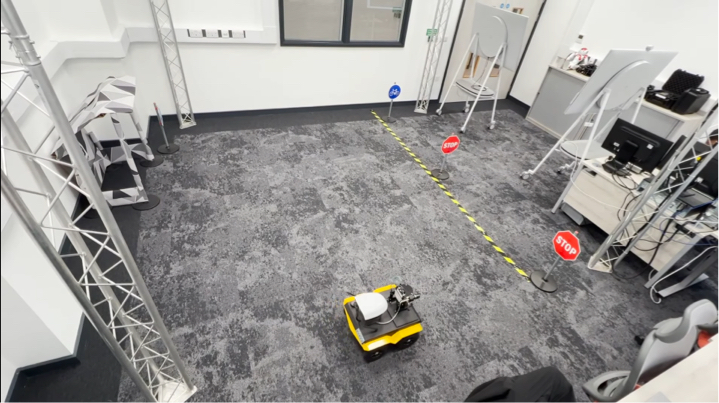}
         \caption{Physical Environment}
         \label{fig:environment}
     \end{subfigure}
     \caption{The \gls{UGV} and its deployed environment.}
\end{figure}

\textcolor{black}{In this case study, we deployed a customised real-world \gls{UGV} robot, named JACKAL, as shown in Fig. \ref{fig:jackal}. A demo video is available on Youtube\footnote{Jackal video demo: \url{https://youtu.be/E95vh5sxs7I}}.}

Similar to the AUV case study, the mission of UGV case study is to detect the different traffic signs. In this experiment, we set 5 different labels, stop sign, park sign, cycle sign, cross sign and one-way sign. We trained a YOLOv3 model with normal training and adversarial training based on the same dataset. All the experiment settings and details are given about the evaluation of the RAM in Section \ref{sec_app_B}. The RAM results are described in the following parts.

\begin{figure*}[ht]
	\centering
	\includegraphics[width=\linewidth]{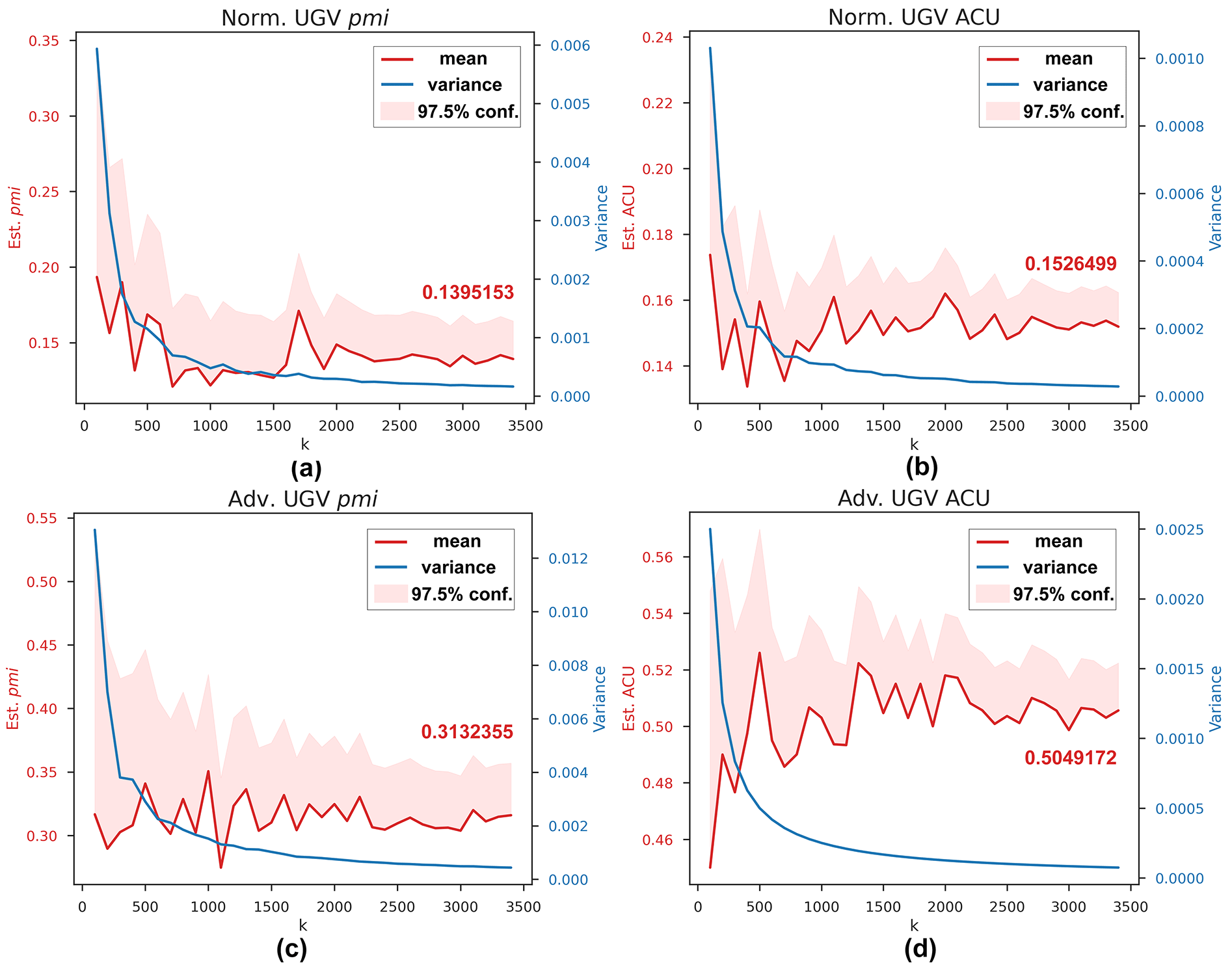}
	\caption{\textcolor{black}{The mean, variance and 97.5\% confidence upper bound of UGV's \textit{pmi} and ACU as functions of $k$ sampled norm balls.}}
	\label{fig_pmi_acu_ugv}
\end{figure*}

As shown in Fig.~\ref{fig_pmi_acu_ugv}, with the increasing amount of sampled data, the final assessment results are all converged with decreasing variance. Comparing sub-figures (a) and (b), or (c) and (d), we can find our $pmi$ evaluation is different from the ACU in consideration of the OP. 
In addition, we estimate $pmi$ and ACU on two models trained with different schemes. The train and test average precision (AP) are presented in Table~\ref{tab_yolov3_performance}. The adversarial training can improve the robustness of YOLO model at the cost of compromising the generalisation (indicated by mAP). The $pmi$ metric can be conceptually perceived as a product of ``generalisation $\times$ robustness'' \cite{zhao_assessing_2021}. Although previous experiments for MNIST, CIFAR10 and AUV shows the effectiveness of adversarial training for improving DL model's reliability by reducing the $pmi$, Fig.~\ref{fig_pmi_acu_ugv} displays the risk that adversarial training may potentially harm the reliability when there exists improper trade-off between generalisation and robustness. mAP for adversarially trained UGV model is significantly decreased from 0.98 to nearly 0.81, shown in Table~\ref{tab_yolov3_performance}. Too large drop-off of generalisation is disadvantageous to the reliability improvement. This motivates us to design new training schemes in the future towards more reliable DL models that are both robust and generalisable.





\section{Discussion}
\label{sec_discussion}

\subsection{Discussions on the Proposed RAM}
\textcolor{black}{In this section, we summarise the \textit{model assumptions} made in our RAM, and discuss if/how they can be validated and which new assumptions and compromises in the solutions are needed to cope with real-world applications with high-dimensional data. 
Finally, we list the \textit{inherent difficulties} of assessing ML reliability uncovered by our RAM.}

\paragraph{$R$-Separation and its Estimation} Assumption \ref{assumption_r_estiamtes_from_data}
derives from Remark \ref{remark_r_sep}. We concur with \cite{yang_closer_2020} and believe that, for any real-world \gls{ML} classification application where the inputs are data-points with ``physical meanings'', there should always exist an $r$-stable ground truth.
Such $r$-stable ground truth varies between applications, and the smaller the $r$ is, the harder the inherent difficulty of the classification problem becomes.
This $r$ is therefore a \textit{difficulty indicator} for the given classification problem.
Indeed, it is hard to estimate the $r$ (either in the input pixel space nor the latent feature space)---the best we can do is to estimate it from the existing dataset.
One way of solving the problem is to keep monitoring the $r$ estimates as more labelled data is collected, e.g.\ during operation, and to redo the cell partition when the estimated $r$ has changed significantly. 
Such a dynamic way of estimating $r$ can be supported by the concept of dynamic assurance cases \cite{asaadi_dynamic_2020}.

\paragraph{Approximation of the \gls{OP} from Data}
Assumption \ref{assumption_dataset_represents_OP} says that the collected dataset statistically represents the \gls{OP}, which may not hold for many practical reasons---e.g., when the future \gls{OP} is uncertain at the training stage and data is therefore collected in a balanced way to perform well in all categories of inputs. 
\textcolor{black}{
Although we demonstrate our \gls{RAM} under this assumption for simplicity, it can be relaxed for the following reasons. Any given dataset that differs from the OP requires a \textit{preprocessing} step to synthesis a new ``operational dataset'' (representing the \gls{OP}), e.g., by generative models, adding weights to data-points, etc. Indeed, such preprocessing step may require domain expert knowledge and/or historical data of similar products which is not an uncommon practice for traditional software \cite{smidts_software_2014}.
That is, we try to fit a \gls{PDF} over the input space from the ``operational dataset'', and data-points in this set can be raw data generated from historical data of previous applications which can then be scaled based on domain expert knowledge and by adding new data from simulation/generative-models.
Obtaining such an operational dataset is an application-specific engineering problem, and manageable thanks to the fact that it does not require manual labelling. }
Notably, the \gls{OP} may also be approximated at \textit{runtime} based on the data stream of operational data. Efficient KDE for data streams \cite{qahtan_kde_track_2017} can be used. If the \gls{OP} was subject to sudden changes, change-point detectors like \cite{zhao_interval_2020} should also be paired with the runtime estimator to robustly approximate the \gls{OP}. Such dynamic way of estimating \gls{OP} can also be supported by dynamic assurance cases \cite{asaadi_dynamic_2020}.

\paragraph{Determination of the Ground Truth of a Cell}
Assumptions \ref{assumption_single_gt_cell} and \ref{assumption_empty_cell_label} are essentially on how to determine the ground truth label for a given cell, which relates to the oracle problem of testing ML software.
While this still remains challenging, we partially solve it by leveraging the $r$-separation property.
Thanks to $r$, it is easy to determine a cell's ground truth when we see that it contains labelled data-points.
However, for an empty cell, it is non-trivial.
We assume the overall performance of the ML model is fairly good (e.g., better than a classifier doing random classifications), thus misclassifications within an empty cell are relatively rare events. 
We can determine the ground truth label of the cell by majority voting of predictions.
Indeed, it is a strong assumption when there are some ``failure regions'' in the input space, within which the ML model performs really badly (even worse than random labelling).
In this case, we need new mechanism to detect such ``really bad failure regions'' or spend more budget on, for example, asking humans to do the labelling.

\paragraph{Efficiency of Cell Robustness Evaluation}
Although we only applied the two methods of \gls{SMC} and \cite{webb_statistical_2019} in our experiments to evaluate the local robustness, we 
believe that other statistical sampling methods designed for estimating the probability of rare-events could be used as well.
Moreover, the cell robustness estimator in our \gls{RAM} works in a ``hot-swappable'' manner: any new and more efficient estimator can easily be incorporated.
Thus, despite being an important question, how to improve the efficiency of the robustness estimation for cells is beyond the scope of our \gls{RAM}.

\paragraph{Conditional OP of a Cell}
We assume that the distribution of inputs (the conditional OP) within each cell is uniform by Assumption \ref{assumption_conditonal_OP_uniform}.
Although we conjecture that this is the common case due to the small size of cells (i.e., those very close/similar inputs within a small region are only subject to noise factors that can be modelled uniformly), the specific situation may vary; this requires justification in safety cases. 
For a real-world dataset, the conditional OP might represent certain distributions of ``natural variations'' \cite{zhong_understanding_2021}, e.g.\ lighting conditions, that obey certain distributions.
Ideally, the conditional OP of cells should capture the distribution of such natural variations. Recent advance on measuring the naturalness/realisticness of \gls{AE}s \cite{harel_canada_is_2020} relates to this assumption and may relax it.

\paragraph{Independent $\lambda_i$s and $\mathsf{Op}_i$s} As per Assumption \ref{assumption_op_lambda_indep}, we assume all $\lambda_i$s and $\mathsf{Op}_i$s are independent when ``assembling'' their estimates via Eqn.~\eqref{eq_expected_pfd} and deriving the variance via Eqn.~\eqref{eq_varaince_pfd}.
This assumption is largely for the mathematical tractability when propagating the confidence in individual estimates at the cell-level to the \textit{pmi}.
Although this independence assumption is hard to justify in practice, it is not unusual in reliability models that do partition, e.g., in \cite{pietrantuono_reliability_2020,miller_estimating_1992}. 
We believe that RAMs are still useful under this assumption, while we envisage that Bayesian estimators leveraging joint priors and conjugacy may relax it.

\paragraph{Uncertainties Raised by Individual OP and Robustness Estimates} This relates to how reliable the chosen OP and robustness estimators themselves are. 
Our RAM is flexible and evolvable in the sense that it does not depend on any specific estimators.
New and more reliable estimators can therefore easily be integrated to reduce the estimation uncertainties. 
Moreover, such uncertainties raised by estimators are propagated and compounded in our overall RAM results, cf.~Eqn.s~\eqref{eq_varaince_pfd} and \eqref{eq_weighted_avg_est_var}. 
Although we ignore them as per Assumption \ref{assumption_k_is_the_major_source_uncertainty}, this is arguably the case when the two estimators are fairly reliable and the number of samples $k$ is much smaller than the sample frame size $n$.

\paragraph{Inherent Difficulties of Reliability Assessment on \gls{ML} Software} Finally, based on our \gls{RAM} and the discussions above, we summarise the inherent difficulties of assessing \gls{ML} reliability as the following questions:
\begin{itemize}
	\item How to accurately learn the \gls{OP} in a potentially high-dimensional input space with relatively sparse data? 
	\item How to build an accurate test oracle (to determine the ground truth label) by, e.g., leveraging the existing labels (done by humans) in the training dataset?
	\item What is the local distribution (i.e.~the conditional OP) over a small input region (which is potentially only subject to subtle natural variations of physical conditions in the environment)?
	\item How to efficiently evaluate the robustness of a small region, given that \gls{AE}s are normally rare events? And how to reduce the risk associated with an AE (e.g., referring to ALARP)?
	\item How to efficiently sample small regions from a large population (due to the high-dimensionality) of regions to test the local robustness in an unbiased and uncertainty informed way, given a limited budget? 
\end{itemize}
\textcolor{black}{We provide solutions in our \gls{RAM} that are practical compromises (cf.\ Section \ref{sec_ram_gen_to_hd_ds}), while the questions above are still challenging and generic. For some domains, say self-driving cars, they are relatively easier to tackle, thanks to the large amount of available data including millions of miles of historical road test data collected in recent years. At this stage, we doubt the existence of other \gls{RAM}s for \gls{ML} software with weaker assumptions that achieve the same level of rigorousness as ours, in which sense our \gls{RAM} advances in this research direction.}

\subsection{Discussions on the Overall Assurance Case Framework and Low-Level Probabilistic Safety Arguments}

With the emphasis on quantitative aspects of assuring \gls{LES} (and thus complementing existing assurance frameworks, e.g., \cite{bloomfield2021safety}), our overall assurance framework and the low-level probabilistic safety arguments together form an ``vertically'' end-to-end assurance case, in which a chain of safety/reliability techniques are integrated. 
However, the assurance case presented is still incomplete ``horizontally''---some sub-cases and (side-)claims are undeveloped. Because, they are either generic claims that have been studied elsewhere (and omitted for simplicity), e.g. in \cite{bloomfield2021safety,ashmore2021assuring}, or are still quite hard to argue in general and thus require specific expert judgement in a case-by-case manner.

The proposed safety analysis activities---HAZOP, hazard scenarios modelling, FTA, our RAM, and the determination of the system-level safety targets based on the performance of human/similar-products---are not exclusive in our assurance framework; rather we concur with \cite{KKB2019} that credible safety cases require a heterogeneous approach.
A dangerous pitfall is that those activities are not performed sufficiently because of, say, the analyser's limited engineering knowledge/experience and the lack of empirical data.
This is, however, not unique to our assurance framework, but rather generic to any assurance studies.

We only present safety arguments for the classification function of the ML component, based on our new RAM for ML classifiers, leaving claims for the other three functions---localisation, detection timing, and object tracking---undeveloped\footnote{Certainly for real safety cases, we also need to develop claims on ``non-ML'' parts (e.g., capability of the development team and quality of the code) which can be addressed by conventional approaches that we omit in this work.}.
The general idea and principles, however, are applicable to the other three functions, too:
we may first define bespoke reliability measures for each (like \textit{pmi} for classification), and then do probabilistic reliability modelling based on statistical testing evidence. This forms important future work.

\subsection{Discussions on the Case Studies}

So far, we have conducted a AUV case study in simulators to validate and demonstrate our proposed methods. 
While defending the role of simulation in certification and regulation is beyond the scope of this work, simulation is arguably necessary for many reasons as long as the simulation satisfies some prerequisites---for example, the fidelity is justifiable, scenario-coverage is sufficiently high, and non-zero real-world testing is conducted to validate the simulation. 
Besides, we plan to conduct a real-world AUV case study in a physical wave tank, in which the conditions may be adjusted to have real-world disturbances, e.g., generating various types of waves in offshore scenarios and changing the lighting conditions.


To further validate and demonstrate the proposed approaches, we extended our experiments to a physical UGV case study and a CORONET case study in real-world (see Section \ref{sec_app_D} \& Appendix \ref{sec_app_C}). The experimental results show that the proposed methods can acclimate to other different real-world applications.


\section{Conclusion and Future Work}
\label{sec_conclusion_future_work}

This article introduces a RAM designed for ML classifiers, extending its initial version of \cite{zhao_assessing_2021} with more practical considerations for real-world applications of high-dimensional data and autonomous systems, e.g., \textcolor{black}{the new estimator Eqn.s~\eqref{eq_weighted_avg_est_mean} and \eqref{eq_weighted_avg_est_var}, alternative solutions discussed in Section \ref{sec_ram_gen_to_hd_ds}, and new experiments on image datasets, AUV/UGV missions, and a real-world healthcare application.}
To the best of our knowledge, it is the first ML RAM that explicitly considers both the OP information and robustness evidence.
It has also allowed us to uncover some inherent challenges when assessing ML reliability. 
Based on the RAM, we present probabilistic safety arguments for ML components incorporating low-level \gls{VnV} evidence.

\textcolor{black}{While the main focus of this work is assessing the reliability for ML components, it is unwise to study safety without considering the \gls{LES} in its application context. Because, an ML model, as software, would not cause physical mishaps on its own. What truly concerns us is the safety of the whole \gls{LES}. Thus, we propose an overall assurance framework, in which a set of safety analysis activities are integrated to identify the whole LES level safety targets and break down them to component-level reliability requirements of ML functions. Such analysis requires the understanding of the interplay between non-ML and ML components.}
Finally, case studies based on \gls{RAS} are conducted. 
The case studies are comprehensive in terms of exercising and demonstrating all proposed methods in our assurance framework and also identifying key challenges with recommendations.

An intuitive way of perceiving our RAM, comparing with the usual accuracy testing, is that we enlarge the test set with more test cases around the ``seeds'' (original data-points in the test set).
We determine the oracle of a new test case according to its seed's label and the $r$-distance.
Those enlarged test results form the robustness evidence, while how much they contribute to the overall reliability is proportional to its OP. 
Consequently, \textit{exposing to more tests (robustness evaluation) and being more representative of how it will be used (the OP)}, our RAM is more informative---and therefore more trustworthy.
In line with the gist of our RAM, we believe that the DL reliability should follow the conceptualised equation of:
$$
DL \,\,\, reliability = generalisability  \times robustness.
$$
Intuitively, this equation says that, when assessing the reliability of ML software, we should not only consider how the DL model generalises to a new data-point (according to the OP), but also how robustness it performs around the new data-point.

Apart from the future works mentioned in the discussion section, we also plan to conduct more real-world case studies to examine the scalability of our methods.
We presume a \textit{trained} ML model for our assessment purpose. 
A natural follow-up question is how to actually improve the reliability when our RAM results indicate that a system is not good enough.
As described in \cite{zhao_detecting_2021}, we plan to investigate integrating ML debug testing (e.g. \cite{huang2021coverage}) and retraining methods \cite{bai_recent_2021} with the RAM, to form a closed loop of debugging-improving-assessing. 
Last but not least, we find the idea of dynamic assurance cases \cite{asaadi_dynamic_2020} may have a great potential for addressing some challenges we currently face in our framework.

\begin{acks}
This work is supported by the UK DSTL (through the project of Safety Argument for Learning-enabled Autonomous Underwater Vehicles) and the UK EPSRC (through the Offshore Robotics for Certification of Assets [EP/W001136/1] and End-to-End Conceptual Guarding of Neural Architectures [EP/T026995/1]). Xingyu Zhao and Alec Banks’ contribution to the work is partially supported through Fellowships at the Assuring Autonomy International Programme. \includegraphics[height=8pt]{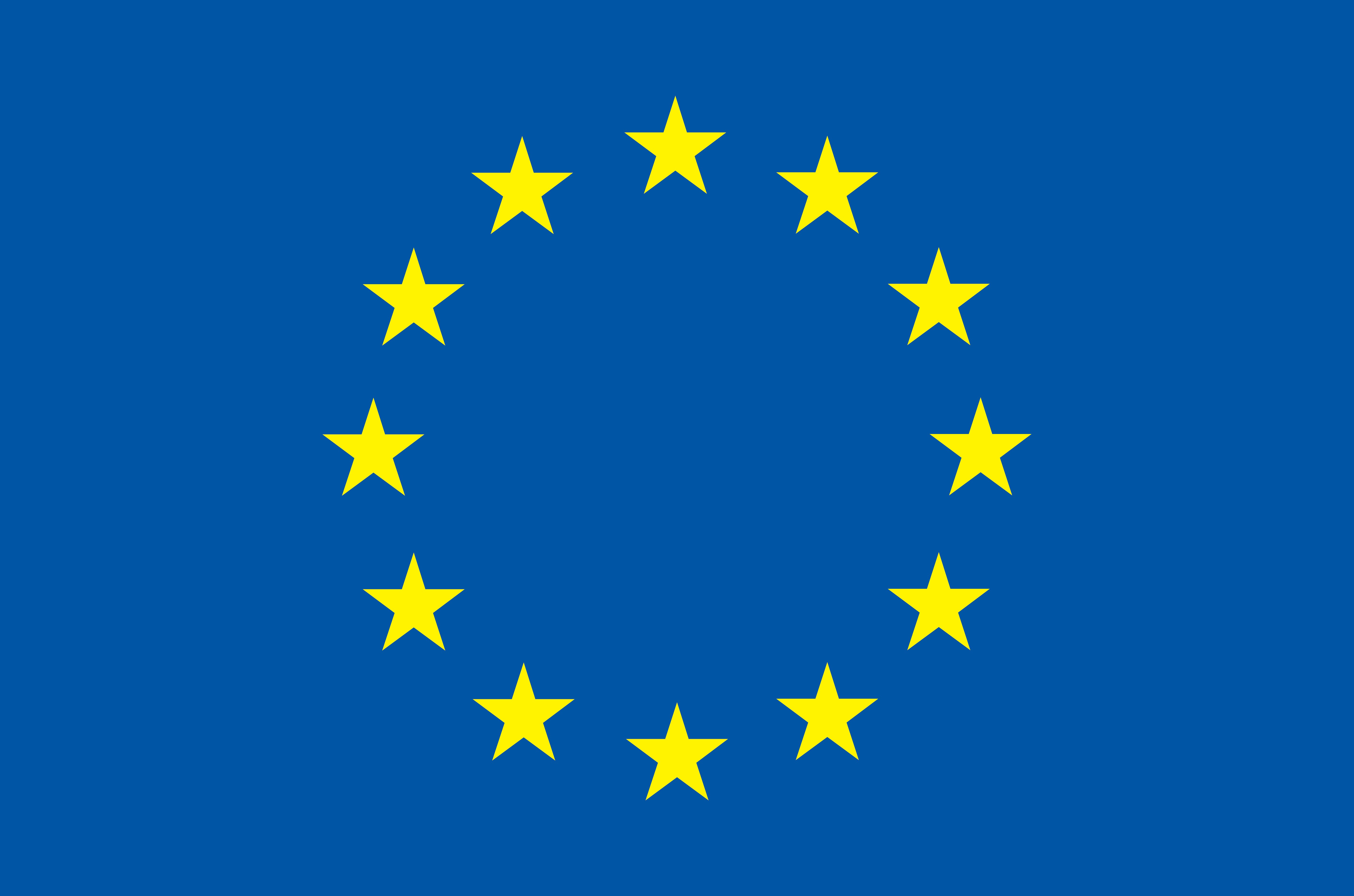} 
This project has received funding from the European Union’s Horizon 2020 research and innovation programme under grant agreement No 956123. We thank Philippa Ryan and anonymous reviewers for insightful comments on earlier versions of this work. \textcolor{black}{We thank TECS editors and the anonymous reviewers whose comments helped us to improve the manuscript.}

This document is an overview of UK MOD (part) sponsored research and is released for informational purposes only. The contents of this document should not be interpreted as representing the views of the UK MOD, nor should it be assumed that they reflect any current or future UK MOD policy. The information contained in this document cannot supersede any statutory or contractual requirements or liabilities and is offered without prejudice or commitment. 
\end{acks}


  \bibliographystyle{ACM-Reference-Format}
\bibliography{ref}


\begin{thebibliography}{95}


\ifx \showCODEN    \undefined \def \showCODEN     #1{\unskip}     \fi
\ifx \showDOI      \undefined \def \showDOI       #1{#1}\fi
\ifx \showISBNx    \undefined \def \showISBNx     #1{\unskip}     \fi
\ifx \showISBNxiii \undefined \def \showISBNxiii  #1{\unskip}     \fi
\ifx \showISSN     \undefined \def \showISSN      #1{\unskip}     \fi
\ifx \showLCCN     \undefined \def \showLCCN      #1{\unskip}     \fi
\ifx \shownote     \undefined \def \shownote      #1{#1}          \fi
\ifx \showarticletitle \undefined \def \showarticletitle #1{#1}   \fi
\ifx \showURL      \undefined \def \showURL       {\relax}        \fi
\providecommand\bibfield[2]{#2}
\providecommand\bibinfo[2]{#2}
\providecommand\natexlab[1]{#1}
\providecommand\showeprint[2][]{arXiv:#2}

\bibitem[\protect\citeauthoryear{Alves, Bhatt, Hall, Driscoll, Murugesan, and
  Rushby}{Alves et~al\mbox{.}}{2018}]%
        {alves_considerations_2018}
\bibfield{author}{\bibinfo{person}{Erin Alves}, \bibinfo{person}{Devesh Bhatt},
  \bibinfo{person}{Brendan Hall}, \bibinfo{person}{Kevin Driscoll},
  \bibinfo{person}{Anitha Murugesan}, {and} \bibinfo{person}{John Rushby}.}
  \bibinfo{year}{2018}\natexlab{}.
\newblock \bibinfo{booktitle}{\emph{Considerations in assuring safety of
  increasingly autonomous systems}}.
\newblock \bibinfo{type}{Technical {Report}} NASA/CR-2018-220080.
  \bibinfo{institution}{NASA}. \bibinfo{pages}{172} pages.
\newblock


\bibitem[\protect\citeauthoryear{Aminifar}{Aminifar}{2020}]%
        {aminifar2020universal}
\bibfield{author}{\bibinfo{person}{Amir Aminifar}.}
  \bibinfo{year}{2020}\natexlab{}.
\newblock \showarticletitle{Universal adversarial perturbations in epileptic
  seizure detection}. In \bibinfo{booktitle}{\emph{2020 International Joint
  Conference on Neural Networks (IJCNN)}}. IEEE, \bibinfo{pages}{1--6}.
\newblock


\bibitem[\protect\citeauthoryear{Asaadi, Denney, Menzies, Pai, and
  Petroff}{Asaadi et~al\mbox{.}}{2020b}]%
        {asaadi_dynamic_2020}
\bibfield{author}{\bibinfo{person}{Erfan Asaadi}, \bibinfo{person}{Ewen
  Denney}, \bibinfo{person}{Jonathan Menzies}, \bibinfo{person}{Ganesh~J. Pai},
  {and} \bibinfo{person}{Dimo Petroff}.} \bibinfo{year}{2020}\natexlab{b}.
\newblock \showarticletitle{Dynamic {Assurance} {Cases}: {A} {Pathway} to
  {Trusted} {Autonomy}}.
\newblock \bibinfo{journal}{\emph{Computer}} \bibinfo{volume}{53},
  \bibinfo{number}{12} (\bibinfo{year}{2020}), \bibinfo{pages}{35--46}.
\newblock
\urldef\tempurl%
\url{https://doi.org/10.1109/MC.2020.3022030}
\showDOI{\tempurl}


\bibitem[\protect\citeauthoryear{Asaadi, Denney, and Pai}{Asaadi
  et~al\mbox{.}}{2020a}]%
        {asaadi_quantifying_2020}
\bibfield{author}{\bibinfo{person}{Erfan Asaadi}, \bibinfo{person}{Ewen
  Denney}, {and} \bibinfo{person}{Ganesh Pai}.}
  \bibinfo{year}{2020}\natexlab{a}.
\newblock \showarticletitle{Quantifying assurance in learning-enabled systems}.
  In \bibinfo{booktitle}{\emph{Computer {Safety}, {Reliability}, and
  {Security}}} \emph{(\bibinfo{series}{{LNCS}})},
  \bibfield{editor}{\bibinfo{person}{António Casimiro}, \bibinfo{person}{Frank
  Ortmeier}, \bibinfo{person}{Friedemann Bitsch}, {and} \bibinfo{person}{Pedro
  Ferreira}} (Eds.), Vol.~\bibinfo{volume}{12234}.
  \bibinfo{publisher}{Springer}, \bibinfo{address}{Cham},
  \bibinfo{pages}{270--286}.
\newblock
\showISBNx{978-3-030-54549-9}
\urldef\tempurl%
\url{https://doi.org/10.1007/978-3-030-54549-9_18}
\showDOI{\tempurl}


\bibitem[\protect\citeauthoryear{Ashmore, Calinescu, and Paterson}{Ashmore
  et~al\mbox{.}}{2021}]%
        {ashmore2021assuring}
\bibfield{author}{\bibinfo{person}{Rob Ashmore}, \bibinfo{person}{Radu
  Calinescu}, {and} \bibinfo{person}{Colin Paterson}.}
  \bibinfo{year}{2021}\natexlab{}.
\newblock \showarticletitle{Assuring the machine learning lifecycle:
  Desiderata, methods, and challenges}.
\newblock \bibinfo{journal}{\emph{ACM Computing Surveys (CSUR)}}
  \bibinfo{volume}{54}, \bibinfo{number}{5} (\bibinfo{year}{2021}),
  \bibinfo{pages}{1--39}.
\newblock


\bibitem[\protect\citeauthoryear{Backurs, Indyk, and Wagner}{Backurs
  et~al\mbox{.}}{2019}]%
        {backurs2019space}
\bibfield{author}{\bibinfo{person}{Arturs Backurs}, \bibinfo{person}{Piotr
  Indyk}, {and} \bibinfo{person}{Tal Wagner}.} \bibinfo{year}{2019}\natexlab{}.
\newblock \showarticletitle{Space and Time Efficient Kernel Density Estimation
  in High Dimensions}. In \bibinfo{booktitle}{\emph{Advances in Neural
  Information Processing Systems}}, Vol.~\bibinfo{volume}{32}.
  \bibinfo{publisher}{Curran Associates, Inc.}, \bibinfo{pages}{15773--15782}.
\newblock


\bibitem[\protect\citeauthoryear{Bai, Luo, Zhao, Wen, and Wang}{Bai
  et~al\mbox{.}}{2021}]%
        {bai_recent_2021}
\bibfield{author}{\bibinfo{person}{Tao Bai}, \bibinfo{person}{Jinqi Luo},
  \bibinfo{person}{Jun Zhao}, \bibinfo{person}{Bihan Wen}, {and}
  \bibinfo{person}{Qian Wang}.} \bibinfo{year}{2021}\natexlab{}.
\newblock \showarticletitle{Recent {Advances} in {Adversarial} {Training} for
  {Adversarial} {Robustness}}. In \bibinfo{booktitle}{\emph{Proc. of the 30th
  {Int.} {Joint} {Conf.} on {Artificial} {Intelligence}}}
  \emph{(\bibinfo{series}{{IJCAI}'21})}. \bibinfo{pages}{4312--4321}.
\newblock
\urldef\tempurl%
\url{https://doi.org/10.24963/ijcai.2021/591}
\showDOI{\tempurl}


\bibitem[\protect\citeauthoryear{Berend}{Berend}{2021}]%
        {DBLP:conf/icse/Berend21}
\bibfield{author}{\bibinfo{person}{David Berend}.}
  \bibinfo{year}{2021}\natexlab{}.
\newblock \showarticletitle{Distribution Awareness for {AI} System Testing}. In
  \bibinfo{booktitle}{\emph{43rd {IEEE/ACM} International Conference on
  Software Engineering: Companion Proceedings, {ICSE} Companion 2021, Madrid,
  Spain, May 25-28, 2021}}. \bibinfo{publisher}{{IEEE}},
  \bibinfo{pages}{96--98}.
\newblock


\bibitem[\protect\citeauthoryear{Bergstra and Bengio}{Bergstra and
  Bengio}{2012}]%
        {bergstra_random_2012}
\bibfield{author}{\bibinfo{person}{James Bergstra} {and}
  \bibinfo{person}{Yoshua Bengio}.} \bibinfo{year}{2012}\natexlab{}.
\newblock \showarticletitle{Random search for hyper-parameter optimization.}
\newblock \bibinfo{journal}{\emph{J. of Machine Learning Research}}
  \bibinfo{volume}{13}, \bibinfo{number}{2} (\bibinfo{year}{2012}),
  \bibinfo{pages}{281--305}.
\newblock


\bibitem[\protect\citeauthoryear{Bertolino, Miranda, Pietrantuono, and
  Russo}{Bertolino et~al\mbox{.}}{2017}]%
        {bertolino_adaptive_2017}
\bibfield{author}{\bibinfo{person}{A. Bertolino}, \bibinfo{person}{B. Miranda},
  \bibinfo{person}{R. Pietrantuono}, {and} \bibinfo{person}{S. Russo}.}
  \bibinfo{year}{2017}\natexlab{}.
\newblock \showarticletitle{Adaptive {Coverage} and {Operational}
  {Profile}-{Based} {Testing} for {Reliability} {Improvement}}. In
  \bibinfo{booktitle}{\emph{2017 {IEEE}/{ACM} 39th {International} {Conference}
  on {Software} {Engineering} ({ICSE})}}. \bibinfo{publisher}{IEEE},
  \bibinfo{address}{Buenos Aires, Argentina}, \bibinfo{pages}{541--551}.
\newblock
\urldef\tempurl%
\url{https://doi.org/10.1109/ICSE.2017.56}
\showDOI{\tempurl}


\bibitem[\protect\citeauthoryear{Bertolino, Miranda, Pietrantuono, and
  Russo}{Bertolino et~al\mbox{.}}{2021}]%
        {bertolino_adaptive_2021}
\bibfield{author}{\bibinfo{person}{Antonia Bertolino}, \bibinfo{person}{Breno
  Miranda}, \bibinfo{person}{Roberto Pietrantuono}, {and}
  \bibinfo{person}{Stefano Russo}.} \bibinfo{year}{2021}\natexlab{}.
\newblock \showarticletitle{Adaptive {Test} {Case} {Allocation}, {Selection}
  and {Generation} {Using} {Coverage} {Spectrum} and {Operational} {Profile}}.
\newblock \bibinfo{journal}{\emph{IEEE Transactions on Software Engineering}}
  \bibinfo{volume}{47}, \bibinfo{number}{5} (\bibinfo{year}{2021}),
  \bibinfo{pages}{881--898}.
\newblock


\bibitem[\protect\citeauthoryear{Bevington, Robinson, Blair, Mallinckrodt, and
  McKay}{Bevington et~al\mbox{.}}{1993}]%
        {bevington_data_1993}
\bibfield{author}{\bibinfo{person}{Philip~R Bevington},
  \bibinfo{person}{D~Keith Robinson}, \bibinfo{person}{J~Morris Blair},
  \bibinfo{person}{A~John Mallinckrodt}, {and} \bibinfo{person}{Susan McKay}.}
  \bibinfo{year}{1993}\natexlab{}.
\newblock \bibinfo{booktitle}{\emph{Data reduction and error analysis for the
  physical sciences}}. Vol.~\bibinfo{volume}{7}.
\newblock \bibinfo{publisher}{American Institute of Physics}.
\newblock


\bibitem[\protect\citeauthoryear{Bishop and Bloomfield}{Bishop and
  Bloomfield}{2000}]%
        {bishop_methodology_2000}
\bibfield{author}{\bibinfo{person}{Peter Bishop} {and} \bibinfo{person}{Robin
  Bloomfield}.} \bibinfo{year}{2000}\natexlab{}.
\newblock \showarticletitle{A methodology for safety case development}.
\newblock \bibinfo{journal}{\emph{Safety and Reliability}}
  \bibinfo{volume}{20}, \bibinfo{number}{1} (\bibinfo{year}{2000}),
  \bibinfo{pages}{34--42}.
\newblock


\bibitem[\protect\citeauthoryear{Bishop, Bloomfield, Littlewood, Povyakalo, and
  Wright}{Bishop et~al\mbox{.}}{2011}]%
        {bishop_toward_2011}
\bibfield{author}{\bibinfo{person}{Peter Bishop}, \bibinfo{person}{Robin
  Bloomfield}, \bibinfo{person}{Bev Littlewood}, \bibinfo{person}{Andrey
  Povyakalo}, {and} \bibinfo{person}{David Wright}.}
  \bibinfo{year}{2011}\natexlab{}.
\newblock \showarticletitle{Toward a formalism for conservative claims about
  the dependability of software-based systems}.
\newblock \bibinfo{journal}{\emph{IEEE Tran. on Software Engineering}}
  \bibinfo{volume}{37}, \bibinfo{number}{5} (\bibinfo{year}{2011}),
  \bibinfo{pages}{708--717}.
\newblock


\bibitem[\protect\citeauthoryear{Bishop and Povyakalo}{Bishop and
  Povyakalo}{2017}]%
        {bishop_deriving_2017}
\bibfield{author}{\bibinfo{person}{Peter Bishop} {and} \bibinfo{person}{Andrey
  Povyakalo}.} \bibinfo{year}{2017}\natexlab{}.
\newblock \showarticletitle{Deriving a frequentist conservative confidence
  bound for probability of failure per demand for systems with different
  operational and test profiles}.
\newblock \bibinfo{journal}{\emph{Reliability Engineering \& System Safety}}
  \bibinfo{volume}{158} (\bibinfo{year}{2017}), \bibinfo{pages}{246--253}.
\newblock


\bibitem[\protect\citeauthoryear{Bloomfield and Bishop}{Bloomfield and
  Bishop}{2010}]%
        {bloomfield_safety_2010}
\bibfield{author}{\bibinfo{person}{Robin Bloomfield} {and}
  \bibinfo{person}{Peter Bishop}.} \bibinfo{year}{2010}\natexlab{}.
\newblock \showarticletitle{Safety and assurance cases: past, present and
  possible future -- an {Adelard} perspective}. In
  \bibinfo{booktitle}{\emph{Making {Systems} {Safer}}},
  \bibfield{editor}{\bibinfo{person}{Chris Dale} {and} \bibinfo{person}{Tom
  Anderson}} (Eds.). \bibinfo{publisher}{Springer London},
  \bibinfo{address}{London}, \bibinfo{pages}{51--67}.
\newblock
\showISBNx{978-1-84996-086-1}


\bibitem[\protect\citeauthoryear{Bloomfield, Fletcher, Khlaaf, Hinde, and
  Ryan}{Bloomfield et~al\mbox{.}}{2021}]%
        {bloomfield2021safety}
\bibfield{author}{\bibinfo{person}{Robin Bloomfield}, \bibinfo{person}{Gareth
  Fletcher}, \bibinfo{person}{Heidy Khlaaf}, \bibinfo{person}{Luke Hinde},
  {and} \bibinfo{person}{Philippa Ryan}.} \bibinfo{year}{2021}\natexlab{}.
\newblock \showarticletitle{Safety Case Templates for Autonomous Systems}.
\newblock \bibinfo{journal}{\emph{arXiv preprint arXiv:2102.02625}}
  (\bibinfo{year}{2021}).
\newblock


\bibitem[\protect\citeauthoryear{Bloomfield, Khlaaf, Conmy, and
  Fletcher}{Bloomfield et~al\mbox{.}}{2019}]%
        {BKCF2019}
\bibfield{author}{\bibinfo{person}{Robin Bloomfield}, \bibinfo{person}{Heidy
  Khlaaf}, \bibinfo{person}{Philippa~Ryan Conmy}, {and} \bibinfo{person}{Gareth
  Fletcher}.} \bibinfo{year}{2019}\natexlab{}.
\newblock \showarticletitle{Disruptive Innovations and Disruptive Assurance:
  Assuring Machine Learning and Autonomy}.
\newblock \bibinfo{journal}{\emph{Computer}} \bibinfo{volume}{52},
  \bibinfo{number}{9} (\bibinfo{year}{2019}), \bibinfo{pages}{82--89}.
\newblock
\showISSN{0018-9162}


\bibitem[\protect\citeauthoryear{Bloomfield and Netkachova}{Bloomfield and
  Netkachova}{2014}]%
        {bloomfield_building_2014}
\bibfield{author}{\bibinfo{person}{R. Bloomfield} {and} \bibinfo{person}{K.
  Netkachova}.} \bibinfo{year}{2014}\natexlab{}.
\newblock \showarticletitle{Building blocks for assurance cases}. In
  \bibinfo{booktitle}{\emph{{IEEE} {International} {Symposium} on {Software}
  {Reliability} {Engineering} {Workshops}}}. \bibinfo{publisher}{IEEE},
  \bibinfo{address}{Naples, Italy}, \bibinfo{pages}{186--191}.
\newblock
\urldef\tempurl%
\url{https://doi.org/10.1109/ISSREW.2014.72}
\showDOI{\tempurl}


\bibitem[\protect\citeauthoryear{Bloomfield and Rushby}{Bloomfield and
  Rushby}{2020}]%
        {bloomfield2020assurance}
\bibfield{author}{\bibinfo{person}{Robin Bloomfield} {and}
  \bibinfo{person}{John Rushby}.} \bibinfo{year}{2020}\natexlab{}.
\newblock \showarticletitle{Assurance 2.0: A Manifesto}.
\newblock \bibinfo{journal}{\emph{arXiv preprint arXiv:2004.10474}}
  (\bibinfo{year}{2020}).
\newblock


\bibitem[\protect\citeauthoryear{Burton, Habli, Lawton, McDermid, Morgan, and
  Porter}{Burton et~al\mbox{.}}{2020}]%
        {burton_mind_2020}
\bibfield{author}{\bibinfo{person}{Simon Burton}, \bibinfo{person}{Ibrahim
  Habli}, \bibinfo{person}{Tom Lawton}, \bibinfo{person}{John McDermid},
  \bibinfo{person}{Phillip Morgan}, {and} \bibinfo{person}{Zoe Porter}.}
  \bibinfo{year}{2020}\natexlab{}.
\newblock \showarticletitle{Mind the gaps: {Assuring} the safety of autonomous
  systems from an engineering, ethical, and legal perspective}.
\newblock \bibinfo{journal}{\emph{Artificial Intelligence}}
  \bibinfo{volume}{279} (\bibinfo{year}{2020}), \bibinfo{pages}{103201}.
\newblock
\showISSN{0004-3702}
\urldef\tempurl%
\url{https://doi.org/10.1016/j.artint.2019.103201}
\showDOI{\tempurl}


\bibitem[\protect\citeauthoryear{Calinescu, Weyns, Gerasimou, Iftikhar, Habli,
  and Kelly}{Calinescu et~al\mbox{.}}{2018}]%
        {calinescu_engineering_2018}
\bibfield{author}{\bibinfo{person}{R. Calinescu}, \bibinfo{person}{D. Weyns},
  \bibinfo{person}{S. Gerasimou}, \bibinfo{person}{M.~U. Iftikhar},
  \bibinfo{person}{I. Habli}, {and} \bibinfo{person}{T. Kelly}.}
  \bibinfo{year}{2018}\natexlab{}.
\newblock \showarticletitle{Engineering trustworthy self-adaptive software with
  dynamic assurance cases}.
\newblock \bibinfo{journal}{\emph{IEEE Tran. on Software Engineering}}
  \bibinfo{volume}{44}, \bibinfo{number}{11} (\bibinfo{year}{2018}),
  \bibinfo{pages}{1039--1069}.
\newblock
\showISSN{0098-5589}


\bibitem[\protect\citeauthoryear{Chen}{Chen}{2017}]%
        {chen2017tutorial}
\bibfield{author}{\bibinfo{person}{Yen-Chi Chen}.}
  \bibinfo{year}{2017}\natexlab{}.
\newblock \showarticletitle{A tutorial on kernel density estimation and recent
  advances}.
\newblock \bibinfo{journal}{\emph{Biostatistics \& Epidemiology}}
  \bibinfo{volume}{1}, \bibinfo{number}{1} (\bibinfo{year}{2017}),
  \bibinfo{pages}{161--187}.
\newblock


\bibitem[\protect\citeauthoryear{Cotroneo, Pietrantuono, and Russo}{Cotroneo
  et~al\mbox{.}}{2016}]%
        {cotroneo_relai_2016}
\bibfield{author}{\bibinfo{person}{Domenico Cotroneo}, \bibinfo{person}{Roberto
  Pietrantuono}, {and} \bibinfo{person}{Stefano Russo}.}
  \bibinfo{year}{2016}\natexlab{}.
\newblock \showarticletitle{{RELAI} {Testing}: {A} {Technique} to {Assess} and
  {Improve} {Software} {Reliability}}.
\newblock \bibinfo{journal}{\emph{IEEE Transactions on Software Engineering}}
  \bibinfo{volume}{42}, \bibinfo{number}{5} (\bibinfo{year}{2016}),
  \bibinfo{pages}{452--475}.
\newblock
\urldef\tempurl%
\url{https://doi.org/10.1109/TSE.2015.2491931}
\showDOI{\tempurl}


\bibitem[\protect\citeauthoryear{Crawley and Tyler}{Crawley and Tyler}{2015}]%
        {crawley2015hazop}
\bibfield{author}{\bibinfo{person}{Frank Crawley} {and} \bibinfo{person}{Brian
  Tyler}.} \bibinfo{year}{2015}\natexlab{}.
\newblock \bibinfo{booktitle}{\emph{HAZOP: Guide to best practice}}.
\newblock \bibinfo{publisher}{Elsevier}.
\newblock


\bibitem[\protect\citeauthoryear{Dola, Dwyer, and Soffa}{Dola
  et~al\mbox{.}}{2021}]%
        {DBLP:conf/icse/DolaDS21}
\bibfield{author}{\bibinfo{person}{Swaroopa Dola}, \bibinfo{person}{Matthew~B.
  Dwyer}, {and} \bibinfo{person}{Mary~Lou Soffa}.}
  \bibinfo{year}{2021}\natexlab{}.
\newblock \showarticletitle{Distribution-Aware Testing of Neural Networks Using
  Generative Models}. In \bibinfo{booktitle}{\emph{43rd {IEEE/ACM}
  International Conference on Software Engineering, {ICSE} 2021, Madrid, Spain,
  22-30 May 2021}}. \bibinfo{publisher}{{IEEE}}, \bibinfo{pages}{226--237}.
\newblock


\bibitem[\protect\citeauthoryear{Fawzi, Fawzi, and Frossard}{Fawzi
  et~al\mbox{.}}{2018}]%
        {fawzi2018analysis}
\bibfield{author}{\bibinfo{person}{Alhussein Fawzi}, \bibinfo{person}{Omar
  Fawzi}, {and} \bibinfo{person}{Pascal Frossard}.}
  \bibinfo{year}{2018}\natexlab{}.
\newblock \showarticletitle{Analysis of classifiers’ robustness to
  adversarial perturbations}.
\newblock \bibinfo{journal}{\emph{Machine learning}} \bibinfo{volume}{107},
  \bibinfo{number}{3} (\bibinfo{year}{2018}), \bibinfo{pages}{481--508}.
\newblock


\bibitem[\protect\citeauthoryear{Frankl, Hamlet, Littlewood, and
  Strigini}{Frankl et~al\mbox{.}}{1998}]%
        {frankl_evaluating_1998}
\bibfield{author}{\bibinfo{person}{P.~G. Frankl}, \bibinfo{person}{R.~G.
  Hamlet}, \bibinfo{person}{B. Littlewood}, {and} \bibinfo{person}{L.
  Strigini}.} \bibinfo{year}{1998}\natexlab{}.
\newblock \showarticletitle{Evaluating testing methods by delivered reliability
  [software]}.
\newblock \bibinfo{journal}{\emph{IEEE Tran. on Softw. Eng.}}
  \bibinfo{volume}{24}, \bibinfo{number}{8} (\bibinfo{year}{1998}),
  \bibinfo{pages}{586--601}.
\newblock


\bibitem[\protect\citeauthoryear{Gehr, Mirman, Drachsler-Cohen, Tsankov,
  Chaudhuri, and Vechev}{Gehr et~al\mbox{.}}{2018}]%
        {gehr2018ai2}
\bibfield{author}{\bibinfo{person}{Timon Gehr}, \bibinfo{person}{Matthew
  Mirman}, \bibinfo{person}{Dana Drachsler-Cohen}, \bibinfo{person}{Petar
  Tsankov}, \bibinfo{person}{Swarat Chaudhuri}, {and} \bibinfo{person}{Martin
  Vechev}.} \bibinfo{year}{2018}\natexlab{}.
\newblock \showarticletitle{Ai2: Safety and robustness certification of neural
  networks with abstract interpretation}. In \bibinfo{booktitle}{\emph{2018
  IEEE symposium on security and privacy (SP)}}. IEEE, \bibinfo{pages}{3--18}.
\newblock


\bibitem[\protect\citeauthoryear{Guerriero}{Guerriero}{2020}]%
        {guerriero_reliability_2020}
\bibfield{author}{\bibinfo{person}{Antonio Guerriero}.}
  \bibinfo{year}{2020}\natexlab{}.
\newblock \showarticletitle{Reliability {Evaluation} of {ML} systems, the
  oracle problem}. In \bibinfo{booktitle}{\emph{Int. Symp. on Software
  Reliability Engineering Workshops (ISSREW)}}. \bibinfo{publisher}{IEEE},
  \bibinfo{address}{Coimbra, Portugal}, \bibinfo{pages}{127--130}.
\newblock
\urldef\tempurl%
\url{https://doi.org/10.1109/ISSREW51248.2020.00050}
\showDOI{\tempurl}


\bibitem[\protect\citeauthoryear{Guerriero, Pietrantuono, and Russo}{Guerriero
  et~al\mbox{.}}{2021}]%
        {guerriero_operation_2021}
\bibfield{author}{\bibinfo{person}{Antonio Guerriero}, \bibinfo{person}{Roberto
  Pietrantuono}, {and} \bibinfo{person}{Stefano Russo}.}
  \bibinfo{year}{2021}\natexlab{}.
\newblock \showarticletitle{Operation is the {Hardest} {Teacher}: {Estimating}
  {DNN} {Accuracy} {Looking} for {Mispredictions}}. In
  \bibinfo{booktitle}{\emph{{IEEE}/{ACM} 43rd {Int}. {Conf}. on {Software}
  {Engineering}}} \emph{(\bibinfo{series}{{ICSE}'21})}.
  \bibinfo{address}{Madrid, Spain}, \bibinfo{pages}{348--358}.
\newblock
\urldef\tempurl%
\url{https://doi.org/10.1109/ICSE43902.2021.00042}
\showDOI{\tempurl}


\bibitem[\protect\citeauthoryear{Guo and Kang}{Guo and Kang}{2015}]%
        {guo_extended_2015}
\bibfield{author}{\bibinfo{person}{Lijie Guo} {and} \bibinfo{person}{Jianxin
  Kang}.} \bibinfo{year}{2015}\natexlab{}.
\newblock \showarticletitle{An extended {HAZOP} analysis approach with dynamic
  fault tree}.
\newblock \bibinfo{journal}{\emph{Journal of Loss Prevention in the Process
  Industries}}  \bibinfo{volume}{38} (\bibinfo{year}{2015}),
  \bibinfo{pages}{224--232}.
\newblock
\showISSN{0950-4230}
\urldef\tempurl%
\url{https://doi.org/10.1016/j.jlp.2015.10.003}
\showDOI{\tempurl}


\bibitem[\protect\citeauthoryear{Hamlet and Taylor}{Hamlet and Taylor}{1990}]%
        {hamlet_partition_1990}
\bibfield{author}{\bibinfo{person}{D. Hamlet} {and} \bibinfo{person}{R.
  Taylor}.} \bibinfo{year}{1990}\natexlab{}.
\newblock \showarticletitle{Partition testing does not inspire confidence}.
\newblock \bibinfo{journal}{\emph{IEEE Tran. on Software Engineering}}
  \bibinfo{volume}{16}, \bibinfo{number}{12} (\bibinfo{year}{1990}),
  \bibinfo{pages}{1402--1411}.
\newblock


\bibitem[\protect\citeauthoryear{Harel-Canada, Wang, Gulzar, Gu, and
  Kim}{Harel-Canada et~al\mbox{.}}{2020}]%
        {harel_canada_is_2020}
\bibfield{author}{\bibinfo{person}{Fabrice Harel-Canada},
  \bibinfo{person}{Lingxiao Wang}, \bibinfo{person}{Muhammad~Ali Gulzar},
  \bibinfo{person}{Quanquan Gu}, {and} \bibinfo{person}{Miryung Kim}.}
  \bibinfo{year}{2020}\natexlab{}.
\newblock \showarticletitle{Is {Neuron} {Coverage} a {Meaningful} {Measure} for
  {Testing} {Deep} {Neural} {Networks}?}. In \bibinfo{booktitle}{\emph{Proc. of
  the 28th {ACM} {Joint} {Meeting} on {European} {Software} {Engineering}
  {Conference} and {Symposium} on the {Foundations} of {Software}
  {Engineering}}}. \bibinfo{publisher}{ACM}, \bibinfo{address}{New York, NY,
  USA}, \bibinfo{pages}{851--862}.
\newblock


\bibitem[\protect\citeauthoryear{Hereau, Godary-Dejean, Guiochet, Robert,
  Claverie, and Crestani}{Hereau et~al\mbox{.}}{2020}]%
        {hereau_testing_2020}
\bibfield{author}{\bibinfo{person}{Adrien Hereau}, \bibinfo{person}{Karen
  Godary-Dejean}, \bibinfo{person}{J{\'e}r{\'e}mie Guiochet},
  \bibinfo{person}{Clément Robert}, \bibinfo{person}{Thomas Claverie}, {and}
  \bibinfo{person}{Didier Crestani}.} \bibinfo{year}{2020}\natexlab{}.
\newblock \showarticletitle{Testing an {Underwater} {Robot} {Executing}
  {Transect} {Missions} in {Mayotte}}. In \bibinfo{booktitle}{\emph{Towards
  {Autonomous} {Robotic} {Systems}}} \emph{(\bibinfo{series}{{LNCS}})},
  \bibfield{editor}{\bibinfo{person}{Abdelkhalick Mohammad},
  \bibinfo{person}{Xin Dong}, {and} \bibinfo{person}{Matteo Russo}} (Eds.),
  Vol.~\bibinfo{volume}{12228}. \bibinfo{publisher}{Springer},
  \bibinfo{address}{Cham}, \bibinfo{pages}{116--127}.
\newblock


\bibitem[\protect\citeauthoryear{Huang, Sun, Zhao, Sharp, Ruan, Meng, and
  Huang}{Huang et~al\mbox{.}}{2021}]%
        {huang2021coverage}
\bibfield{author}{\bibinfo{person}{Wei Huang}, \bibinfo{person}{Youcheng Sun},
  \bibinfo{person}{Xingyu Zhao}, \bibinfo{person}{James Sharp},
  \bibinfo{person}{Wenjie Ruan}, \bibinfo{person}{Jie Meng}, {and}
  \bibinfo{person}{Xiaowei Huang}.} \bibinfo{year}{2021}\natexlab{}.
\newblock \showarticletitle{Coverage Guided Testing for Recurrent Neural
  Networks}.
\newblock \bibinfo{journal}{\emph{IEEE Tran. on Reliability}}
  (\bibinfo{year}{2021}).
\newblock
\urldef\tempurl%
\url{https://doi.org/0.1109/TR.2021.3080664}
\showDOI{\tempurl}
\newblock
\shownote{early access.}


\bibitem[\protect\citeauthoryear{Huang, Kroening, Ruan, and et~al}{Huang
  et~al\mbox{.}}{2020}]%
        {huang_survey_2020}
\bibfield{author}{\bibinfo{person}{Xiaowei Huang}, \bibinfo{person}{Daniel
  Kroening}, \bibinfo{person}{Wenjie Ruan}, {and} \bibinfo{person}{et al}.}
  \bibinfo{year}{2020}\natexlab{}.
\newblock \showarticletitle{A survey of safety and trustworthiness of deep
  neural networks: {Verification}, testing, adversarial attack and defence, and
  interpretability}.
\newblock \bibinfo{journal}{\emph{Computer Science Review}}
  \bibinfo{volume}{37} (\bibinfo{year}{2020}), \bibinfo{pages}{100270}.
\newblock
\showISSN{1574-0137}


\bibitem[\protect\citeauthoryear{Huang, Kwiatkowska, Wang, and Wu}{Huang
  et~al\mbox{.}}{2017}]%
        {huang_safety_2017}
\bibfield{author}{\bibinfo{person}{Xiaowei Huang}, \bibinfo{person}{Marta
  Kwiatkowska}, \bibinfo{person}{Sen Wang}, {and} \bibinfo{person}{Min Wu}.}
  \bibinfo{year}{2017}\natexlab{}.
\newblock \showarticletitle{Safety verification of deep neural networks}. In
  \bibinfo{booktitle}{\emph{Computer {Aided} {Verification}}}
  \emph{(\bibinfo{series}{{LNCS}})}, Vol.~\bibinfo{volume}{10426}.
  \bibinfo{publisher}{Springer International Publishing},
  \bibinfo{address}{Cham}, \bibinfo{pages}{3--29}.
\newblock
\showISBNx{978-3-319-63387-9}


\bibitem[\protect\citeauthoryear{Ishikawa and Matsuno}{Ishikawa and
  Matsuno}{2018}]%
        {ishikawa_continuous_2018}
\bibfield{author}{\bibinfo{person}{Fuyuki Ishikawa} {and}
  \bibinfo{person}{Yutaka Matsuno}.} \bibinfo{year}{2018}\natexlab{}.
\newblock \showarticletitle{Continuous argument engineering: {Tackling}
  uncertainty in machine learning based systems}. In
  \bibinfo{booktitle}{\emph{SafeComp'18}} \emph{(\bibinfo{series}{{LNCS}})},
  \bibfield{editor}{\bibinfo{person}{Barbara Gallina}, \bibinfo{person}{Amund
  Skavhaug}, \bibinfo{person}{Erwin Schoitsch}, {and}
  \bibinfo{person}{Friedemann Bitsch}} (Eds.), Vol.~\bibinfo{volume}{11094}.
  \bibinfo{publisher}{Springer}, \bibinfo{address}{Cham},
  \bibinfo{pages}{14--21}.
\newblock


\bibitem[\protect\citeauthoryear{Javed, Muram, Hansson, Punnekkat, and
  Thane}{Javed et~al\mbox{.}}{2021}]%
        {javed_towards_2021}
\bibfield{author}{\bibinfo{person}{Muhammad~Atif Javed},
  \bibinfo{person}{Faiz~Ul Muram}, \bibinfo{person}{Hans Hansson},
  \bibinfo{person}{Sasikumar Punnekkat}, {and} \bibinfo{person}{Henrik Thane}.}
  \bibinfo{year}{2021}\natexlab{}.
\newblock \showarticletitle{Towards dynamic safety assurance for {Industry}
  4.0}.
\newblock \bibinfo{journal}{\emph{Journal of Systems Architecture}}
  \bibinfo{volume}{114} (\bibinfo{year}{2021}), \bibinfo{pages}{101914}.
\newblock
\showISSN{1383-7621}
\urldef\tempurl%
\url{https://doi.org/10.1016/j.sysarc.2020.101914}
\showDOI{\tempurl}


\bibitem[\protect\citeauthoryear{{Johnson, C. W.}}{{Johnson, C. W.}}{2018}]%
        {johnson_increasing_2018}
\bibfield{author}{\bibinfo{person}{{Johnson, C. W.}}}
  \bibinfo{year}{2018}\natexlab{}.
\newblock \showarticletitle{The increasing risks of risk assessment: {On} the
  rise of artificial intelligence and non-determinism in safety-critical
  systems}. In \bibinfo{booktitle}{\emph{the 26th {Safety}-{Critical} {Systems}
  {Symposium}}}. \bibinfo{publisher}{Safety-Critical Systems Club},
  \bibinfo{address}{York, UK.}, \bibinfo{pages}{15}.
\newblock


\bibitem[\protect\citeauthoryear{Kalra and Paddock}{Kalra and Paddock}{2016}]%
        {kalra_driving_2016}
\bibfield{author}{\bibinfo{person}{Nidhi Kalra} {and} \bibinfo{person}{Susan~M.
  Paddock}.} \bibinfo{year}{2016}\natexlab{}.
\newblock \showarticletitle{Driving to safety: {How} many miles of driving
  would it take to demonstrate autonomous vehicle reliability?}
\newblock \bibinfo{journal}{\emph{Transportation Research Part A: Policy and
  Practice}}  \bibinfo{volume}{94} (\bibinfo{year}{2016}), \bibinfo{pages}{182
  -- 193}.
\newblock
\showISSN{0965-8564}


\bibitem[\protect\citeauthoryear{Katz, Barrett, Dill, Julian, and
  Kochenderfer}{Katz et~al\mbox{.}}{2017}]%
        {katz2017reluplex}
\bibfield{author}{\bibinfo{person}{Guy Katz}, \bibinfo{person}{Clark Barrett},
  \bibinfo{person}{David~L. Dill}, \bibinfo{person}{Kyle Julian}, {and}
  \bibinfo{person}{Mykel~J. Kochenderfer}.} \bibinfo{year}{2017}\natexlab{}.
\newblock \showarticletitle{Reluplex: {An} efficient {SMT} solver for verifying
  deep neural networks}. In \bibinfo{booktitle}{\emph{CAV'17}}
  \emph{(\bibinfo{series}{{LNCS}})}, Vol.~\bibinfo{volume}{10426}.
  \bibinfo{publisher}{Springer}, \bibinfo{address}{Cham},
  \bibinfo{pages}{97--117}.
\newblock
\showISBNx{978-3-319-63387-9}


\bibitem[\protect\citeauthoryear{Kelly}{Kelly}{1999}]%
        {kelly_arguing_1999}
\bibfield{author}{\bibinfo{person}{Timothy~Patrick Kelly}.}
  \bibinfo{year}{1999}\natexlab{}.
\newblock \emph{\bibinfo{title}{Arguing safety: {A} systematic approach to
  managing safety cases}}.
\newblock {PhD} {Thesis}. \bibinfo{school}{University of York}.
\newblock


\bibitem[\protect\citeauthoryear{Kl{\"a}s, Adler, J{\"o}ckel, Gro{\ss}, and
  Reich}{Kl{\"a}s et~al\mbox{.}}{2021}]%
        {klas2021using}
\bibfield{author}{\bibinfo{person}{Michael Kl{\"a}s}, \bibinfo{person}{Rasmus
  Adler}, \bibinfo{person}{Lisa J{\"o}ckel}, \bibinfo{person}{Janek Gro{\ss}},
  {and} \bibinfo{person}{Jan Reich}.} \bibinfo{year}{2021}\natexlab{}.
\newblock \showarticletitle{Using Complementary Risk Acceptance Criteria to
  Structure Assurance Cases for Safety-Critical AI Components}. In
  \bibinfo{booktitle}{\emph{{AISafety}'21 {Workshop} at {IJCAI}'21}}.
\newblock


\bibitem[\protect\citeauthoryear{Knight}{Knight}{2015}]%
        {knight_importance_2015}
\bibfield{author}{\bibinfo{person}{John Knight}.}
  \bibinfo{year}{2015}\natexlab{}.
\newblock \showarticletitle{The importance of security cases: {Proof} is good,
  but not enough}.
\newblock \bibinfo{journal}{\emph{IEEE Security Privacy}} \bibinfo{volume}{13},
  \bibinfo{number}{4} (\bibinfo{year}{2015}), \bibinfo{pages}{73--75}.
\newblock
\showISSN{1540-7993}
\urldef\tempurl%
\url{https://doi.org/10.1109/MSP.2015.68}
\showDOI{\tempurl}


\bibitem[\protect\citeauthoryear{Koopman, Kane, and Black}{Koopman
  et~al\mbox{.}}{2019}]%
        {KKB2019}
\bibfield{author}{\bibinfo{person}{Philip Koopman}, \bibinfo{person}{Aaron
  Kane}, {and} \bibinfo{person}{Jen Black}.} \bibinfo{year}{2019}\natexlab{}.
\newblock \showarticletitle{Credible autonomy safety argumentation}. In
  \bibinfo{booktitle}{\emph{27th {Safety}-{Critical} {Systems} {Symp.}}}
  \bibinfo{publisher}{Safety-Critical Systems Club}, \bibinfo{address}{Bristol,
  UK}.
\newblock


\bibitem[\protect\citeauthoryear{Kurakin, Goodfellow, and Bengio}{Kurakin
  et~al\mbox{.}}{2018}]%
        {kurakin2018adversarial}
\bibfield{author}{\bibinfo{person}{Alexey Kurakin}, \bibinfo{person}{Ian~J
  Goodfellow}, {and} \bibinfo{person}{Samy Bengio}.}
  \bibinfo{year}{2018}\natexlab{}.
\newblock \showarticletitle{Adversarial examples in the physical world}.
\newblock In \bibinfo{booktitle}{\emph{Artificial intelligence safety and
  security}}. \bibinfo{publisher}{Chapman and Hall/CRC},
  \bibinfo{pages}{99--112}.
\newblock


\bibitem[\protect\citeauthoryear{Lane, Bisset, Buckingham, Pegman, and
  Prescott}{Lane et~al\mbox{.}}{2016}]%
        {lane_new_2016}
\bibfield{author}{\bibinfo{person}{David Lane}, \bibinfo{person}{David Bisset},
  \bibinfo{person}{Rob Buckingham}, \bibinfo{person}{Geoff Pegman}, {and}
  \bibinfo{person}{Tony Prescott}.} \bibinfo{year}{2016}\natexlab{}.
\newblock \bibinfo{booktitle}{\emph{New foresight review on robotics and
  autonomous systems}}.
\newblock \bibinfo{type}{{T}echnical {R}eport} No. 2016.1.
  \bibinfo{institution}{Lloyd’s Register Foundation},
  \bibinfo{address}{London, U.K.} \bibinfo{pages}{65} pages.
\newblock


\bibitem[\protect\citeauthoryear{Lee, Wysocki, Zhou, Shotton, Tivey, Lever,
  Woodcock, Albiges, Angelakas, Arnold, et~al\mbox{.}}{Lee
  et~al\mbox{.}}{2022}]%
        {lee2022establishment}
\bibfield{author}{\bibinfo{person}{Rebecca~J Lee}, \bibinfo{person}{Oskar
  Wysocki}, \bibinfo{person}{Cong Zhou}, \bibinfo{person}{Rohan Shotton},
  \bibinfo{person}{Ann Tivey}, \bibinfo{person}{Louise Lever},
  \bibinfo{person}{Joshua Woodcock}, \bibinfo{person}{Laurence Albiges},
  \bibinfo{person}{Angelos Angelakas}, \bibinfo{person}{Dirk Arnold},
  {et~al\mbox{.}}} \bibinfo{year}{2022}\natexlab{}.
\newblock \showarticletitle{Establishment of CORONET, COVID-19 Risk in Oncology
  Evaluation Tool, to Identify Patients With Cancer at Low Versus High Risk of
  Severe Complications of COVID-19 Disease On Presentation to Hospital}.
\newblock \bibinfo{journal}{\emph{JCO Clinical Cancer Informatics}}
  \bibinfo{volume}{6} (\bibinfo{year}{2022}), \bibinfo{pages}{e2100177}.
\newblock


\bibitem[\protect\citeauthoryear{Lee, Grosh, Tillman, and Lie}{Lee
  et~al\mbox{.}}{1985}]%
        {lee_fault_1985}
\bibfield{author}{\bibinfo{person}{W.~S. Lee}, \bibinfo{person}{D.~L. Grosh},
  \bibinfo{person}{F.~A. Tillman}, {and} \bibinfo{person}{C.~H. Lie}.}
  \bibinfo{year}{1985}\natexlab{}.
\newblock \showarticletitle{Fault {Tree} {Analysis}, {Methods}, and
  {Applications} - {A} {Review}}.
\newblock \bibinfo{journal}{\emph{IEEE Tran. on Reliability}}
  \bibinfo{volume}{R-34}, \bibinfo{number}{3} (\bibinfo{year}{1985}),
  \bibinfo{pages}{194--203}.
\newblock
\urldef\tempurl%
\url{https://doi.org/10.1109/TR.1985.5222114}
\showDOI{\tempurl}


\bibitem[\protect\citeauthoryear{Li, Ma, Xu, Cao, Xu, and Lü}{Li
  et~al\mbox{.}}{2019}]%
        {li_boosting_2019}
\bibfield{author}{\bibinfo{person}{Zenan Li}, \bibinfo{person}{Xiaoxing Ma},
  \bibinfo{person}{Chang Xu}, \bibinfo{person}{Chun Cao},
  \bibinfo{person}{Jingwei Xu}, {and} \bibinfo{person}{Jian Lü}.}
  \bibinfo{year}{2019}\natexlab{}.
\newblock \showarticletitle{Boosting operational {DNN} testing efficiency
  through conditioning}. In \bibinfo{booktitle}{\emph{Proc. of the 27th {ACM}
  {Joint} {Meeting} on {European} {Software} {Engineering} {Conference} and
  {Symposium} on the {Foundations} of {Software} {Engineering}}}
  \emph{(\bibinfo{series}{{ESEC}/{FSE} 2019})}. \bibinfo{publisher}{ACM},
  \bibinfo{address}{New York, NY, USA}, \bibinfo{pages}{499--509}.
\newblock
\showISBNx{978-1-4503-5572-8}
\urldef\tempurl%
\url{https://doi.org/10.1145/3338906.3338930}
\showDOI{\tempurl}


\bibitem[\protect\citeauthoryear{Littlewood and Rushby}{Littlewood and
  Rushby}{2012}]%
        {littlewood_reasoning_2012}
\bibfield{author}{\bibinfo{person}{B. Littlewood} {and} \bibinfo{person}{J.
  Rushby}.} \bibinfo{year}{2012}\natexlab{}.
\newblock \showarticletitle{Reasoning about the reliability of diverse
  two-channel systems in which one channel is ``possibly perfect''}.
\newblock \bibinfo{journal}{\emph{IEEE Tran. on Software Engineering}}
  \bibinfo{volume}{38}, \bibinfo{number}{5} (\bibinfo{year}{2012}),
  \bibinfo{pages}{1178--1194}.
\newblock


\bibitem[\protect\citeauthoryear{Littlewood and Strigini}{Littlewood and
  Strigini}{1993}]%
        {littlewood_validation_1993}
\bibfield{author}{\bibinfo{person}{Bev Littlewood} {and}
  \bibinfo{person}{Lorenzo Strigini}.} \bibinfo{year}{1993}\natexlab{}.
\newblock \showarticletitle{Validation of ultra-high dependability for
  software-based systems}.
\newblock \bibinfo{journal}{\emph{Commun. ACM}} \bibinfo{volume}{36},
  \bibinfo{number}{11} (\bibinfo{year}{1993}), \bibinfo{pages}{69--80}.
\newblock
\showISSN{0001-0782}


\bibitem[\protect\citeauthoryear{Littlewood and Strigini}{Littlewood and
  Strigini}{2000}]%
        {littlewood_software_2000}
\bibfield{author}{\bibinfo{person}{Bev Littlewood} {and}
  \bibinfo{person}{Lorenzo Strigini}.} \bibinfo{year}{2000}\natexlab{}.
\newblock \showarticletitle{Software reliability and dependability: {A}
  roadmap}. In \bibinfo{booktitle}{\emph{Proc. of the {Conference} on {The}
  {Future} of {Software} {Engineering}}} \emph{(\bibinfo{series}{{ICSE} '00})}.
  \bibinfo{publisher}{ACM}, \bibinfo{address}{New York, NY, USA},
  \bibinfo{pages}{175--188}.
\newblock
\showISBNx{1-58113-253-0}
\urldef\tempurl%
\url{https://doi.org/10.1145/336512.336551}
\showDOI{\tempurl}


\bibitem[\protect\citeauthoryear{Liu, Yang, and Xu}{Liu et~al\mbox{.}}{2019}]%
        {liu_how_2019}
\bibfield{author}{\bibinfo{person}{Peng Liu}, \bibinfo{person}{Run Yang}, {and}
  \bibinfo{person}{Zhigang Xu}.} \bibinfo{year}{2019}\natexlab{}.
\newblock \showarticletitle{How safe is safe enough for self-driving vehicles?}
\newblock \bibinfo{journal}{\emph{Risk Analysis}} \bibinfo{volume}{39},
  \bibinfo{number}{2} (\bibinfo{year}{2019}), \bibinfo{pages}{315--325}.
\newblock


\bibitem[\protect\citeauthoryear{Madry, Makelov, Schmidt, Tsipras, and
  Vladu}{Madry et~al\mbox{.}}{2018}]%
        {madry2018towards}
\bibfield{author}{\bibinfo{person}{Aleksander Madry},
  \bibinfo{person}{Aleksandar Makelov}, \bibinfo{person}{Ludwig Schmidt},
  \bibinfo{person}{Dimitris Tsipras}, {and} \bibinfo{person}{Adrian Vladu}.}
  \bibinfo{year}{2018}\natexlab{}.
\newblock \showarticletitle{Towards Deep Learning Models Resistant to
  Adversarial Attacks}. In \bibinfo{booktitle}{\emph{International Conference
  on Learning Representations}}.
\newblock


\bibitem[\protect\citeauthoryear{Matsuno, Ishikawa, and Tokumoto}{Matsuno
  et~al\mbox{.}}{2019}]%
        {matsuno_tackling_2019}
\bibfield{author}{\bibinfo{person}{Yutaka Matsuno}, \bibinfo{person}{Fuyuki
  Ishikawa}, {and} \bibinfo{person}{Susumu Tokumoto}.}
  \bibinfo{year}{2019}\natexlab{}.
\newblock \showarticletitle{Tackling uncertainty in safety assurance for
  machine learning: {Continuous} argument engineering with attributed tests}.
  In \bibinfo{booktitle}{\emph{SafeComp'19}} \emph{(\bibinfo{series}{{LNCS}})},
  Vol.~\bibinfo{volume}{11699}. \bibinfo{publisher}{Springer},
  \bibinfo{address}{Cham}, \bibinfo{pages}{398--404}.
\newblock
\showISBNx{978-3-030-26250-1}


\bibitem[\protect\citeauthoryear{Micouin}{Micouin}{2008}]%
        {Micouin2008}
\bibfield{author}{\bibinfo{person}{Patrice Micouin}.}
  \bibinfo{year}{2008}\natexlab{}.
\newblock \showarticletitle{Toward a property based requirements theory: System
  requirements structured as a semilattice}.
\newblock \bibinfo{journal}{\emph{Systems Engineering}} \bibinfo{volume}{11},
  \bibinfo{number}{3} (\bibinfo{year}{2008}), \bibinfo{pages}{235--245}.
\newblock


\bibitem[\protect\citeauthoryear{Miller, Morell, Noonan, Park, Nicol, Murrill,
  and Voas}{Miller et~al\mbox{.}}{1992}]%
        {miller_estimating_1992}
\bibfield{author}{\bibinfo{person}{Keith~W. Miller}, \bibinfo{person}{Larry~J.
  Morell}, \bibinfo{person}{Robert~E. Noonan}, \bibinfo{person}{Stephen~K.
  Park}, \bibinfo{person}{David~M. Nicol}, \bibinfo{person}{Branson~W.
  Murrill}, {and} \bibinfo{person}{M Voas}.} \bibinfo{year}{1992}\natexlab{}.
\newblock \showarticletitle{Estimating the probability of failure when testing
  reveals no failures}.
\newblock \bibinfo{journal}{\emph{IEEE Tran. on Software Engineering}}
  \bibinfo{volume}{18}, \bibinfo{number}{1} (\bibinfo{year}{1992}),
  \bibinfo{pages}{33--43}.
\newblock
\showISSN{0098-5589}


\bibitem[\protect\citeauthoryear{Moosavi-Dezfooli, Fawzi, Fawzi, and
  Frossard}{Moosavi-Dezfooli et~al\mbox{.}}{2017}]%
        {moosavi2017universal}
\bibfield{author}{\bibinfo{person}{Seyed-Mohsen Moosavi-Dezfooli},
  \bibinfo{person}{Alhussein Fawzi}, \bibinfo{person}{Omar Fawzi}, {and}
  \bibinfo{person}{Pascal Frossard}.} \bibinfo{year}{2017}\natexlab{}.
\newblock \showarticletitle{Universal adversarial perturbations}. In
  \bibinfo{booktitle}{\emph{Proceedings of the IEEE conference on computer
  vision and pattern recognition}}. \bibinfo{pages}{1765--1773}.
\newblock


\bibitem[\protect\citeauthoryear{Moosavi-Dezfooli, Fawzi, and
  Frossard}{Moosavi-Dezfooli et~al\mbox{.}}{2016}]%
        {moosavi2016deepfool}
\bibfield{author}{\bibinfo{person}{Seyed-Mohsen Moosavi-Dezfooli},
  \bibinfo{person}{Alhussein Fawzi}, {and} \bibinfo{person}{Pascal Frossard}.}
  \bibinfo{year}{2016}\natexlab{}.
\newblock \showarticletitle{Deepfool: a simple and accurate method to fool deep
  neural networks}. In \bibinfo{booktitle}{\emph{Proceedings of the IEEE
  Conference on Computer Vision and Pattern Recognition}}.
  \bibinfo{pages}{2574--2582}.
\newblock


\bibitem[\protect\citeauthoryear{Musa}{Musa}{1993}]%
        {musa_operational_1993}
\bibfield{author}{\bibinfo{person}{John Musa}.}
  \bibinfo{year}{1993}\natexlab{}.
\newblock \showarticletitle{Operational profiles in software-reliability
  engineering}.
\newblock \bibinfo{journal}{\emph{IEEE Software}} \bibinfo{volume}{10},
  \bibinfo{number}{2} (\bibinfo{year}{1993}), \bibinfo{pages}{14--32}.
\newblock
\showISSN{0740-7459}


\bibitem[\protect\citeauthoryear{Picardi, Hawkins, Paterson, and Habli}{Picardi
  et~al\mbox{.}}{2019}]%
        {picardi_pattern_2019}
\bibfield{author}{\bibinfo{person}{Chiara Picardi}, \bibinfo{person}{Richard
  Hawkins}, \bibinfo{person}{Colin Paterson}, {and} \bibinfo{person}{Ibrahim
  Habli}.} \bibinfo{year}{2019}\natexlab{}.
\newblock \showarticletitle{A pattern for arguing the assurance of machine
  learning in medical diagnosis systems}. In \bibinfo{booktitle}{\emph{Computer
  {Safety}, {Reliability}, and {Security}}} \emph{(\bibinfo{series}{{LNCS}})},
  \bibfield{editor}{\bibinfo{person}{Alexander Romanovsky},
  \bibinfo{person}{Elena Troubitsyna}, {and} \bibinfo{person}{Friedemann
  Bitsch}} (Eds.), Vol.~\bibinfo{volume}{11698}. \bibinfo{publisher}{Springer},
  \bibinfo{address}{Cham}, \bibinfo{pages}{165--179}.
\newblock


\bibitem[\protect\citeauthoryear{Pietrantuono, Popov, and Russo}{Pietrantuono
  et~al\mbox{.}}{2020}]%
        {pietrantuono_reliability_2020}
\bibfield{author}{\bibinfo{person}{Roberto Pietrantuono},
  \bibinfo{person}{Peter Popov}, {and} \bibinfo{person}{Stefano Russo}.}
  \bibinfo{year}{2020}\natexlab{}.
\newblock \showarticletitle{Reliability assessment of service-based software
  under operational profile uncertainty}.
\newblock \bibinfo{journal}{\emph{Reliability Engineering \& System Safety}}
  \bibinfo{volume}{204} (\bibinfo{year}{2020}), \bibinfo{pages}{107193}.
\newblock


\bibitem[\protect\citeauthoryear{Qahtan, Wang, and Zhang}{Qahtan
  et~al\mbox{.}}{2017}]%
        {qahtan_kde_track_2017}
\bibfield{author}{\bibinfo{person}{Abdulhakim Qahtan}, \bibinfo{person}{Suojin
  Wang}, {and} \bibinfo{person}{Xiangliang Zhang}.}
  \bibinfo{year}{2017}\natexlab{}.
\newblock \showarticletitle{{KDE}-{Track}: {An} {Efficient} {Dynamic} {Density}
  {Estimator} for {Data} {Streams}}.
\newblock \bibinfo{journal}{\emph{IEEE Tran. on Knowledge and Data
  Engineering}} \bibinfo{volume}{29}, \bibinfo{number}{3}
  (\bibinfo{year}{2017}), \bibinfo{pages}{642--655}.
\newblock
\urldef\tempurl%
\url{https://doi.org/10.1109/TKDE.2016.2626441}
\showDOI{\tempurl}


\bibitem[\protect\citeauthoryear{Qi, Conmy, Huang, Zhao, and Huang}{Qi
  et~al\mbox{.}}{2022}]%
        {hills2022}
\bibfield{author}{\bibinfo{person}{Yi Qi}, \bibinfo{person}{Philippa~Ryan
  Conmy}, \bibinfo{person}{Wei Huang}, \bibinfo{person}{Xingyu Zhao}, {and}
  \bibinfo{person}{Xiaowei Huang}.} \bibinfo{year}{2022}\natexlab{}.
\newblock \showarticletitle{A Hierarchical {HAZOP}-Like Safety Analysis for
  Learning-Enabled Systems}. In \bibinfo{booktitle}{\emph{{AISafety}'22
  {Workshop} at {IJCAI}'22}}.
\newblock


\bibitem[\protect\citeauthoryear{Redmon and Farhadi}{Redmon and
  Farhadi}{2018}]%
        {redmon2018yolov3}
\bibfield{author}{\bibinfo{person}{Joseph Redmon} {and} \bibinfo{person}{Ali
  Farhadi}.} \bibinfo{year}{2018}\natexlab{}.
\newblock \showarticletitle{Yolov3: An incremental improvement}.
\newblock \bibinfo{journal}{\emph{arXiv preprint arXiv:1804.02767}}
  (\bibinfo{year}{2018}).
\newblock


\bibitem[\protect\citeauthoryear{Robert, Sotiropoulos, Waeselynck, Guiochet,
  and Vernhes}{Robert et~al\mbox{.}}{2020}]%
        {robert_virtual_2020}
\bibfield{author}{\bibinfo{person}{Cl{\'e}ment Robert},
  \bibinfo{person}{Thierry Sotiropoulos}, \bibinfo{person}{H{\'e}lène
  Waeselynck}, \bibinfo{person}{J{\'e}r{\'e}mie Guiochet}, {and}
  \bibinfo{person}{Simon Vernhes}.} \bibinfo{year}{2020}\natexlab{}.
\newblock \showarticletitle{The virtual lands of {Oz}: Testing an agribot in
  simulation}.
\newblock \bibinfo{journal}{\emph{Empirical Software Engineering}}
  \bibinfo{volume}{25}, \bibinfo{number}{3} (\bibinfo{date}{May}
  \bibinfo{year}{2020}), \bibinfo{pages}{2025--2054}.
\newblock
\urldef\tempurl%
\url{https://doi.org/10.1007/s10664-020-09800-3}
\showDOI{\tempurl}


\bibitem[\protect\citeauthoryear{Ruijters and Stoelinga}{Ruijters and
  Stoelinga}{2015}]%
        {ruijters_fault_2015}
\bibfield{author}{\bibinfo{person}{Enno Ruijters} {and}
  \bibinfo{person}{Mariëlle Stoelinga}.} \bibinfo{year}{2015}\natexlab{}.
\newblock \showarticletitle{Fault tree analysis: {A} survey of the
  state-of-the-art in modeling, analysis and tools}.
\newblock \bibinfo{journal}{\emph{Computer Science Review}}
  \bibinfo{volume}{15-16} (\bibinfo{year}{2015}), \bibinfo{pages}{29--62}.
\newblock
\showISSN{1574-0137}


\bibitem[\protect\citeauthoryear{{S. Toulmin}}{{S. Toulmin}}{1958}]%
        {s_toulmin_uses_1958}
\bibfield{author}{\bibinfo{person}{{S. Toulmin}}.}
  \bibinfo{year}{1958}\natexlab{}.
\newblock \bibinfo{booktitle}{\emph{The {Uses} of {Argument}}}.
\newblock \bibinfo{publisher}{Cambridge University Press}.
\newblock


\bibitem[\protect\citeauthoryear{Scott}{Scott}{2015}]%
        {scott2015multivariate}
\bibfield{author}{\bibinfo{person}{David~W Scott}.}
  \bibinfo{year}{2015}\natexlab{}.
\newblock \bibinfo{booktitle}{\emph{Multivariate density estimation: theory,
  practice, and visualization}}.
\newblock \bibinfo{publisher}{John Wiley \& Sons}.
\newblock


\bibitem[\protect\citeauthoryear{Silverman}{Silverman}{1986}]%
        {silverman1986density}
\bibfield{author}{\bibinfo{person}{Bernard~W Silverman}.}
  \bibinfo{year}{1986}\natexlab{}.
\newblock \bibinfo{booktitle}{\emph{Density estimation for statistics and data
  analysis}}. Vol.~\bibinfo{volume}{26}.
\newblock \bibinfo{publisher}{CRC press}.
\newblock


\bibitem[\protect\citeauthoryear{Singh, Gehr, P{\"u}schel, and Vechev}{Singh
  et~al\mbox{.}}{2019}]%
        {singh2019abstract}
\bibfield{author}{\bibinfo{person}{Gagandeep Singh}, \bibinfo{person}{Timon
  Gehr}, \bibinfo{person}{Markus P{\"u}schel}, {and} \bibinfo{person}{Martin
  Vechev}.} \bibinfo{year}{2019}\natexlab{}.
\newblock \showarticletitle{An abstract domain for certifying neural networks}.
\newblock \bibinfo{journal}{\emph{Proceedings of the ACM on Programming
  Languages}} \bibinfo{volume}{3}, \bibinfo{number}{POPL}
  (\bibinfo{year}{2019}), \bibinfo{pages}{1--30}.
\newblock


\bibitem[\protect\citeauthoryear{Smidts, Mutha, Rodríguez, and Gerber}{Smidts
  et~al\mbox{.}}{2014}]%
        {smidts_software_2014}
\bibfield{author}{\bibinfo{person}{Carol Smidts}, \bibinfo{person}{Chetan
  Mutha}, \bibinfo{person}{Manuel Rodríguez}, {and}
  \bibinfo{person}{Matthew~J. Gerber}.} \bibinfo{year}{2014}\natexlab{}.
\newblock \showarticletitle{Software {Testing} with an {Operational} {Profile}:
  {OP} {Definition}}.
\newblock \bibinfo{journal}{\emph{Comput. Surveys}} \bibinfo{volume}{46},
  \bibinfo{number}{3} (\bibinfo{year}{2014}).
\newblock


\bibitem[\protect\citeauthoryear{Strigini and Littlewood}{Strigini and
  Littlewood}{1997}]%
        {strigini_guidelines_1997}
\bibfield{author}{\bibinfo{person}{Lorenzo Strigini} {and} \bibinfo{person}{Bev
  Littlewood}.} \bibinfo{year}{1997}\natexlab{}.
\newblock \bibinfo{booktitle}{\emph{Guidelines for Statistical Testing}}.
\newblock \bibinfo{type}{{T}echnical {R}eport}. \bibinfo{institution}{City,
  University of London}.
\newblock
\urldef\tempurl%
\url{http://openaccess.city.ac.uk/254/}
\showURL{%
\tempurl}


\bibitem[\protect\citeauthoryear{Strigini and Povyakalo}{Strigini and
  Povyakalo}{2013}]%
        {strigini_software_2013}
\bibfield{author}{\bibinfo{person}{Lorenzo Strigini} {and}
  \bibinfo{person}{Andrey Povyakalo}.} \bibinfo{year}{2013}\natexlab{}.
\newblock \showarticletitle{Software fault-freeness and reliability
  predictions}. In \bibinfo{booktitle}{\emph{Computer {Safety}, {Reliability},
  and {Security}}} \emph{(\bibinfo{series}{{LNCS}})},
  \bibfield{editor}{\bibinfo{person}{Friedemann Bitsch},
  \bibinfo{person}{Jérémie Guiochet}, {and} \bibinfo{person}{Mohamed
  Kaâniche}} (Eds.), Vol.~\bibinfo{volume}{8153}. \bibinfo{publisher}{Springer
  Berlin Heidelberg}, \bibinfo{address}{Berlin, Heidelberg},
  \bibinfo{pages}{106--117}.
\newblock
\showISBNx{978-3-642-40793-2}
\urldef\tempurl%
\url{https://doi.org/10.1007/978-3-642-40793-2_10}
\showDOI{\tempurl}


\bibitem[\protect\citeauthoryear{Swann and Preston}{Swann and Preston}{1995}]%
        {swann_twenty_five_1995}
\bibfield{author}{\bibinfo{person}{C.~D. Swann} {and} \bibinfo{person}{M.~L.
  Preston}.} \bibinfo{year}{1995}\natexlab{}.
\newblock \showarticletitle{Twenty-five years of {HAZOPs}}.
\newblock \bibinfo{journal}{\emph{Journal of Loss Prevention in the Process
  Industries}} \bibinfo{volume}{8}, \bibinfo{number}{6} (\bibinfo{year}{1995}),
  \bibinfo{pages}{349--353}.
\newblock
\showISSN{0950-4230}
\urldef\tempurl%
\url{https://doi.org/10.1016/0950-4230(95)00041-0}
\showDOI{\tempurl}


\bibitem[\protect\citeauthoryear{{UK Office for Nuclear Regulation}}{{UK Office
  for Nuclear Regulation}}{2019}]%
        {uk_office_for_nuclear_regulation_purpose_2019}
\bibfield{author}{\bibinfo{person}{{UK Office for Nuclear Regulation}}.}
  \bibinfo{year}{2019}\natexlab{}.
\newblock \bibinfo{booktitle}{\emph{The purpose, scope and content of safety
  cases}}.
\newblock \bibinfo{type}{Nuclear {Safety} {Technical} {Assessment} {Guide}}
  NS-TAST-GD-051. \bibinfo{institution}{Office for Nuclear Regulation}.
  \bibinfo{pages}{39} pages.
\newblock
\urldef\tempurl%
\url{https://www.onr.org.uk/operational/tech_asst_guides/ns-tast-gd-051.pdf}
\showURL{%
\tempurl}


\bibitem[\protect\citeauthoryear{Walter and Augustin}{Walter and
  Augustin}{2009}]%
        {walter_imprecision_2009}
\bibfield{author}{\bibinfo{person}{Gero Walter} {and} \bibinfo{person}{Thomas
  Augustin}.} \bibinfo{year}{2009}\natexlab{}.
\newblock \showarticletitle{Imprecision and prior-data conflict in generalized
  {Bayesian} inference}.
\newblock \bibinfo{journal}{\emph{Journal of Statistical Theory \& Practice}}
  \bibinfo{volume}{3}, \bibinfo{number}{1} (\bibinfo{year}{2009}),
  \bibinfo{pages}{255--271}.
\newblock


\bibitem[\protect\citeauthoryear{Wang, Webb, and Rainforth}{Wang
  et~al\mbox{.}}{2021}]%
        {wang_statistically_2021}
\bibfield{author}{\bibinfo{person}{Benjie Wang}, \bibinfo{person}{Stefan Webb},
  {and} \bibinfo{person}{Tom Rainforth}.} \bibinfo{year}{2021}\natexlab{}.
\newblock \showarticletitle{Statistically robust neural network
  classification}. In \bibinfo{booktitle}{\emph{Proc. of the 37th {Conf.} on
  {Uncertainty} in {Artificial} {Intelligence}}}, Vol.~\bibinfo{volume}{161}.
  \bibinfo{publisher}{PMLR}, \bibinfo{pages}{1735--1745}.
\newblock


\bibitem[\protect\citeauthoryear{Webb, Rainforth, Teh, and Kumar}{Webb
  et~al\mbox{.}}{2019}]%
        {webb_statistical_2019}
\bibfield{author}{\bibinfo{person}{Stefan Webb}, \bibinfo{person}{Tom
  Rainforth}, \bibinfo{person}{Yee~Whye Teh}, {and} \bibinfo{person}{M.~Pawan
  Kumar}.} \bibinfo{year}{2019}\natexlab{}.
\newblock \showarticletitle{A statistical approach to assessing neural network
  robustness}. In \bibinfo{booktitle}{\emph{7th {Int}. {Conf}. {Learning}
  {Representations} (ICLR'19)}}. \bibinfo{publisher}{OpenReview.net},
  \bibinfo{address}{New Orleans, LA, USA}.
\newblock


\bibitem[\protect\citeauthoryear{Weng, Chen, Nguyen, Squillante, Boopathy,
  Oseledets, and Daniel}{Weng et~al\mbox{.}}{2019}]%
        {weng2019proven}
\bibfield{author}{\bibinfo{person}{Lily Weng}, \bibinfo{person}{Pin-Yu Chen},
  \bibinfo{person}{Lam Nguyen}, \bibinfo{person}{Mark Squillante},
  \bibinfo{person}{Akhilan Boopathy}, \bibinfo{person}{Ivan Oseledets}, {and}
  \bibinfo{person}{Luca Daniel}.} \bibinfo{year}{2019}\natexlab{}.
\newblock \showarticletitle{PROVEN: Verifying robustness of neural networks
  with a probabilistic approach}. In \bibinfo{booktitle}{\emph{Int. Conf. on
  Machine Learning}}. PMLR, \bibinfo{pages}{6727--6736}.
\newblock


\bibitem[\protect\citeauthoryear{{Weng}, {Zhang}, {Chen}, {Yi}, {Su}, {Gao},
  {Hsieh}, and {Daniel}}{{Weng} et~al\mbox{.}}{2018}]%
        {WZCYSGHD2018}
\bibfield{author}{\bibinfo{person}{T.-W. {Weng}}, \bibinfo{person}{H. {Zhang}},
  \bibinfo{person}{P.-Y. {Chen}}, \bibinfo{person}{J. {Yi}},
  \bibinfo{person}{D. {Su}}, \bibinfo{person}{Y. {Gao}}, \bibinfo{person}{C.-J.
  {Hsieh}}, {and} \bibinfo{person}{L. {Daniel}}.}
  \bibinfo{year}{2018}\natexlab{}.
\newblock \showarticletitle{{Evaluating the Robustness of Neural Networks: An
  Extreme Value Theory Approach}}. In \bibinfo{booktitle}{\emph{International
  Conference on Learning Representations (ICLR)}}.
\newblock


\bibitem[\protect\citeauthoryear{Yang, Rashtchian, Zhang, Salakhutdinov, and
  Chaudhuri}{Yang et~al\mbox{.}}{2020}]%
        {yang_closer_2020}
\bibfield{author}{\bibinfo{person}{Yao-Yuan Yang}, \bibinfo{person}{Cyrus
  Rashtchian}, \bibinfo{person}{Hongyang Zhang}, \bibinfo{person}{Russ~R
  Salakhutdinov}, {and} \bibinfo{person}{Kamalika Chaudhuri}.}
  \bibinfo{year}{2020}\natexlab{}.
\newblock \showarticletitle{A {Closer} {Look} at {Accuracy} vs. {Robustness}}.
  In \bibinfo{booktitle}{\emph{Advances in {Neural} {Information} {Processing}
  {Systems}}} \emph{(\bibinfo{series}{{NeurIPS}'20})},
  \bibfield{editor}{\bibinfo{person}{H.~Larochelle},
  \bibinfo{person}{M.~Ranzato}, \bibinfo{person}{R.~Hadsell},
  \bibinfo{person}{M.~F. Balcan}, {and} \bibinfo{person}{H.~Lin}} (Eds.),
  Vol.~\bibinfo{volume}{33}. \bibinfo{publisher}{Curran Associates, Inc.},
  \bibinfo{pages}{8588--8601}.
\newblock


\bibitem[\protect\citeauthoryear{Yu, Wang, Shelhamer, and Darrell}{Yu
  et~al\mbox{.}}{2018}]%
        {yu2018deep}
\bibfield{author}{\bibinfo{person}{Fisher Yu}, \bibinfo{person}{Dequan Wang},
  \bibinfo{person}{Evan Shelhamer}, {and} \bibinfo{person}{Trevor Darrell}.}
  \bibinfo{year}{2018}\natexlab{}.
\newblock \showarticletitle{Deep layer aggregation}. In
  \bibinfo{booktitle}{\emph{Proceedings of the IEEE conference on computer
  vision and pattern recognition}}. \bibinfo{pages}{2403--2412}.
\newblock


\bibitem[\protect\citeauthoryear{Zhao, Banks, Sharp, Robu, Flynn, Fisher, and
  Huang}{Zhao et~al\mbox{.}}{2020a}]%
        {zhao_safety_2020}
\bibfield{author}{\bibinfo{person}{Xingyu Zhao}, \bibinfo{person}{Alec Banks},
  \bibinfo{person}{James Sharp}, \bibinfo{person}{Valentin Robu},
  \bibinfo{person}{David Flynn}, \bibinfo{person}{Michael Fisher}, {and}
  \bibinfo{person}{Xiaowei Huang}.} \bibinfo{year}{2020}\natexlab{a}.
\newblock \showarticletitle{A {Safety} {Framework} for {Critical} {Systems}
  {Utilising} {Deep} {Neural} {Networks}}. In
  \bibinfo{booktitle}{\emph{Computer {Safety}, {Reliability}, and {Security}}}
  \emph{(\bibinfo{series}{{LNCS}})},
  \bibfield{editor}{\bibinfo{person}{António Casimiro}, \bibinfo{person}{Frank
  Ortmeier}, \bibinfo{person}{Friedemann Bitsch}, {and} \bibinfo{person}{Pedro
  Ferreira}} (Eds.), Vol.~\bibinfo{volume}{12234}.
  \bibinfo{publisher}{Springer}, \bibinfo{pages}{244--259}.
\newblock
\showISBNx{978-3-030-54549-9}
\urldef\tempurl%
\url{https://doi.org/10.1007/978-3-030-54549-9_16}
\showDOI{\tempurl}


\bibitem[\protect\citeauthoryear{Zhao, Calinescu, Gerasimou, Robu, and
  Flynn}{Zhao et~al\mbox{.}}{2020b}]%
        {zhao_interval_2020}
\bibfield{author}{\bibinfo{person}{Xingyu Zhao}, \bibinfo{person}{Radu
  Calinescu}, \bibinfo{person}{Simos Gerasimou}, \bibinfo{person}{Valentin
  Robu}, {and} \bibinfo{person}{David Flynn}.}
  \bibinfo{year}{2020}\natexlab{b}.
\newblock \showarticletitle{Interval {Change}-{Point} {Detection} for {Runtime}
  {Probabilistic} {Model} {Checking}}. In \bibinfo{booktitle}{\emph{Proc. of
  the 35th {IEEE}/{ACM} {Int}. {Conf}. on {Automated} {Software}
  {Engineering}}} \emph{(\bibinfo{series}{{ASE}'20})}.
  \bibinfo{publisher}{ACM}, \bibinfo{pages}{163--174}.
\newblock
\urldef\tempurl%
\url{https://doi.org/10.1145/3324884.3416565}
\showDOI{\tempurl}


\bibitem[\protect\citeauthoryear{Zhao, Huang, Banks, Cox, Flynn, Schewe, and
  Huang}{Zhao et~al\mbox{.}}{2021a}]%
        {zhao_assessing_2021}
\bibfield{author}{\bibinfo{person}{Xingyu Zhao}, \bibinfo{person}{Wei Huang},
  \bibinfo{person}{Alec Banks}, \bibinfo{person}{Victoria Cox},
  \bibinfo{person}{David Flynn}, \bibinfo{person}{Sven Schewe}, {and}
  \bibinfo{person}{Xiaowei Huang}.} \bibinfo{year}{2021}\natexlab{a}.
\newblock \showarticletitle{Assessing the {Reliability} of {Deep} {Learning}
  {Classifiers} {Through} {Robustness} {Evaluation} and {Operational}
  {Profiles}}. In \bibinfo{booktitle}{\emph{{AISafety}'21 {Workshop} at
  {IJCAI}'21}}, Vol.~\bibinfo{volume}{2916}.
\newblock


\bibitem[\protect\citeauthoryear{Zhao, Huang, Schewe, Dong, and Huang}{Zhao
  et~al\mbox{.}}{2021b}]%
        {zhao_detecting_2021}
\bibfield{author}{\bibinfo{person}{Xingyu Zhao}, \bibinfo{person}{Wei Huang},
  \bibinfo{person}{Sven Schewe}, \bibinfo{person}{Yi Dong}, {and}
  \bibinfo{person}{Xiaowei Huang}.} \bibinfo{year}{2021}\natexlab{b}.
\newblock \showarticletitle{Detecting Operational Adversarial Examples for
  Reliable Deep Learning}. In \bibinfo{booktitle}{\emph{51th Annual IEEE-IFIP
  Int. Conf. on Dependable Systems and Networks (DSN'21)}},
  Vol.~\bibinfo{volume}{Fast Abstract}.
\newblock


\bibitem[\protect\citeauthoryear{Zhao, Littlewood, Povyakalo, Strigini, and
  Wright}{Zhao et~al\mbox{.}}{2017}]%
        {zhao_modeling_2017}
\bibfield{author}{\bibinfo{person}{Xingyu Zhao}, \bibinfo{person}{Bev
  Littlewood}, \bibinfo{person}{Andrey Povyakalo}, \bibinfo{person}{Lorenzo
  Strigini}, {and} \bibinfo{person}{David Wright}.}
  \bibinfo{year}{2017}\natexlab{}.
\newblock \showarticletitle{Modeling the probability of failure on demand (pfd)
  of a 1-out-of-2 system in which one channel is ``quasi-perfect''}.
\newblock \bibinfo{journal}{\emph{Reliability Engineering \& System Safety}}
  \bibinfo{volume}{158} (\bibinfo{year}{2017}), \bibinfo{pages}{230--245}.
\newblock


\bibitem[\protect\citeauthoryear{Zhao, Robu, Flynn, Dinmohammadi, Fisher, and
  Webster}{Zhao et~al\mbox{.}}{2019a}]%
        {zhao_probabilistic_2019}
\bibfield{author}{\bibinfo{person}{Xingyu Zhao}, \bibinfo{person}{Valentin
  Robu}, \bibinfo{person}{David Flynn}, \bibinfo{person}{Fateme Dinmohammadi},
  \bibinfo{person}{Michael Fisher}, {and} \bibinfo{person}{Matt Webster}.}
  \bibinfo{year}{2019}\natexlab{a}.
\newblock \showarticletitle{Probabilistic model checking of robots deployed in
  extreme environments}. In \bibinfo{booktitle}{\emph{Proc. of the {AAAI}
  {Conference} on {Artificial} {Intelligence}}}, Vol.~\bibinfo{volume}{33}.
  \bibinfo{address}{Honolulu, Hawaii, USA}, \bibinfo{pages}{8076--8084}.
\newblock


\bibitem[\protect\citeauthoryear{Zhao, Robu, Flynn, Salako, and Strigini}{Zhao
  et~al\mbox{.}}{2019b}]%
        {zhao_assessing_2019}
\bibfield{author}{\bibinfo{person}{Xingyu Zhao}, \bibinfo{person}{Valentin
  Robu}, \bibinfo{person}{David Flynn}, \bibinfo{person}{Kizito Salako}, {and}
  \bibinfo{person}{Lorenzo Strigini}.} \bibinfo{year}{2019}\natexlab{b}.
\newblock \showarticletitle{Assessing the safety and reliability of autonomous
  vehicles from road testing}. In \bibinfo{booktitle}{\emph{the 30th {Int}.
  {Symp}. on {Software} {Reliability} {Engineering}}}.
  \bibinfo{publisher}{IEEE}, \bibinfo{address}{Berlin, Germany},
  \bibinfo{pages}{13--23}.
\newblock


\bibitem[\protect\citeauthoryear{Zhao, Salako, Strigini, Robu, and Flynn}{Zhao
  et~al\mbox{.}}{2020c}]%
        {zhao_assessing_2020}
\bibfield{author}{\bibinfo{person}{Xingyu Zhao}, \bibinfo{person}{Kizito
  Salako}, \bibinfo{person}{Lorenzo Strigini}, \bibinfo{person}{Valentin Robu},
  {and} \bibinfo{person}{David Flynn}.} \bibinfo{year}{2020}\natexlab{c}.
\newblock \showarticletitle{Assessing safety-critical systems from operational
  testing: {A} study on autonomous vehicles}.
\newblock \bibinfo{journal}{\emph{Information and Software Technology}}
  \bibinfo{volume}{128} (\bibinfo{year}{2020}), \bibinfo{pages}{106393}.
\newblock
\showISSN{0950-5849}


\bibitem[\protect\citeauthoryear{Zhong, Tian, and Ray}{Zhong
  et~al\mbox{.}}{2021}]%
        {zhong_understanding_2021}
\bibfield{author}{\bibinfo{person}{Ziyuan Zhong}, \bibinfo{person}{Yuchi Tian},
  {and} \bibinfo{person}{Baishakhi Ray}.} \bibinfo{year}{2021}\natexlab{}.
\newblock \showarticletitle{Understanding {Local} {Robustness} of {Deep}
  {Neural} {Networks} under {Natural} {Variations}}. In
  \bibinfo{booktitle}{\emph{Fundamental {Approaches} to {Software}
  {Engineering}}} \emph{(\bibinfo{series}{{LNCS}})},
  \bibfield{editor}{\bibinfo{person}{Esther Guerra} {and}
  \bibinfo{person}{Mariëlle Stoelinga}} (Eds.), Vol.~\bibinfo{volume}{12649}.
  \bibinfo{publisher}{Springer International Publishing},
  \bibinfo{address}{Cham}, \bibinfo{pages}{313--337}.
\newblock
\showISBNx{978-3-030-71500-7}


\end{thebibliography}
\appendix

\section{\gls{KDE} Bootstrapping}
\label{sec_app_A}
Bootstrapping is a statistical approach to estimate any sampling distribution by a random sampling method. 
We sample with replacement from the original data points $(X,Y)$ to obtain a new bootstrap dataset $(X^b, Y^b)$ and train the \gls{KDE} on the bootstrap dataset. 
Assume the bootstrapping process is repeated $B$ times, leading to $B$ bootstrap \gls{KDE}s, denoted as
$\widehat{\mathsf{Op}}^1(x), \dots, \widehat{\mathsf{Op}}^B(x)$.
Then we can estimate the variance of $\hat{f}(x)$ by the sample variance of the bootstrap \gls{KDE} \cite{chen2017tutorial}:
$$
\label{eq_bootstrap_var}
\hat{\sigma}_B^2(x) = \frac{1}{B-1} \sum_{b = 1}^B (\widehat{\mathsf{Op}}^b(x) - \mu_B)^2 \ ,
$$
where the $\mu_B$ can be approximated by
$$
\hat{\mu}_B(x) = \frac{1}{B} \sum_{b = 1}^B \widehat{\mathsf{Op}}^b(x) \ .
$$

\section{Experiment Setup}
In this section, the details of different experiments are summarised. All the cases are performed using Python on a computer equipped with an AMD core $EPYC\ 7452$ and NVIDIA's GPU $A100$.

\subsection{Details of the MNIST and CIFAR10 Models}
\label{sec:mnist_cifar_model_details}
\begin{figure} [ht]
    \centering
    \begin{minipage}[b]{0.4\textwidth}\centering%
        \includegraphics[width=\textwidth]{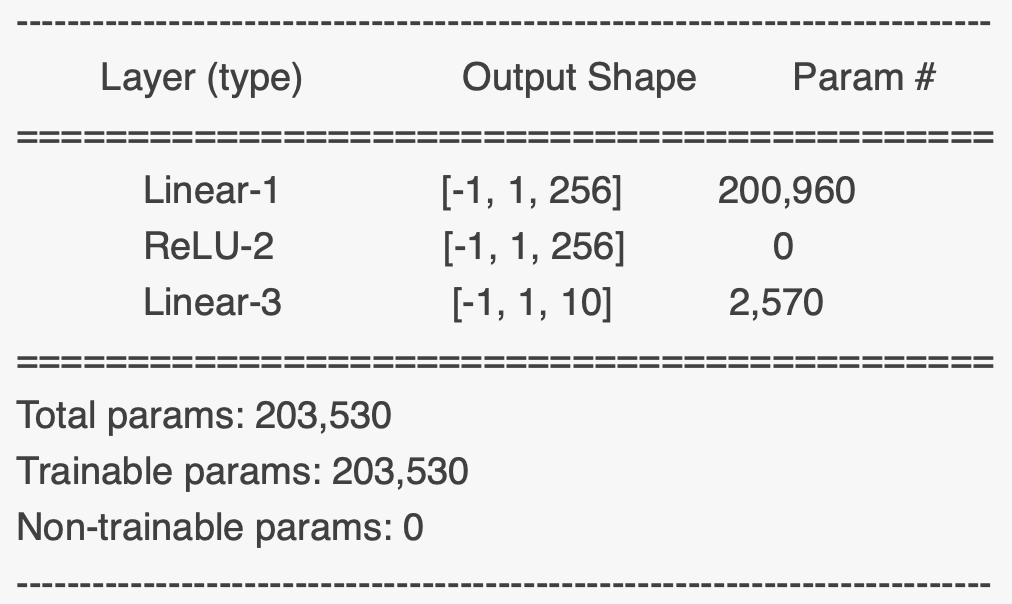}
        \caption{\textcolor{black}{The architecture of DNN trained on MNIST dataset}}
         \label{mnist_model_details}
    \end{minipage}\hfill
    \begin{minipage}[b]{0.55\textwidth}\centering%
        \includegraphics[width=\textwidth]{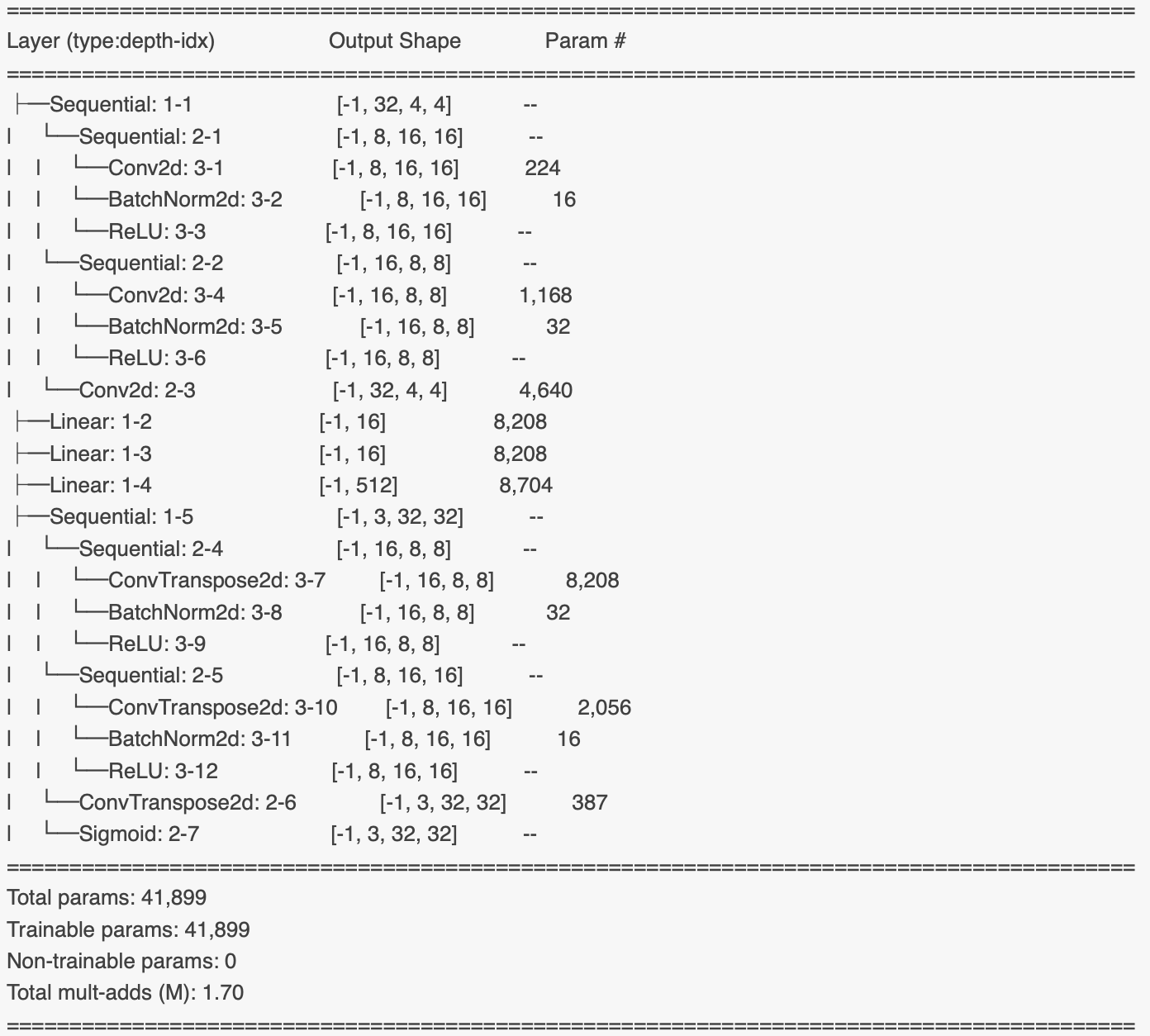} 
        \caption{\textcolor{black}{The architecture of convolutional variational autoencoder (VAE)}}
        \label{vae_model_details}
    \end{minipage}
\end{figure}

\begin{figure*}[ht]
	\centering
	\includegraphics[width=\linewidth]{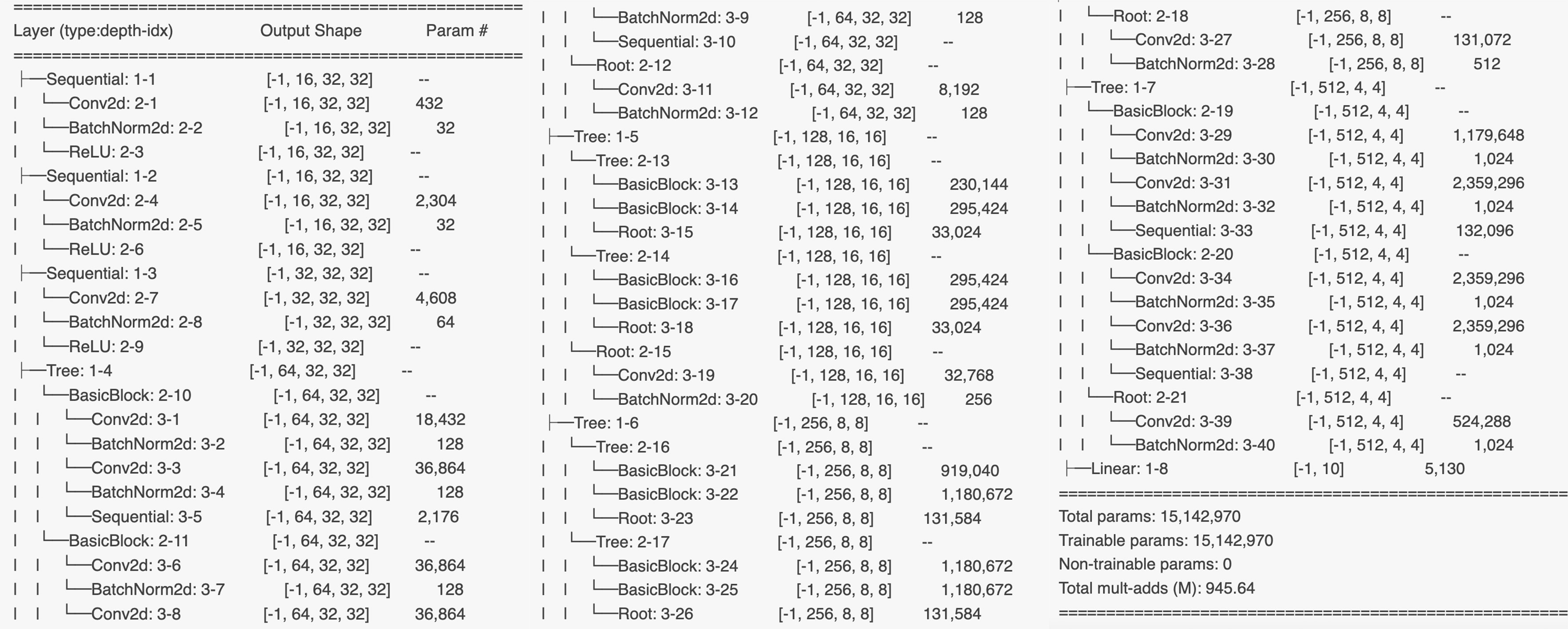}
	\caption{\textcolor{black}{The architecture of Deep Layer Aggregation model trained on CIFAR10 dataset}}
	\label{cifar10_model_details}
\end{figure*}

Fig.~\ref{mnist_model_details} and \ref{cifar10_model_details} present the architecture of DNNs trained on MNIST and CIFAR10 datasets. The Deep Layer Aggregation (DLA) model \cite{yu2018deep} is especially used for the CIFAR10 dataset to achieve very high train and test accuracy.

MNIST dataset contains 70000 handwritten digit images, among which 60000 images are used for training while 10000 are used for testing the model's generalisation. We use Adam optimiser with learning rate $10^{-3}$, cross-entropy loss as training loss function and train 20 epochs to get the result.

CIFAR10 dataset contains 60000 32x32 colour images in 10 classes, with 6000 images per class. There are 50000 train images and 10000 test images. We use Stochastic gradient descent (SGD) optimiser with a learning rate of 0.01, momentum 0.9, and weight decay $5\times10^{-4}$ to train the model with 50 epochs.

\textcolor{black}{We adopt adversarial training to obtain a more robust model for reliability comparison. We use PGD-based adversarial training \cite{madry2018towards} with the same training parameters as normal training to generate attacked data (AEs when successfully attacked) in replacement of all original training data to improve robustness. To be specific, the PGD attack calculates the gradient of training loss\footnote{We usually use cross entropy loss for classification problems. The training loss for object detection models like YoloV3 includes the objectness score of a bounding box.} with respect to the input. It is deployed with 10 steps and 2/255 step size to generate pixel-level perturbation for training data. The perturbed training data will be utilised over the whole training process.}

We use the VAE model to reduce the dimension of data, the architecture of which is shown in Fig.~\ref{vae_model_details}. To train the VAE model, we use an Adam optimiser with a learning rate $10^{-3}$ and train 100 epochs.

\subsection{Details of the YOLOv3 and VAE Models Trained in the Case Studies}
\label{sec_app_B}
\begin{figure}[!h]
     \centering
     \begin{subfigure}[b]{0.48\textwidth}
         \centering
         \includegraphics[width=\textwidth]{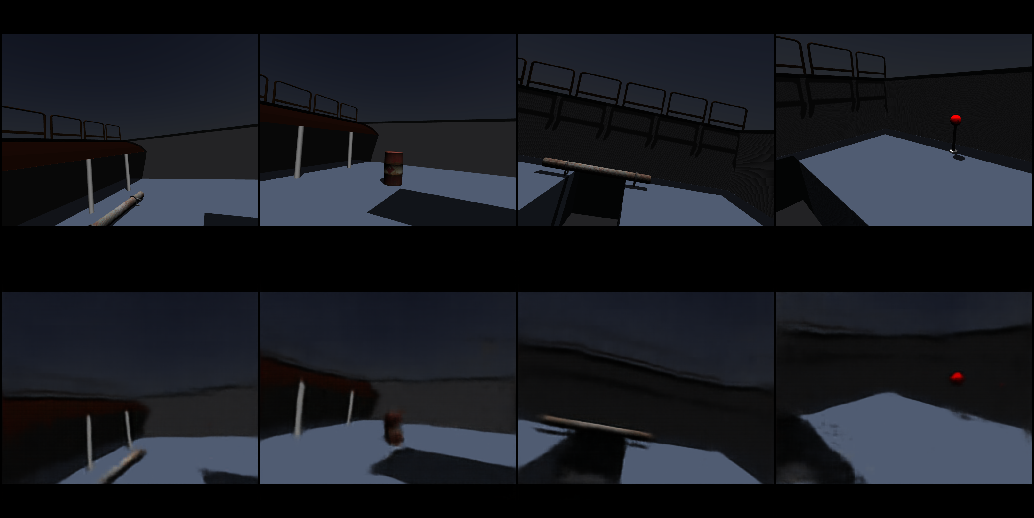}
         \caption{AUV}
         \label{fig:auv_recon}
     \end{subfigure}
     \hfill
     \begin{subfigure}[b]{0.48\textwidth}
         \centering
         \includegraphics[width=\textwidth]{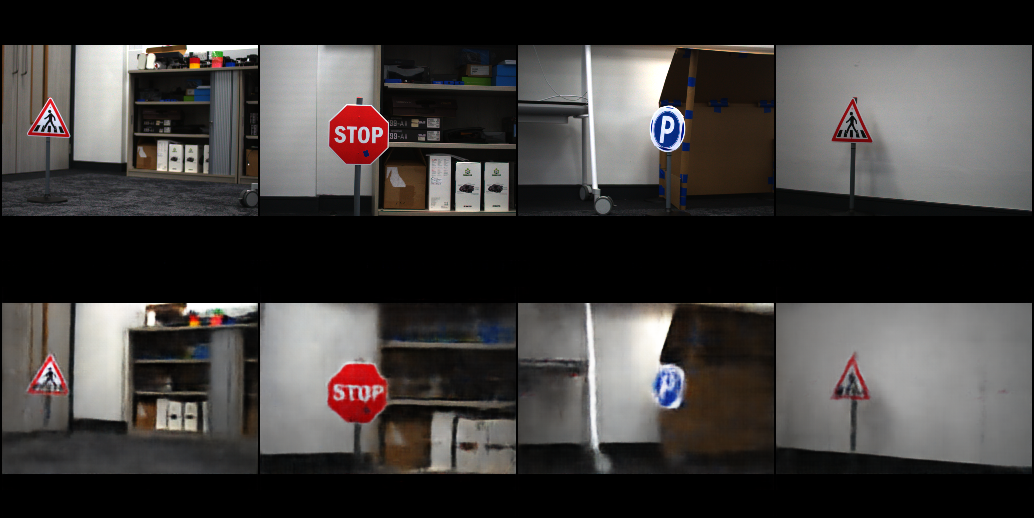}
         \caption{\textcolor{black}{UGV}}
         \label{fig:ugv_recon}
     \end{subfigure}
        \caption{Original images (top) and corresponding reconstructed images (bottom) by the VAE model.}
        \label{fig_vae_examples}
\end{figure}

We present more details of the YOLOv3 and VAE models trained in the AUV and UGV case studies, respectively in Table \ref{tab_yolov3_performance} and \ref{tab_VAE_performance}, while in Fig. \ref{fig_vae_examples} eight images reconstructed based on the VAE models are shown as examples. For more detailed \gls{ML} models, datasets and experimental results, please refer to our public repository \url{https://github.com/Solitude-SAMR}.


\begin{table}[!h]
    \begin{subtable}[h]{0.47\textwidth}
        \centering
        \begin{tabular}{ccccc}
			\hline
			\multirow{2}{*}{Class} & \multicolumn{2}{c}{Normal Training} & \multicolumn{2}{c}{Adversarial Training} \\ \cline{2-5} 
			& Train & Test & Train & Test \\ \hline
			Pipe & 0.98343 & 0.97131 & 0.92225 & 0.91380 \\
			Floating Ball & 0.85765 & 0.90912 & 0.83785 & 0.88291 \\
			Gas Canister & 0.87230 & 0.87406 & 0.83245 & 0.84319 \\
			Gas Tank & 0.98930 & 0.99346 & 0.93623 & 0.93769 \\
			Oil Barrel & 0.84578 & 0.84258 & 0.82811 & 0.82973 \\
			Docking Cage & 0.88771 & 0.91076 & 0.72786 & 0.71478 \\
			mAP & 0.90603 & 0.91688 & 0.84746 & 0.85368 \\ \hline
		\end{tabular}
       \caption{AUV}
       \label{tab:week1}
    \end{subtable}
    \hfill
    \begin{subtable}[h]{0.47\textwidth}
        \centering
        \begin{tabular}{ccccc}
			\hline
			\multirow{2}{*}{Class} & \multicolumn{2}{c}{Normal Training} & \multicolumn{2}{c}{Adversarial Training} \\ \cline{2-5} 
			& Train & Test & Train & Test \\ \hline
			Stop Sign & 0.97958 & 0.99471 & 0.76873 & 0.79065 \\
			Park Sign & 0.98182 & 0.98158 & 0.84459 & 0.82143 \\
			Cycle Sign & 0.99911 & 0.98648 & 0.92750 & 0.88669 \\
			Cross Sign & 0.99067 & 0.98878 & 0.98850 & 0.98496 \\
			One Way Sign & 0.98921 & 0.99023 & 0.53972 & 0.52041 \\
			mAP & 0.98808 & 0.98836 & 0.81381 & 0.80083 \\ \hline
		\end{tabular}
        \caption{\textcolor{black}{UGV}}
        \label{tab:week2}
     \end{subtable}
     \caption{Average Precision (AP) of YOLOv3 for object detection.}
     \label{tab_yolov3_performance}
\end{table}

\begin{table}[!h]
\begin{subtable}[h]{0.45\textwidth}
	\centering
		\begin{tabular}{ccc}
			\hline
			VAE model & Train & Test \\ \hline
			Recon. Loss & 0.002601 & 0.003048 \\
			KL Div. Loss & 1.732866 & 1.729756 \\ \hline
		\end{tabular}
	\caption{AUV}
\end{subtable}
\hfill
\begin{subtable}[h]{0.45\textwidth}
	\centering
		\begin{tabular}{ccc}
			\hline
			VAE model & Train & Test \\ \hline
			Recon. Loss & 0.011333 & 0.016665 \\
			KL Div. Loss & 4.781964 & 4.810094 \\ \hline
		\end{tabular}
	\caption{\textcolor{black}{UGV}}
\end{subtable}
\caption{Reconstruction Loss \& KL Divergence Loss of VAE model}
\label{tab_VAE_performance}
\end{table}

\section{Comparison between RAM and state-of-the-art Robustness Metrics}
\label{sec_app_ram_comparison}
\begin{table}[ht]
\centering
\caption{Local robustness evaluation around a selected test seed}
\begin{tabular}{cccccccc}
\hline
\multirow{2}{*}{} & \multirow{2}{*}{Model} & \multicolumn{2}{c}{RAM} & \multicolumn{2}{c}{DeepFool} & \multicolumn{2}{c}{ERAN (AI2)} \\ \cline{3-8} 
 &  & Radius $\epsilon$ & $\hat{\lambda}_i$ & Radius $\epsilon$ & \begin{tabular}[c]{@{}c@{}}Minimal\\ Perturbation\end{tabular} & Radius $\epsilon$ & \begin{tabular}[c]{@{}c@{}}Verified \\ Robustness\end{tabular} \\ \hline
\multirow{2}{*}{MNIST} & Norm. & 0.3 & $5.85 \times 10^{-19}$ & / & 0.3151 & 0.3 & 0 \\
 & Adv. & 0.3 & $1.93 \times 10^{-22}$ & / & 0.7912 & 0.3 & 0 \\
\multirow{2}{*}{CIFAR-10} & Norm. & 0.1 & $0.4177$ & / & 0.0189 & 0.1 & 0 \\
 & Adv. & 0.1 & $1.92 \times 10^{-22}$ & / & 0.2145 & 0.1 & 0 \\ \hline
\end{tabular}
\label{metrics_comp_local}
\end{table}

\begin{table}[ht]
\centering
\caption{Network-wise reliability and robustness evaluation}
\begin{tabular}{ccclcclc}
\hline
\multirow{2}{*}{} & \multirow{2}{*}{Model} & \multicolumn{3}{c}{RAM} & \multicolumn{3}{c}{DeepFool} \\ \cline{3-8} 
 &  & Radius $\epsilon$ & $k$ & \textit{pmi} & Radius $\epsilon$ & $k$ & \begin{tabular}[c]{@{}c@{}}Avg. Minimal\\ Perturbation\end{tabular} \\ \hline
\multirow{2}{*}{MNIST} & Norm. & 0.3 & 70000 & 4.74 e-4 & / & 70000 & 0.1979 \\
 & Adv. & 0.3 & 70000 & 1.77 e-4 & / & 70000 & 0.5141 \\
\multirow{2}{*}{CIFAR-10} & Norm. & 0.1 & 25000 & 0.3152 & / & 60000 & 0.0318 \\
 & Adv. & 0.1 & 25000 & 0.0968 & / & 60000 & 0.3368 \\ \hline
\end{tabular}
\label{metrics_comp_global}
\end{table}

\textcolor{black}{As shown in Table~\ref{metrics_comp_local}, as before, we study both normally and adversarially trained DL models on the two datasets. In our RAM, the norm ball radius $\epsilon$ is determined by the $r$-separation distance and the local probabilistic robustness evaluation is returned by the estimator \cite{webb_statistical_2019}. For a fair comparison, we reuse the same norm ball for applying AI2 which indeed returns the answer ``unverified'' to the binary question. In contrast, DeepFool aims at finding a minimal perturbation to detect AEs around the same seed. Although DeepFool formulates it as an optimisation problem, its solution has no guarantees to be optimal---the returned solutions are larger than the $r$-separation distance when the AEs are quite rare (the 1st, 2nd and 4th rows). While, at the global level of networks, Table~\ref{metrics_comp_global} ``compares'' the reliability metric $pmi$ of our RAM with the \textit{averaged} minimal perturbation by applying DeepFool. We omit the experiments based on AI2 at such a network-wise level due to the high computational cost of formally verifying all $k$ norm balls. Both the assessments by RAM and DeepFool are reflective of the robustness improvement from a normally trained model to an adversarially trained model. The averaged minimal perturbation derived by DeepFool (or other similar tools concerning worst-case metrics) has limited practical values \cite{wang_statistically_2021,webb_statistical_2019}.
At the same time,  our reliability \gls{pmi} concerns the overall robustness and generalisability, that predicts the probability of failure for the next input drawn from the OP. 
}

\section{Additional Case Study on Healthcare Models}\label{sec_app_C}
$CORONET$\footnote{\url{https://coronet.manchester.ac.uk/}} is a decision-support tool validated in healthcare systems worldwide can aid admission decisions and predict COVID-19 severity in patients with cancer \cite{lee2022establishment}. The dependability of healthcare AI/ML systems needs to be assured to increase their trustworthiness. In this section, a case of applying our approach on $CORONET$ is studied to validate our reliability assessment methods.
 
\begin{figure}[htbp]
     \centering
     \begin{subfigure}[b]{0.53\textwidth}
         \centering
         \includegraphics[width=0.8\textwidth]{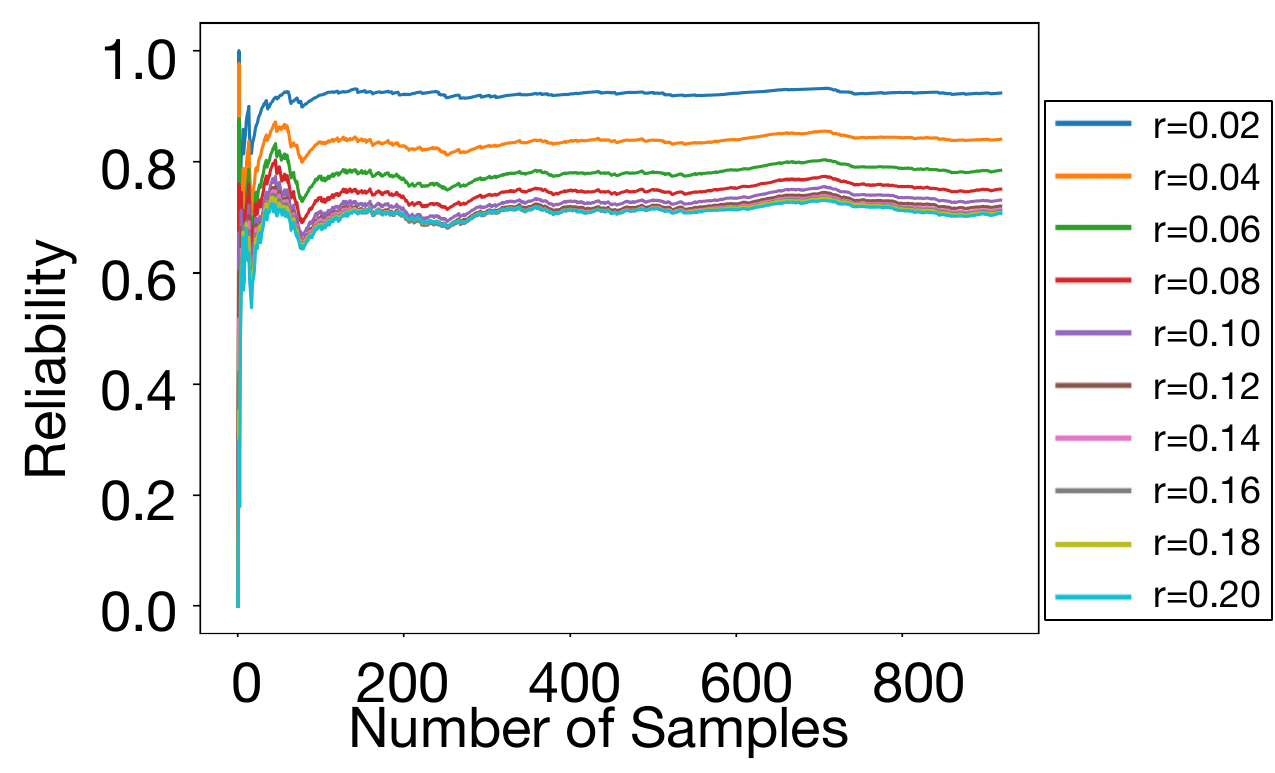}
         \caption{\textcolor{black}{PMI of CORONET dataset with different norm ball radius.}}
         \label{coronet}
     \end{subfigure}
     \hfill
     \begin{subfigure}[b]{0.44\textwidth}
         \centering
         \includegraphics[width=0.8\textwidth]{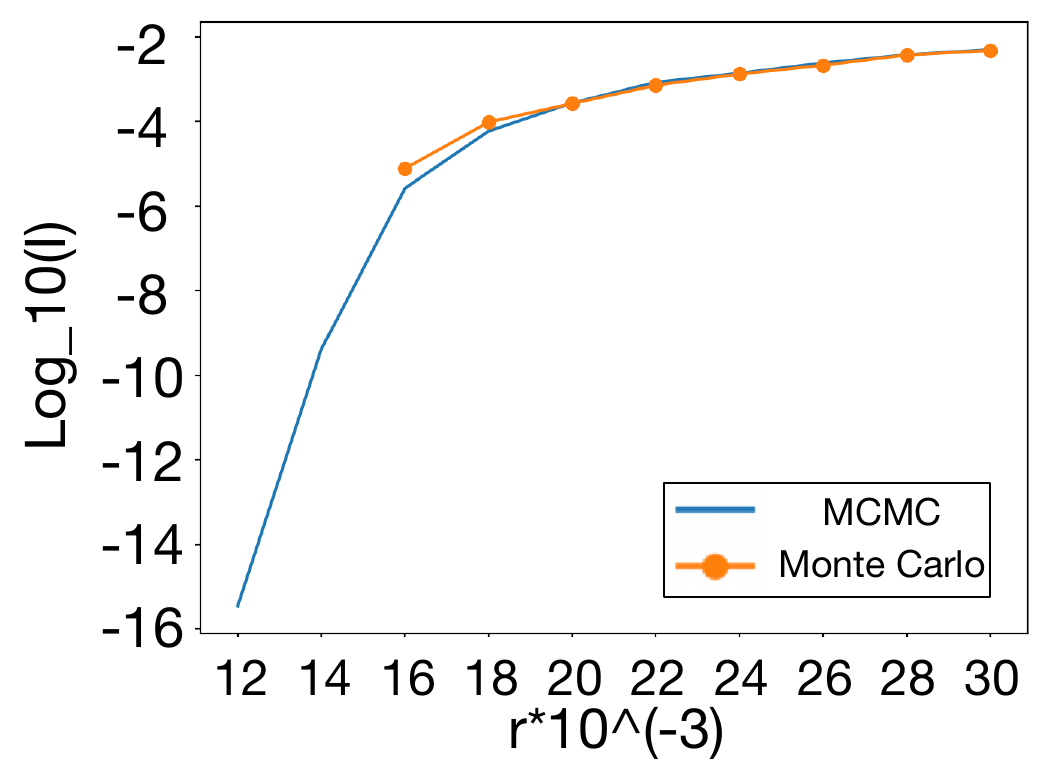}
         \caption{\textcolor{black}{Comparison between the proposed method and Monte-Carlo sampling.}}
         \label{compareMC}
     \end{subfigure}
        \caption{\textcolor{black}{Experiment results of CORONET.}}
\end{figure}

The $CORONET$ model is a decision tree model with 11 dimensions inputs, and three different output labels. Here, we first assess the reliability of a given $CORONET$ model among different $r$-separation radius. The experiment result is shown in Fig. \ref{coronet}.
 
As can be seen from Fig. \ref{coronet}, the reliability of the $CORONET$ model decreases as the disturbance radius increases, which is in accordance with our intuition. The higher disturbance would cause more uncertainty.

To further evaluate the effectiveness of our reliability assessment model, we compare our method with naive Monte-Carlo sampling methods. The results are shown in Fig. \ref{compareMC}.
 
In this experiment, we collect $10^5$ samples based on naive Monte-Carlo sampling method to assess the reliability of the $CORONET$ prediction model. From the experiment results, we can see that the proposed method has an equivalent estimate to the unbiased estimation results of Monte-Carlo sampling. Furthermore, the proposed method can make it much easier to find rare events than the naive Monte Carlo method when the disturbance radius is tiny.

\section{Probabilistic Safety Arguments for \gls{ML} Components}
\label{sec_prob_safe_argument}

At this lower level of ML components, cf.~the \textbf{SubC7} in Fig. \ref{fig_overview_ac}, we further decompose and organise our safety arguments in two levels---\textit{decomposing sub-functionalities of ML components doing object detection} and \textit{claiming the reliability of the classification function}. In the following sections, we discuss both in details, while focusing more on the latter.

\subsection{Arguments for Top Claims on Object Detection  at the \gls{ML} Component-Level}
\label{sec_prob_safe_argu_top_ml_claims}

\begin{figure*}[h!]
	\centering
	\includegraphics[width=0.9\textwidth]{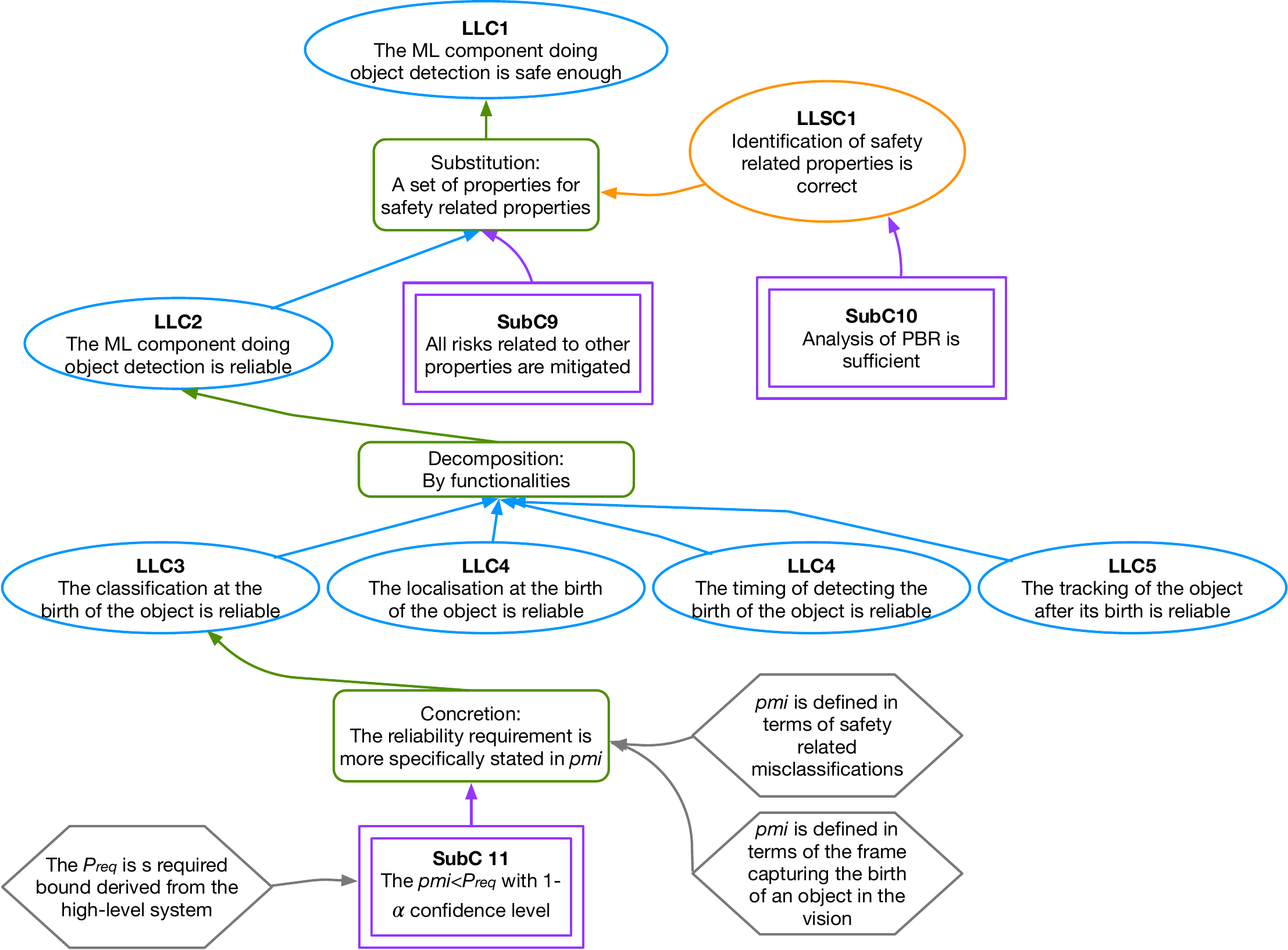}
	\caption{ML component-level arguments breaking down the claim ``The object detection component is safe enough'' \textbf{LLC1} to reliability claims of the classification function stated in specific reliability measures \textbf{SubC11}.}
	\label{fig_cae_low_level_top_claims}
\end{figure*}

In Fig. \ref{fig_cae_low_level_top_claims}, we present an argument template, again in the CAE blocks at the ML component-level. 
It aims at breaking down the claim ``The object detection is safe enough'' \textbf{LLC1} to a reliability claim stated in the specified measure.
The first argument is over all safety related properties, and presented by a CAE block of substitution. 
The list of all properties of interest for the given application can be obtained by utilising the Property Based Requirements (PBR) approach \cite{Micouin2008}, forming the side-claim \textbf{LLSC1}, which is supported by the sub-case \textbf{SubC10}.
The PBR analysis, recommended in \cite{alves_considerations_2018} as a method for safety arguments of autonomous systems, is a way to specify requirements as a set of properties of system objects either in a structured language or formal notations.
In this work, we focus on the main quantitative property---reliability---while other properties like security and interpretability are omitted and remain an undeveloped sub-case \textbf{SubC9} in the CAE template.

Starting from \textbf{LLC2}, we then argue over the decomposition by four sub-functionalities of object detection. At the ``birth'' of an object in the system's vision (e.g., the total number of pixels is greater than a threshold), the ML component should accurately classify it, localise it (normally measured by the Intersection over Union (IoU) of bounding boxes) and in good timing (e.g., no later than some frames after its birth).
Once initially detected at its birth time, the tracking function on that object should be reliable enough to make decision making by other control components safe.
The four sub-functionalities of object detection forms the claims \textbf{LLC3}-\textbf{LLC5}.

To support the reliability of classification at the birth time of the objects \textbf{LLC3}, we concretise the reliability requirements in terms of specific reliability measures, in our case \textit{pmi}.
The ``misclassification'' and ``per input'' in \textit{pmi} need to be clearly defined:
(i) we only consider safety-related misclassification events; 
(ii) an input refers to the image frame capturing the ``birth'' of an object in the camera's vision (so that images can be treated as independent conforming to the definition of \textit{pmi}).
We are then interested in the claim of a bound on \textit{pmi} with $(1-\alpha)$ confidence, where $P_{req}$ is a required bound derived from higher level safety analysis.

While the reliability of the other three sub-functionalities can be similarly concretised by some quantitative measures, e.g.\ IoU for localisation, they remain undeveloped in this article and form important future work.

\subsection{Low-Level Arguments for Classification Based on the \gls{RAM}}
\label{sec_prob_safe_argu_low_ram_steps}

\begin{figure*}[h!]
	\centering
	\includegraphics[width=0.9\textwidth]{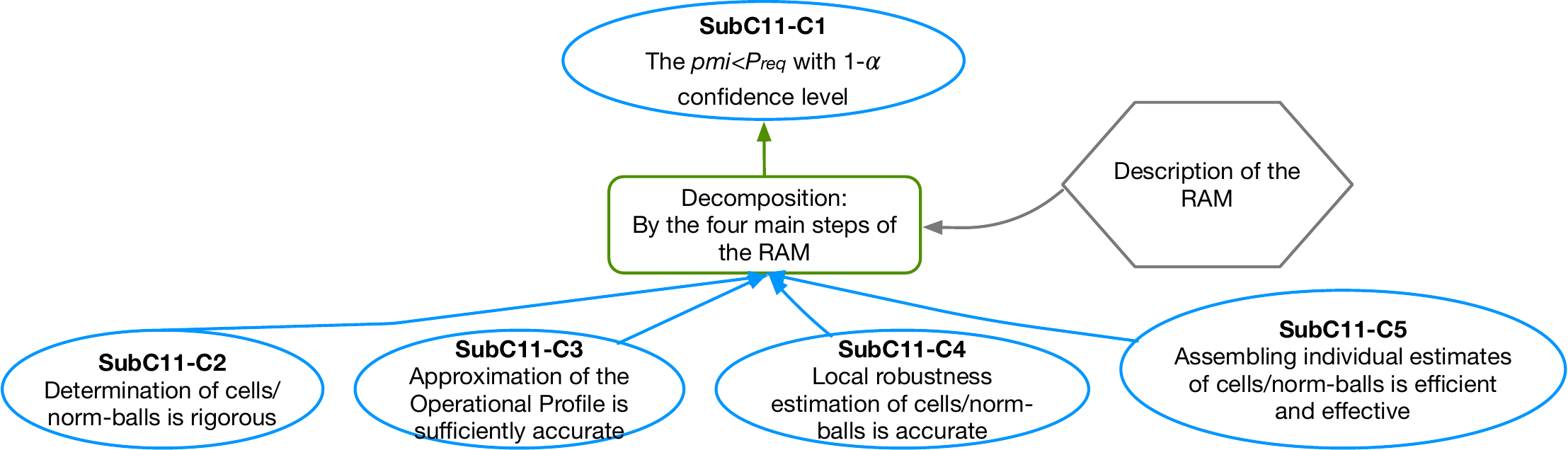}
	\caption{Arguments over the four main steps in the proposed RAM.}
	\label{fig_cae_decomp_ram_steps}
\end{figure*}

In this section, we present \textbf{SubC11} and show how to support a reliability claim stated in \textit{pmi} based on our RAM developed in Section \ref{sec_ram}---the ``backbone'' of the probabilistic arguments at this lower level.
Essentially, we argue over the four main steps of our RAM as shown in Fig. \ref{fig_cae_decomp_ram_steps}. 
Note that, depending on the data dimensionality of the specific application, we may either use the ``low-dimensional'' version of our \gls{RAM}, where the whole input space is partitioned into cells, or apply the ``high-dimensional'' version, in which norm balls (of relatively spare data) are determined instead (cf.\ Remark \ref{remark_map}) based on the collected data to form the sample frame (representing the population of all norm balls partitioning the whole input space). Indeed, the method of exhaustively partitioning cells is also applicable to high-dimensional data, but it would yield an extremely large number of cells that is not only infeasible to examine exhaustively but also quite difficult to index for sampling. That said, for high-dimensional datasets, we determine norm balls from the data instead, forming a smaller and more practical sampling frame. However, the price paid is at introducing two more noise factors in the assurance---the bias/error from the construction of the sampling frame and the relatively low sample rate. The former can be mitigated by conventional ways of checking (and rebuilding if necessary) the sampling frame, while the latter has been captured and quantified by the variance of the point estimate (cf. Eqn.~\eqref{eq_weighted_avg_est_var} and Assumption \ref{assumption_k_is_the_major_source_uncertainty}).

\begin{figure*}[h!]
	\centering
	\includegraphics[width=0.9\textwidth]{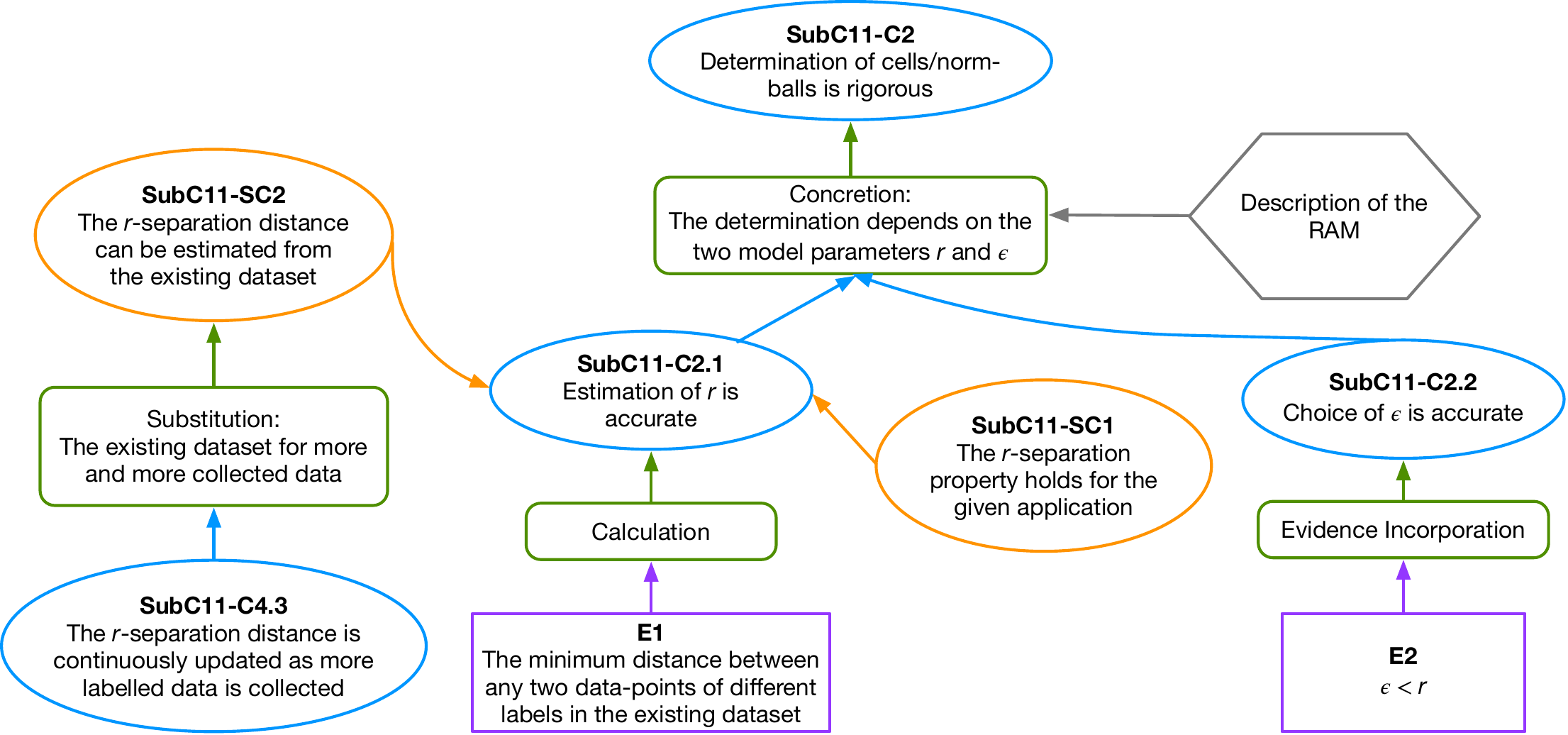}
	\caption{Arguments based on the step 1 of the RAM.}
	\label{fig_cae_step1}
\end{figure*}

\begin{figure*}[h!]
	\centering
	\includegraphics[width=0.85\textwidth]{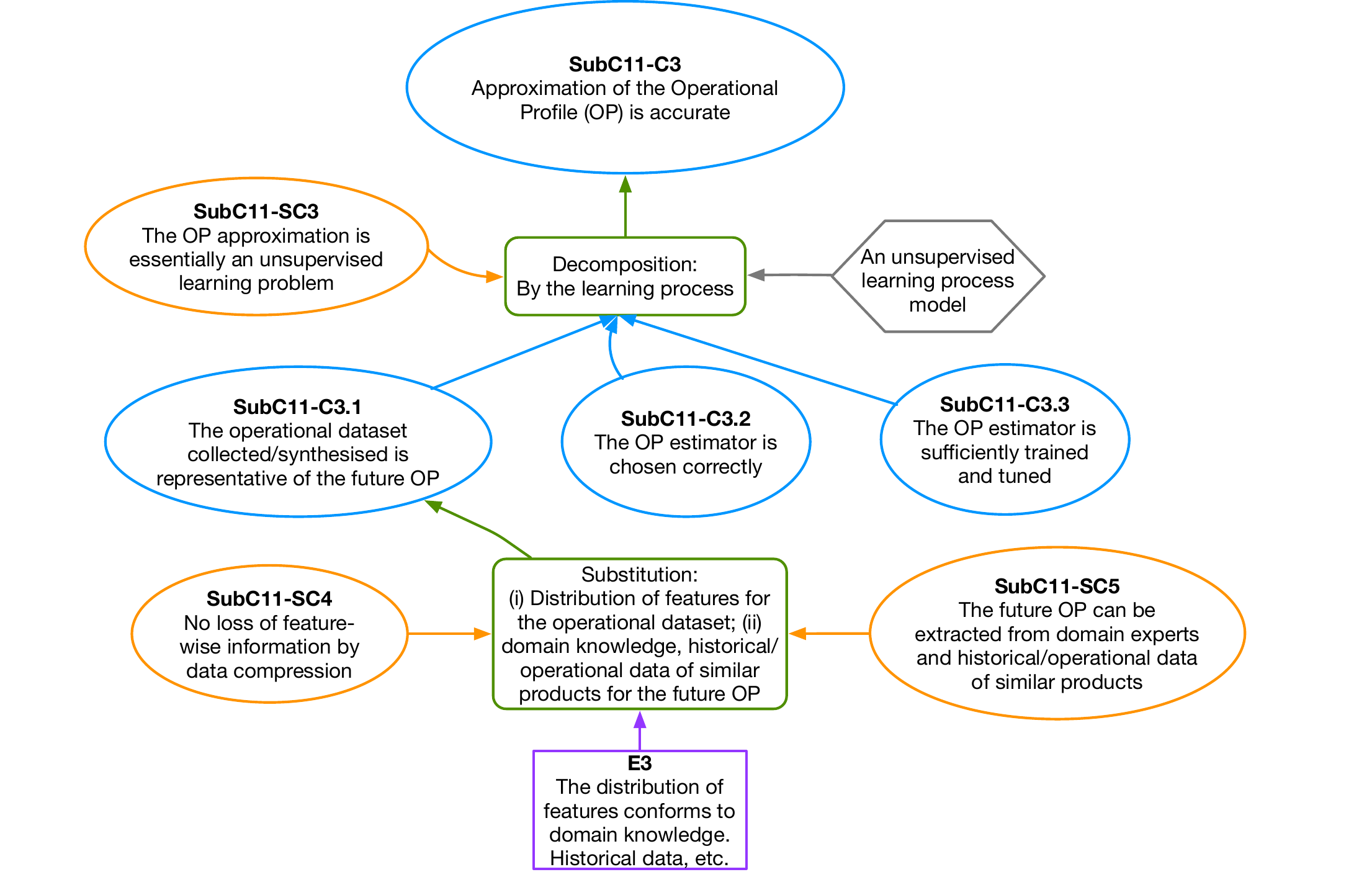}
	\caption{Arguments based on the step 2 of the RAM.}
	\label{fig_cae_step2}
\end{figure*}

\begin{figure*}[h!]
	\centering
	\includegraphics[width=0.85\textwidth]{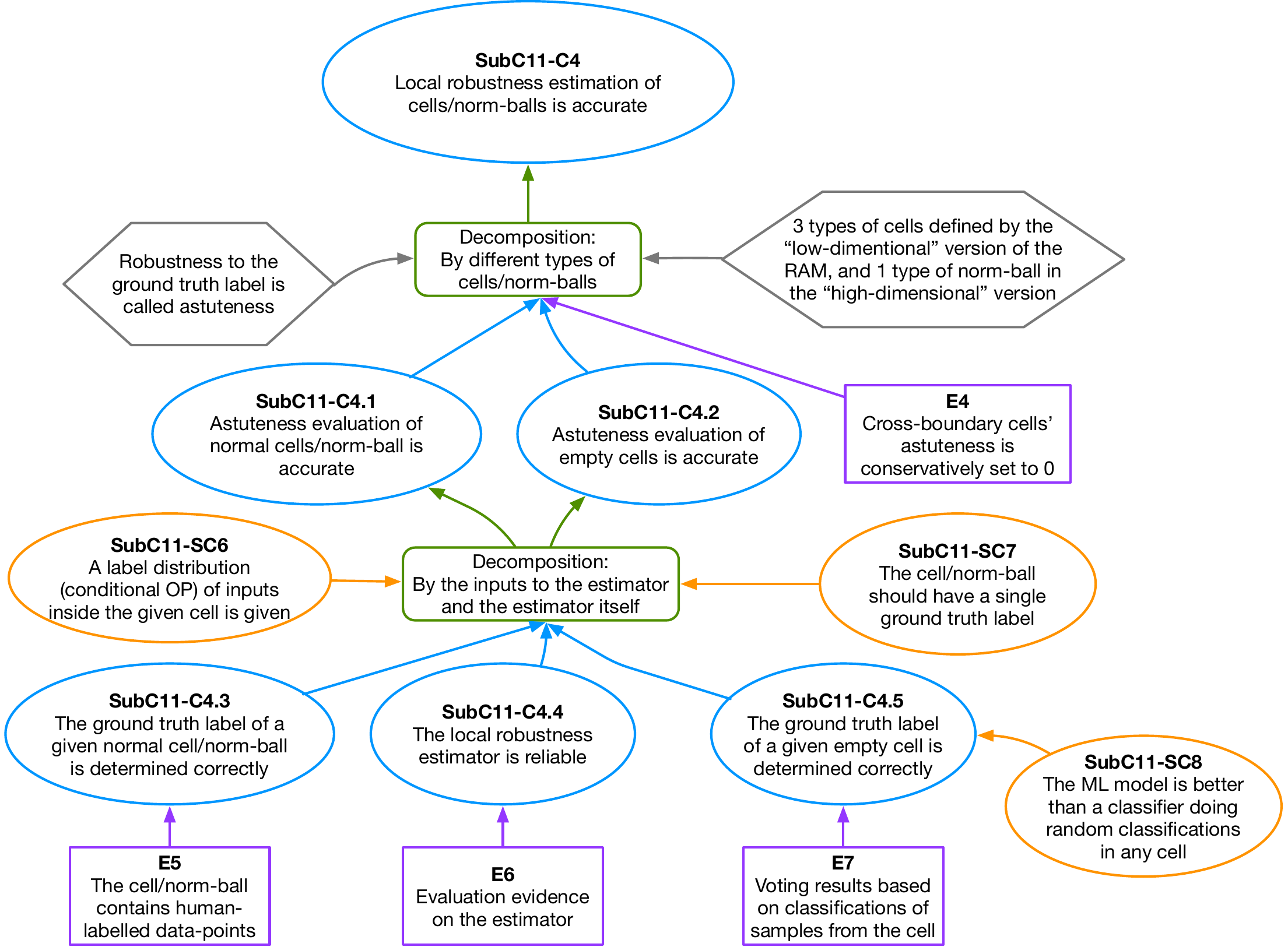}
	\caption{Arguments based on the step 3 of the RAM.}
	\label{fig_cae_step3}
\end{figure*}

\begin{figure*}[h!]
	\centering
	\includegraphics[width=0.9\textwidth]{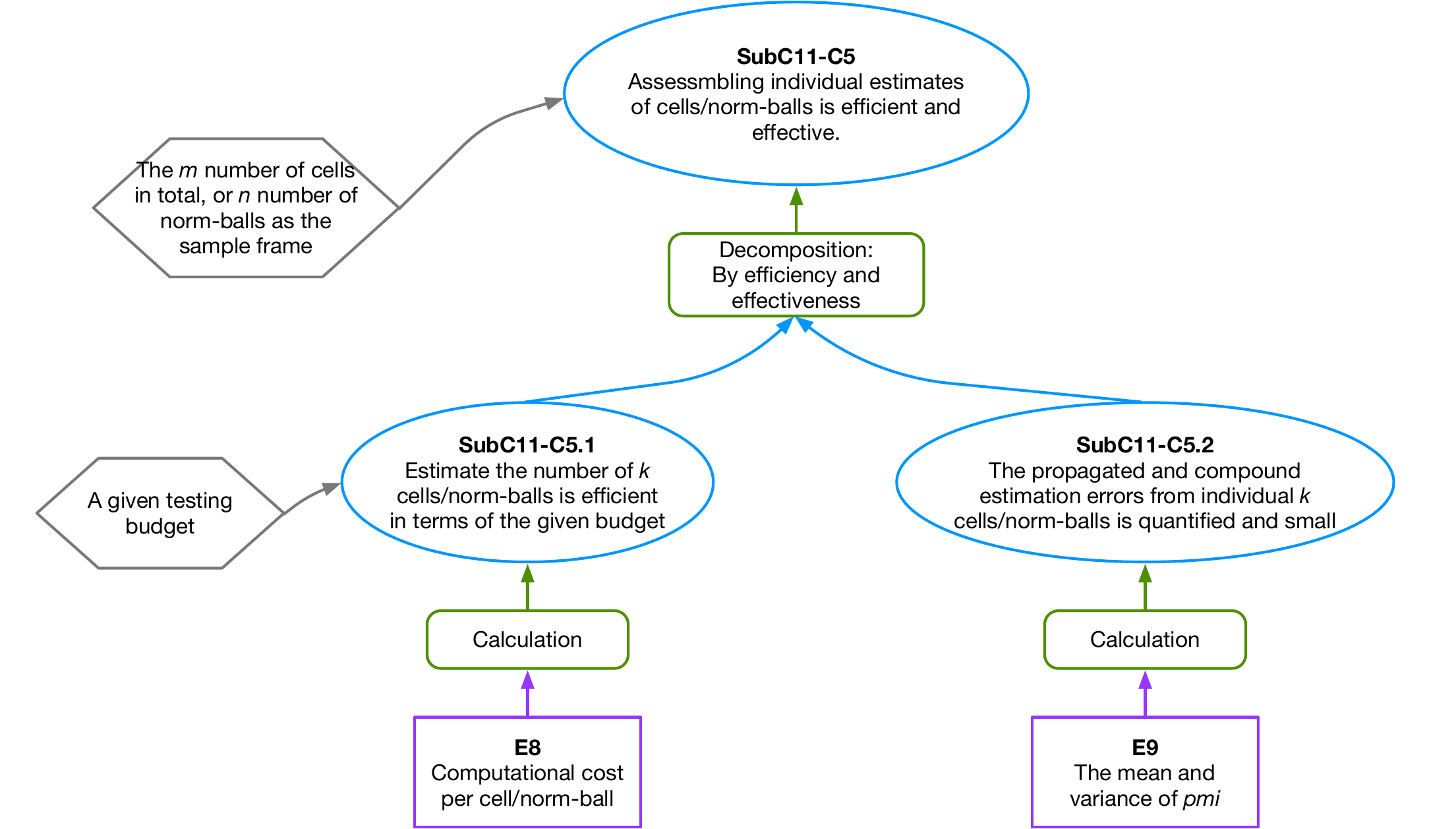}
	\caption{Arguments based on the step 4 of the RAM.}
	\label{fig_cae_step4}
\end{figure*}

Figures \ref{fig_cae_step1} to \ref{fig_cae_step4} show the arguments based on steps 1 to 4 of our RAM, respectively.
While the arguments presented in CAE are self-explanatory together with the technical details articulated in Section \ref{sec_ram}, we note that
i) all modelling assumptions are presented as side-claims of arguments that need more application-specific development and justification; and
ii) the development of some claims is omitted for brevity, because they are generic claims and thus can be referred to other works, e.g.\ \cite{ashmore2021assuring}, for \textbf{SubC11-C3.2} and \textbf{SubC11-C3.3} when we treat the OP estimator as a common data-driven learning model.

\end{document}
\endinput